HYBRID SIMULATION MODEL FOR FAILURE AWARENESS IN OPEN
SOFTWARE DEVELOPMENT

by

Razieh Lotfalian Saremi

A DISSERTATION

Submitted to Faculty of Stevens Institute of Technology
In partial fulfillment of the requirements for the degree of

DOCTOR OF PHILOSOPHY

_______________________________________
Razieh Lotfalian Saremi, Candidate        Date

<u>ADVISORY COMMITTEE</u>

_______________________________________
Dr. Ye Yang, ChaiPerson                   Date

_______________________________________
Dr. Somayeh Moazeni                       Date

_______________________________________
Dr. Guenther Ruhe                         Date

_______________________________________
Dr. Gregg Vesonder                        Date

_______________________________________
Dr. Lu Xiao                               Date

STEVENS INSTITUTE OF TECHNOLOGY
Castle Point on Hudson
Hoboken, NJ 07030
2018

# A Hybrid Simulation Model for Open Software Development Processes

## *Abstract*

Open software development provides software organizations access to infinite online resource supply. The resource supply is a pool of unknown workers who work from different location and time zone and are interested in performing various type of tasks. Improper task execution in such dynamic and competitive environment leads to zero task registration, zero task submissions or low qualified submissions due to unforeseen reasons such as uncertainty in workers' behavior and performance. Therefore, to ensure effectiveness of open software development, there is a need for improved understanding and visibility into characteristics associated with attracting reliable workers in making qualified submissions and reducing task failure.

Existing studies showed that workers are more interested in performing on tasks with similarity in terms of concepts, monitory prize, required technologies, complexities, priorities, and durations. Task similarity is one of the most important factors in a worker's decision on registering for a task and provides the worker's individual knowledge. Also, workers' decision to make a submission is based on workers' experience level, task competition level and probability of winning the competition. Dependencies among tasks and workers makes decreasing task failure in open software development very challenging.

To date various task scheduling methods such as HitBundle, Round Robin, Game with a purpose, and QOS have been introduced with the aim of reducing task failure rate in open production. Most of the existed methods only focused on the static aspect of scheduling such as task status and neglect the dynamic aspect of it such as worker's interest and preference on arrival tasks. To reflect the dynamic aspect of scheduling, there is a need to focus on impact of workers' decision-making in open software environment.

To that end, this dissertation presents three empirical studies and a hybrid simulation model for open software development processes. These empirical studies lead to the understanding and development of empirically derived worker behavior patterns, task completion patterns, and open team performance patterns. The simulation model then integrates these empirically-learned knowledges into its three components for modeling the processes of open software development. More specifically, the three component layers are the meso level, discrete event simulation, representing the task completion, the micro level, agent-based simulation, illustrating the crowd workers' decision-making processes, and the macro level, systems dynamic simulation, reflecting the platform dynamics. At meso level, the discrete event simulation component provides two different failure prediction models for task registration and submission phases respectively, in order to forecast failure probability of the task execution plan. The performance of the failure prediction models was evaluated using empirical data extracted from TopCoder platform. The hybrid simulation model is evaluated through two decision scenarios to demonstrate its effectiveness. The first scenario is evaluating the impact of task similarity in the pool of open tasks on task failure awareness in different state of task completion. The second scenario evaluates the impact of workers' experience level on reliability of making a qualified submissions and failure awareness in different state of task completion. The proposed simulation model empowers managers to explore potential outcomes of open software development and impacts of different level of uncertainties and task configuration strategies.

Author: Razieh L Saremi

Advisor: Dr. Ye Yang

Date: 07 / 23 / 2018

School of Systems and Enterprises

Department: Systems Engineering

Degree: Doctor of Philosophy

# *Dedication*

To my parents, who have devoted their life to raise me, have supported me in all stages of my life, and have encouraged me to always pursue my passions.

# *Acknowledgement*

I would like to acknowledge the people who have helped me complete this dissertation. I truly appreciate non-stop help and support of my advisor Dr. Ye Yang of the School of Systems and Enterprises, for her sustained attention to my research and the priceless supervision through which I have learned how to conduct research to reach lofty objectives. I thank her for sharing her knowledge with me and showing me how to think when I want to solve a problem. I express my gratitude to my dissertation committee members, Dr. Gregg Vesonder, Dr. Guenther Ruhe, Dr. Somayeh Moazeni and Dr. Lu Xiao for their very constructive comments to align our research methodology with the goals. At the end, I send my sincere regards to my parents to whom I owe everything I have. Their constant influence has taken me to this level of education. I express my earnest appreciation to my siblings for their endless love.

# *Table of Contents*

# *List of Figures*

# *List of Tables*

# *Chapter 1     Introduction*

   Software has become an unarguably important part of the modern life and its production processes have also been transforming rapidly. From traditional small in-house development teams to large and globally distributed teams and online communities, software development has increasingly become more intelligent and agile by leveraging external resources across enterprise boundaries. Open market software development (OMSD) [1] is proposed for making changes in the common traditional software development ways in assigning the task to developers. OMSD is built by potentially large number of unknown workers who have access to the internet and are collaborate and coordinate on the tasks. Potentially, workers can perform in parallel, and output of a task can be used as input of the other to producing faster results. Open sourced software development (OSSD) and crowdsourced software development (CSD) are two examples of this emerging paradigm that attracts increasing attention in the past few years. In this dissertation we analyze crowdsourced software development (CSD) as the instance of open market software development.

   Crowdsourcing techniques were initially utilized centuries ago. Although the term crowdsourcing has been academically popularized by Howe [2] in 2006. Initially, crowdsourcing defined as "taking a function once performed by employees and outsourcing it to an undefined (and large) network of people in the form of an open call across a large network of potential laborers" [2]. This definition suggested that crowdsourcing is a form of outsourcing [3], however, the duplication of the worker performing in parallel does not apply to outsourcing. Therefore, to specify outsourcing and crowdsourcing, Brabham [3] introduced a more specific definition of crowdsourcing as "an online model for distributed production and problem-solving, which develops an open call format to recruit global online software engineers to work on various types of software engineering tasks, such as requirements extraction, design, coding, and testing". Recently, Stole et al [4], presented a specific definition for software crowdsourcing based on the key elements of crowdsourcing software development [4] which are: nature of the work, the locus of control and nature of the workforce. Software crowdsourcing is "the accomplishment of specified software development tasks on behalf of an organization by a large and typically undefined group of external people with the requisite specialist knowledge through an open call" [4].

Crowdsourcing provides the benefit of easy access to a wide range of workers, including diverse sets of problem solving and creativity, lower costs and defect rates with flexible development capabilities, and reduced time-to-market by increase parallelism [5] [6]. Many major corporations such as Google, Facebook, Microsoft, NASA, etc. are aware of the beneficial values of crowdsourcing, which has persuaded them to take advantage of this approach frequently in their projects. Generally, in compare with general crowdsourcing tasks, software crowdsourcing tasks are typically more challenging and complex, which may take a longer time to complete. Software crowdsourcing tasks require various specialized skills, significant time, and efforts. Moreover, complex tasks require considerably higher cooperative work amongst co-workers than the independent and self-contained tasks [7]. Therefore, in comparison with in-house software developers, it may be more challenging to predict the behavior of a large and unknown population of crowd workers and their impact on task failure in the crowdsourced platform.

## 1.1    Problem Overview

Preparing appropriate micro-tasks in a crowdsourcing platform is very important. Understanding the crowd workers' sensitivity in behavior and performance to the arrival demand and rate of demand failure is crucial since it is reported that an improper task scheduling would lead to task starvation and consequently project failure [8]. Not knowing workers in person, working from different time zone, and workers interest in other tasks among pool of open tasks lead to one of the most important challenges in CSD, which is choosing the best task execution time to assure minimizing task failure rate and project duration in the platform.

in addition, workers' reliability in making submissions links to the task description, competition level per task, number of required technology and task-worker network modularity as the most pervasive factors on task failure in CSD [9]. Also, according to available empirical studies, more than half of crowd workers are responding to the task call in the first day, while, only 24% of crowd workers who register early for a task will make a submission [10]. Therefore, in order to leverageing crowdsourcing as a viable solution in mass production in software engineering, it is worth to investigate the effective task execution methods to integrate the effectiveness of CSD with the efficiency of resource allocation in terms of minimizing failure ratio and process durtaion.

> *This dissertation aim to tackele the task failure challenge in the CSD platform by presenting a more effective task execution plan.*

The proposed method will decreasing task failure ratio while reducing project duration and increasing available workers reliability in submitting qulified submissions.

## 1.2 Research Question

In the mass production environment with unknown resources, proposing a task execution plan to assure attracting skillful and motivated workers to compete on tasks and return qualified submissions in order to minimized failure ratio is critical. Therefore, to develop a better understanding of crowdsourcing-based software projects, this dissertation focused to answer the following question:

> *How can we provide a task execution model to improve the understanding about CSD environment in order to reduce task failure ratio in the platform?*

To successfully propose such platform there is a need to understand the answers of following sub-questions:

1. How does task attributes impact on crowd workers' behavior?
2. How reliable crowd workers are in performing tasks?

3. How does CSD benefit schedule reduction?

4. How can a more effective CSD platform be achieved via hybrid simulation?

## 1.3 Research Methodology

In comparison with general crowdsourcing tasks, software crowdsourcing tasks are typically more challenging and complex which may take a longer time to complete. They require various specialized skills, significant time, and efforts. Moreover, complex tasks require considerably higher cooperative work amongst co-workers than the independent and self-contained tasks [7]. Therefore, in comparison with in-house software developers, it may be more challenging to predict the decision-making process of a large and unknown population of crowd workers and their impact on task failure ratio in the crowdsourced platform. This challenge urges more investigations on the dynamic of software crowdsourcing in order to develop a better understanding of crowdsourcing decision-making to execution a task and reduce failure ratio in the platform. Combination of empirical analysis and simulation in software engineering provides one of the greatest tools to address this challenge in crowdsourcing.

Untrustworthy workers may cause task starvation or failure, due to high chance of task drop out. Many workers may register for more than one task at a time and drop ones they cannot complete. Up to date, several different methodologies have been used to predict workers' motivation and decision-making process in crowdsourcing platforms. In recent years, researchers have attempted to create a bridge between workers action and crowdsourcing platforms' success rate. One of the most useful practices can be Agent-Based Modeling (ABM) [11]. Also, workers' decision-making process would cause different challenges in project success due to restricted time frames and task sequences. One method to solve such a problem and to understand the task completion and mutual impact between workers' availability and failure ratio is Discrete Event Simulation (DES) [12]. Finally, crowdsourcing platform is a system created by multiple daemon actions; this means a platform cannot be run by only one single worker. Also, different workers performance history will impact on the other workers decision. Therefore, following workers reaction and tracking the available utility algorithm based on workers and tasks history can lead to a better task execution practice. The most effective method to understand task failure ratio in the platform based on workers distribution is applying systems dynamic (SD) models to the platform.

In this research, we aim to study a hybrid simulation model for crowdsourcing platforms to minimize task failure rates by minimizing the number of untrustworthy workers per task. For evaluating the proposed model, we will use data from Topcoder, the largest competitive software crowdsourced platform with over 1 million-member workers from 190 countries, averaging 30 thousand logins every 90 days, 7 thousand challenges hosted per year and 70 million dollars in challenge payouts. Figure 1.3-1 illustrates the overview of the research methodology.

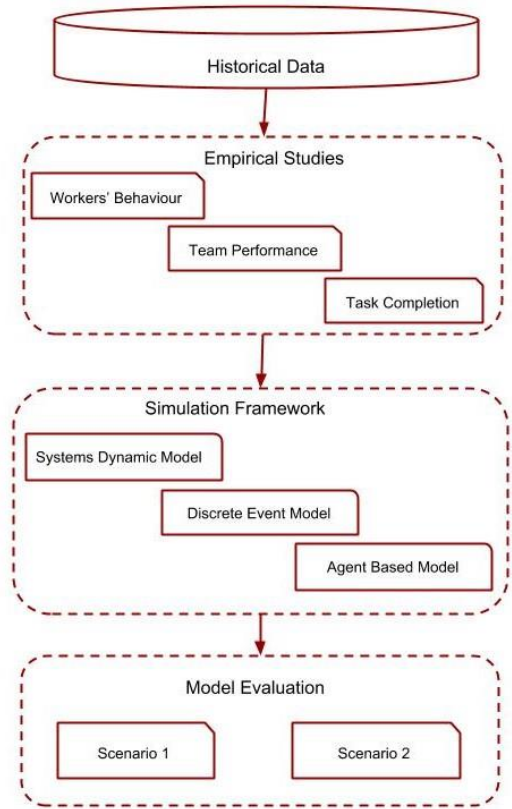

*Figure 1.3-1: Over view of Research Methodology*

In section 2, literature review and background, details of the difference between outsourcing and crowdsourcing, crowdsourcing platform type and design, task preparation in crowdsourcing, workers incentives and applied simulation methods will be discussed. In section 3, research design, and problem statement will be discussed. In Chapters 4,5 and 6, details of Empirical evaluation and uniqueness of this research will be explained. In chapter 7, details of hybrid simulation model will be reported. And, the research contribution and future direction will be introduced in chapter 8.

# Chapter 2      Literature Review and Background

Open market software development (OMSD) offers the advantage of significant savings in costs and time by bringing expert communities together [5] [6]. However, there are unknown factors in OMSD method to motivate workers to put the highest effort in performing tasks either individually or cooperatively.

To apply OMSD, task requestors have difficulty clearly describing the task while only being able to provide limited information due to security reasons [4]. Task requestors also lack knowledge about workers' trust network [13]. Intellectual property issues may also arise when transferring the tasks into deliverables. In the following sections, we will describe these challenges in more detail.

## 2.1   Competitive Open Software Development

The general purpose of an open market development platform is to provide a market in which: 1) requestors can post tasks to be completed, 2) specify prices paid for completing them and 3) workers perform tasks, which are difficult for computers to perform [14]. Clearly, OMSD is the future of open market development as software industry are going toward these leads. Therefore, there is a need to better understand these concepts to make better decisions.

### 2.1.1     Crowdsourcing

Although the term 'crowdsourcing' has recently attracted significant attention, the original concept of crowdsourcing can be found many centuries ago. For example, the origin of crowdsourcing goes back to the Longitude competition in 1714, when the British government announced an open call (with monetary prizes), for developing a method to measure a ship's longitude precisely [15]. Also, internet based crowdsourcing activities can be found as early as 2001, when 'Inno Centive' [16]was funded by Eli Lilly to attract a crowd-based workforce from outside the company to assist with drug development. In the same year, the TopCoder [17] platform was launched by Jack Hughes as a marketplace using crowdsourcing for software development. To facilitate online distributed software development activities, the TopCoder development method and system was proposed [17].

In 2006, Howe argued that crowdsourced work can be done by cooperation or by sole individuals [2]. The main characterization of crowdsourcing in the above definition is that crowdsourcing is "outsourcing on steroids" [3], which suggests that crowdsourcing is merely a form of outsourcing. Brabham discussed that CSD is not an extension model of open sourced software [3] and crowdsourcing is "an online model for distributed production and problem-solving, which develops an open call format to recruit global online software engineers to work on various types of software engineering tasks, such as requirements extraction, design, coding, and testing".

Later, Stole et. al. [4] introduced a definition for crowdsourcing software development (CSD). Based on the new definition, CSD is not a form of OSS and contains the following key concerns in a software development context:

- Nature of the work - software development tasks are quite complex with many interdependencies.
- Locus of control - the customer organization who must specify the tasks and integrate the resulting output into the organization's software development process.
- Nature of the workforce - a large and typically undefined group of external people, but with the requisite 'wide and deep' specialized knowledge to accomplish the task successfully.

Therefore, CSD is defined as: "The accomplishment of specified software development tasks on behalf of an organization by a large and typically undefined group of external people with the requisite specialist knowledge through an open call" [4].

Based on the different definitions of crowdsourcing, four common features can be identified and related to this paradigm:

- Open access in production,
- Flexibility in workforce,

- Free will in participation and,
- Mutual benefits among stakeholders [5].

These features provide benefits in applying crowdsourcing such as: easy access to a wide range of workers, diverse solutions, lower labor rates, and reduced time-to-market [5]. However, to achieve both effective task completion and software worker pleasure, it is recommended that the requestor should:

- Clearly recognize software workers' motivation,
- Understand software workers' behavior, and
- Design structures and models of the platform [8].

These factors influence the workers' arrival to the platform and task competition level.

### 2.1.2    Open Source

Open-source software (OSS) development is a popular method with technical users, who are most often expert software developers [18]. OSS provides open standards, shared source code, and collaborative development in the software development process. Some OSS platforms such as GitHub, Sourceforge, and Google Code provide an open platform for users to create, share, download, and publish OSS for free. IT workers are encouraged to contribute to open source projects as co-developers by submitting additions such as code fixes, bug reports, and feedbacks. While the source code of an OSS product is publicly available, the rights to examine, change, and distribute the code are usually limited by a license.

The arrangements to ensure freely available source code have led to a development process that is radically different, according to OSS proponents, from the usual, industrial style of development. The main differences usually mentioned are:

- OSS systems are built by potentially large numbers of volunteers.
- Work is not assigned; people take the work they choose.
- There is no explicit system-level design, or even detailed design.
- There is no project plan, schedule, or list of deliverables [19].

OSS platforms provide comprehensive support for communication and collaboration by providing various communication mechanisms such as mailing lists, forums, blogs, and wikis. They also integrate version control systems and issue trackers to support collaboration. In contrast to traditional centralized software development, organization structure and roles in an OSS project aren't clearly defined. Coordination such as conflict mediation is conducted democratically, for example, by voting or using moderator mechanisms. Also, open source platforms don't support transfer of business value between requestors and providers. Developers involved in OSS projects don't seek monetary rewards but do pursue technical challenges [20].

There is usually little commercial pressure to keep to any hard schedule. While this may entail longer development cycles, this is also an advantage since OSS projects are largely immune from "time-to-market" pressures; a system need not be released until the project owners are satisfied that the system is mature and stable [21].

### 2.1.3    CSD Platform Type

There are many different crowdsourcing platform types including non-competitive, competitive, and collaborative [8]. Software crowdsourcing is mostly performed in two forms: competitive crowdsourcing, and collaborative crowdsourcing.   In competition form, each participant independently creates a solution and at the end, a winner will be chosen [22]. This method demonstrates great potential to help solve problems rapidly and globally. Commercial platforms such as Topcoder implement this model. Competition methods have been found to be helpful in obtaining high-quality solutions by encouraging redundancy [23] [24] .

In the collaboration model, multiple micro-tasks are frequently chained together into workflows [25]. In these workflows, larger tasks may be decomposed into smaller ones and subsequently the subtask solutions are recomposed into an overall work product. In this method, the requestor is able to monitor and edit the resulting workflows as they are produced. A good example of such a platform is the Turkomatic [25]. Therefore, in the competition method, the quality of submission only depends on the effort of each worker, while in the collaboration method the quality depends on the average effort of all the workers working on the task.

The result of studying the interaction between competition and collaboration in crowdsourcing based design contests suggests that competitive participation should be employed to stimulate the crowd's motivation of making contributions and performance, without disabling the climate for knowledge sharing and collaboration. Cooperative games often tend to have a positive influence on player creativity as they are willing to help each other to improve their performance [26] [27]. Non-cooperative platforms such as Topcoder have fewer stimuli for people to learn from each other. Furthermore, Topcoder workflow is a game of "Chicken" [28], where each participant will try their best to beat their competitors, and thus each party may be mutually destroyed by each other as the worst outcome for all participants. However, the worst possible outcome will happen if both do not yield, and both are killed in the competition.

### 2.1.4    CSD Platform Design

Changing software development processes from the traditional waterfall model, spiral model, and model-driven processes to recent agile methods to component-based methods, open-source approach, and service-oriented computing provides the possibility of using the crowdsourced market to have the project done with less time and budget [29]. These processes differ significantly from the development steps as well as intermediate deliverable products. For example, the Waterfall model requires considerable documentation efforts and each document is cross-validated with other documents during the process, while modern agile processes are light on specifications but heavy on code development.

Software engineering has begun to adopt crowdsourcing in many different contexts. Platform design offers the opportunity to change the understanding of crowdsourcing projects and forms the relationships and practical communication between software workers and requestors [30]. In other words, crowdsourced platforms can be viewed as large distributed computing systems in which each software worker is similar to a processor/agent that can solve a task requiring human intelligence [30].

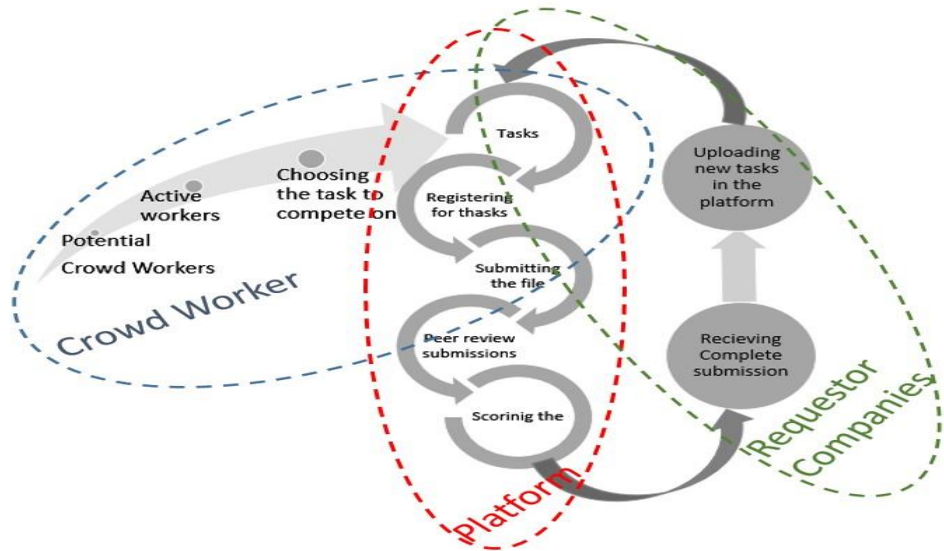

*Figure 2.1-1: Crowdsourcing Software Development Flows*

Figure 2.1-1 illustrates typical competitive software crowdsourcing processes based on the Topcoder platform model [17]. Three key stakeholders involved in this CSD model are the requestor company, which is the project sponsor or client; the crowdsourcing platform that acts as a service provider to the client; and the crowd worker, i.e. the internet developers who are the main agents in the model [31].

Generally, the requestor company divides the project into many small tasks, prepares task descriptions, and distributes tasks through the platform. Each task is tagged with a pre-specified prize as the award to winners and a required schedule deadline to complete. On average, most of the tasks have a lifespan of 2-4 weeks from the first day of registration.   Figure 2.1-2 illustrates task organization in a crowdsourced platform.

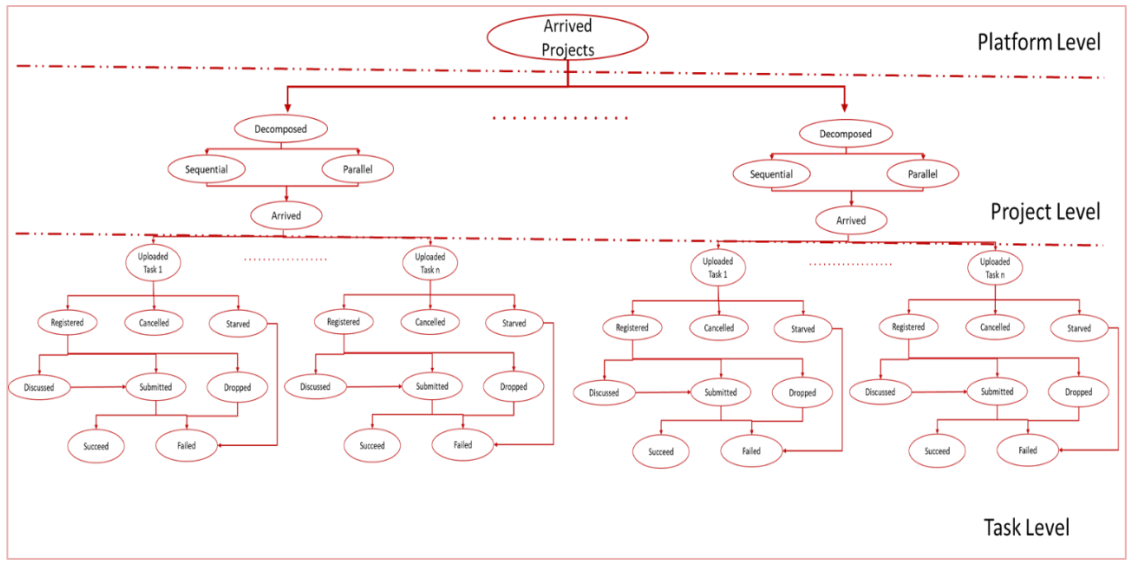

*Figure 2.1-2: Illustration of Task Organization in CSD Platform*

Crowd software workers browse and register to work on selected tasks, and then submit the work products once completed. After the workers submit the final submissions, the files will be evaluated by experts and experienced workers, through a peer review process, to check the code quality and/or document quality [31]. The number of submissions and the associated evaluated scores replicate the level of task success. In Topcoder, usually, the award goes to the top one or two winners with the highest scores. If there are zero submissions or no qualified submissions, the task will be treated as starved or canceled.

Tajedin and Nevo [31] discussed that a successful crowdsourcing platform contains three determinants: the characteristics of the project; the composition of the crowd; and the relationship among key players [31]. A systematic development process in such a platform starts from a requirements phase, where the project goals, task plan, and budget estimation are recognized. This will be performed through communication between the project manager, who may come from the crowd or the platform, and the requestor, who pays for the solutions offered by the crowd. This phase involves several types of competitions such as Conceptualization, Wireframe, Storyboards, UI Prototype, and Specification. The outcome will be a set of requirements and specifications. These requirements are used as inputs for the future architecture phase, where the application is decomposed into multiple components [5].

## 2.2   Open Software Workers' Incentives

It is not wrong if someone claims the most important factors in the success of crowdsourcing software development, is workers' motivation and statistical behavior.  In this section, we will discuss both motivations and behaviors of CSD workers.

### 2.2.1      Workers' Motivation

Due to the diversity of software workers in crowdsourcing, motivational factors are typically divided into two categories: intrinsic factors and extrinsic factors. Intrinsic clusters comprise enjoyment factors and community values that are associated with age, location, personal career, society and even task identity [32]. In addition, participation motivation categories [33] such as learning, and self-marketing are considered as some of the main intrinsic factors for workers which are mentioned in the concept of 'activation enabling'. Extrinsic clusters include financial and social aspects, which are the direct effect of educational background, household income, and task award or payment [32]. It is reported that [33], motivations such as organizer appreciation, prizes, and knowledge experts are the highest ranked among workers. Another high-ranking motivation factor is the presence of requestors with a prestige brand name, such as Google or NASA, which attracts software workers to apply for tasks as it can potentially be used to strengthen resumes and affect software workers' ratings either indirectly or directly [34].

Different studies on motivation patterns of crowdsourcing workers reported that the award received is one of the top motivating factors to attract and involve potential workers in task competition in a crowdsourcing market [4] [8] [36]. The award amount typically correlates with the degree of task complexity and required competition levels as well as task priority in the project development [8] [35].  The second motivation factor which affects award is the worker's skill level [33]. Higher skilled level workers generally have higher motivation level of winning an award and gaining communal identification while the top motivational factors of occasional workers are typically a pastime human capital investment and skill variety. Therefore, task requestors usually concentrate to distribute awards for tasks with a good scheduling plan [8]. The objective of scheduling is getting the best possible task submissions in terms of the tasks' type based on workers' historical activities and submissions

### 2.2.2    Workers' Behavior

Software workers' arrival to the platform and the pattern of taking tasks to completion are factors that shape the worker supply and demand distribution in crowdsourcing platforms. Many software workers tend to optimize their personal utility of choosing a task based on different attributes [8]. Also, most experienced workers are interested in either competing for famous branded companies' tasks or targeting tasks based on the award.

For newcomers or beginners, there is a time window required to improve and to develop into an active worker [8]. Therefore, most focus on registering and gaining experience by competing with peers, even though the likelihood of winning the competition is rather low for them. It is also typical that the workers need to communicate with the requestors in order to better understand the problems to be solved [36]. Existing studies show that over time, registrants gain more experience, exhibit better performance, and consequently gain higher scores[8] [36] [37]. Still, there are workers who manage not only to win but also to raise their submission-to-win ratio [38]. However, issues including different time zones and geographical distributions, and software workers' native language may cause a worker to drop a task after registration or become inactive [12].

Archak [35] [37] presented an empirical analysis of developers' strategic behavior on Topcoder. The results showed the "cheap talk" phenomenon during the registration phase of the contest. Highly rated developers would tend to register for the competition early, therefore intimating their opponents from participating in the marketplace and consequently softening the competition. The "Cheap talk" phenomenon [39] and the ranking mechanisms used by Topcoder contribute to the efficiency of real-time online competition [35]. In CSD, higher rated workers have more freedom of choice in comparison with lower rated workers and can successfully affect the registration of lower rated workers. To assure an easier competition level, higher rated workers register early for some specific projects while lower rated workers must wait for higher rated workers to make their choice.

Moreover, not only would the award associated with the task influence the workers' interests in competitions, the number of registrants for the task, the number of submissions by individual workers, and of course the workers' historical score rate would directly affect their final performance [6]. Therefore, to win in such a mutually destructive contest, a worker must do his/her best to avoid losing to other workers and secure a high-ranking score in the system to scare off other potential competitors. That is the nature of this kind of competition, often called the "Chicken Game" or "Hawk-Dove" [28]. In this game theory, less aggressive players (chicken or dove) will yield to aggressive players.

### 2.2.3    Trust Factor in CSD

Crowdsourcing a project inherently involves a concern for how reliable and trustworthy the unknown crowd workers are [40]. It is important to attract trustworthy workers to work on new tasks in order to receive qualified and successful submissions.

Eickhoff et al [41] discussed that unreliable workers are not very interested in taking novel tasks that require creativity and abstract thinking. Therefore, the best way to avoid unreliable workers [42], and assuring good quality work, is to upload the task in a way that both trustworthy and unreliable workers take the task at the same time. It seems more expensive tasks would result in higher quality submissions by attracting reliable and dedicated workers [43] due to the complexity and novelty of the task. However, at some point, increasing the award would attract more unreliable workers.

Due to the diversity of workers with different individual skill levels, it is not practical for the requestor to evaluate all the workers' trustworthiness [40], nor is there a clear record of workers' interaction in the pool of workers [41]. This fact creates two trust networks in a platform. First, the general trust network among requestor and workers which is based upon requestor company brand, workers' rating and skill set and worker availability. The second trust network is among the worker community itself; this network is a result of workers' rating, skill set and history of winning a task.

In Amazon Mechanical Turk, the overall evaluation rate for workers is the percentage of accepted submissions to the total submissions per task [41]. This type of trust evaluation algorithm causes several issues. One primary example of an issue in Amazon Mechanical Turk is rank-boosting [40] [44], where workers mostly register for easy tasks or fake tasks that they themselves are uploading in order to increase their rating. On the other hand, Topcoder adopts both a numeric worker rating system based on the Elo rating algorithm, and a 5-level rating scheme to divide the worker community into five groups. The numeric ratings are with respect to three different task categories including algorithm, marathon matches and development [45], followed by the sophisticated computational algorithm.

The above systems ignore the fact that workers have different reliability ratings for different task types. Moreover, there should also be a trust network among workers themselves. To solve this issue, Ye et al [40] proposed a quantitative trust evaluation model based on both task type-based trust, and reward amount-based trust. The model can differentiate trustworthy workers from unreliable workers when both have high overall ratings in the system. In addition, Yu et al [46] proposed a reputation task sub-delegation system based on workers' reputation, task load, and price and trust relationship with others.

## 2.3  Task Execution

Crowdsourcing projects may help organizations use external human resources. This fact would help reduce cost from internal employment and help explore the distributed production model to speed up the development process [5]. In contrast, the traditional in-house project management process involves planning, monitoring and controlling the activities [47], with the direct dependency on release and resource decisions [48].

To have a successful crowdsourced project, the first step is decomposing the project into an optimum number of possible tasks then, based on project requirement, planning scheduled tasks and uploading them to the platform. In this section, we will discuss CSD task decomposition methods as well as available crowdsourcing scheduling techniques.

### 2.3.1 *Task Decomposing*

In general, projects in crowdsourcing can be decomposed and executed in both independent and dependent tasks. There are two methods of task decomposition: 1) horizontal task decomposition for independent subtasks and 2) vertical task decomposition for dependent subtasks [49]. In the horizontal decomposition method, workers dedicate efforts independently to their own subtasks for individual utility maximization, while in the vertical decomposition method, each subtask takes the output from the previous subtasks as input, and therefore the quality of each task is not only related to the worker's effort but also associated with the quality of the previous tasks [49].

Crowdsourcing complex tasks, i.e. software development tasks, generate heavy workloads and require dedicated resources with high skill sets, which limit the pool of potential workers. In order to expand the qualified labor pool, it is essential to decompose software engineering tasks into smaller pieces. However, software engineering tasks are often concerned with their specific contexts, for which decomposition may be complicated. This fact opens a discussion on different ways of decomposition based on a hierarchy of workflow [4]. The key factors of decompositions considered in this research are: smaller size of micro-tasks, larger parallelism, reducing time to market, and a higher probability of communication overhead. The most common method in decomposing microtasks is asking individual workers to work on a task of specific artifact. This method will lead to some natural boundaries for software workers and there may be a need of defining new boundaries as well.

### 2.3.2    *Task Flow*

It is important to understand the current scheduling method in both project level and platform level and determine a better understanding and results, in order to understand the most effective scheduling method for the platform. The result of empirical analysis [50]confirm that parallelism in scheduling tasks would positively affect the task completion and project success. In order to have a better understanding about the impact of parallelism on task execution, we will introduce lifecycle scheduling in project level. The result of this component will help with better resource availability management.

Before starting the execution, it is important to understand different types of task dependencies in a project. From a project manager point of view there are 4 types of dependency [51]:

1- Finish-to-Start (FS): Task 1 of the project must be totally completed before task 2 can begin: making task 2 dependent on the completion of task 1.

2- Start-to-Start (SS): Task 2 can be started while task 1 is under process. They may occur simultaneously, as long as task 1 started first. A benefit of this dependency type is that work overlaps, moving the project along more quickly.

3- Finish-to-Finish (FF): Task 2 cannot be finished unless task 1 is finished at the same time. With simultaneous tasks in motion, the completion of these tasks must take place at the same time.

4- Start-to-Finish (SF): Task 2 cannot have finished unless task 1 is starts. This approach in the most complex and so it is used infrequently.

In this research, we are using FS dependency method to understand task sequence in a project. We adopt the task model introduced in [52] to our research. Challenges of scheduling of real time tasks on multi processors under fork joint structure based on parallel programing in OpenMP, a mature system for parallel programming, is analyzed [52].

First, we need to represent each task in project:

$$T_i = (ID_i, ESD_i, LSD_i, EED_i\ LED_i\ , A_i, Tt_i, Tech_i, Req_i, S_i, P_i\ )$$

$$S_i = \{S_i^1, S_i^2, \dots, S_i^k\ \}, \text{ where k is the total number of sequential tasks per task.}$$

$$P_i = \{P_i^1, P_i^2, \dots, P_i^l\}, \text{ where l is the total number of parallel tasks per task.}$$

In which:

$ID_i$           represents task ID in the project,

$ESD_i$           represents task earliest start date, and $LSD_i$ represents task latest start date,

$EED_i$           represents task earliest end date, and $LED_i$ represents task latest end date,

$S_i^n$           represents the sequential tasks,

$P_i^n$           represents parallel tasks,

$A_i$           is associated award with the tasks,

$Tt_i$           represents task type,

$Tech_i$           represents required technology,

$Req_i$           represents detailed requirement described in task description associated with uploaded tasks.

Considering the task model, the Maximum execution time per task ($METT_i$) can be defined as:

$$METT_i = LED_i - ESD_i$$

Total execution time per sequential tasks(ETST) in a project is the sum of execution time of all sequential tasks:

$$ETST_i = \begin{cases} \sum_{i=1}^{k} METT_i & S_i^n \neq 0 \\ 0 & S_i^n = 0 \end{cases}, \text{ where k is number of sequential tasks per project}$$

$$\textit{Subject to:}$$

$$ESD_i + t_i \leq LSD_j$$

And total execution time for parallel task (ETPT) in a project is the maximum execution time of all set of parallel tasks:

$$ETPT_i = \begin{cases} Min(ESD1, ESD2, \dots, ESDl) + Max(LED1, LED2, \dots, LEDl) & P_i^l \neq 0 \\ 0 & P_i^l = 0 \end{cases}$$

$$\textit{Where l is number of parallel tasks per project}$$

$$\textit{Subject to:}$$

$$ESD_i + t_i > LSD_j$$

Therefore, total project duration ($PD_j$) is:

$$PD_j = min(\sum_{i=1}^{k} ETSTi + \sum_{i=1}^{l} ETPTi \ )$$

*Where k is number of Sequential task and l is number of parallel tasks*

*Subject to:*

$$ESD_i + t_i \leq Max\ (ESD_i + T_i)$$

$$0 \leq t_i \leq 24\ hr$$

### *2.3.3    Task Scheduling*

Due to different characteristics of machine and human, delays can occur in product release and lack of systematic processes to balance the appropriate delivery of features with the available resources [48]. Therefore, improper scheduling would result in task starvation [8]. Parallelism in scheduling is a great method to create the chance of utilizing a greater pool of workers [47] as this method helps workers to specialize and complete the task in less time and drive solutions, and helps the requestor to clearly understand how workers decide to compete on a task and analyze the crowd workers' performance [8]. Shorter schedule planning can be one of the biggest advantages of using CSD for managers [6].

Complex crowdsourced projects cannot be performed based on simple available parallel approaches. Complex projects have more dependencies and multiple changing requirements [36]; they require different workers with different levels of expertise. Therefore, this is one of the main challenges in applying an effective method to schedule decomposed projects in crowdsourcing [53]. Considering the fact that coordinating workers is difficult among a distributed global crowd, in most cases, organizational coordination techniques such as programming and feedback as general coordination methods can be applied to crowd work as well [54] [55].

Batching tasks is another effective method to reduce the complexity of tasks and helps dramatically reduce cost [56]. Batching crowdsourcing tasks would lead to a faster result than approaches, which keep workers separate and is also faster than the average of the fastest individual worker [57]. There is a theoretical minimum batch size for every project as one of the principles of product development flow [58]. To some extent, the success of software crowdsourcing is associated with reduced batch size in small tasks.

*Table 2-1: Task Scheduling Methods in CSD*

| Method | definition | Worker suitability | Workers availability | Task Required skill | tasks/batch size | tasks/batch priority | Task from same batch | Min Switch context |
|---|---|---|---|---|---|---|---|---|
| Delay Scheduling [12] | the maximized probability of a worker receiving tasks from the same batch | | | | | | X | X |
| QOS [59] | minimizing scheduling while maximizing quality | X | X | X | | | | |
| Fair Scheduling [60] | resources would be shared among all tasks with different demands | | X | | X | | | |
| Weighted fair scheduling [61] | schedule batches with higher priority first | | X | | X | X | | |
| First in First out [62] | All workers concentrate on single batch until it is done | | | | | | X | |
| Shorter job fair [62] | Fast turnaround for small batches | | | | | | X | X |
| Round Robin [38] [63] | even distribution of tasks and avoiding starvation without prioritizing them | | | | | | X | |
| Game with a Purpose [62] | Task will not be started unless a certain number of workers register for it | | X | | | X | | |
| HITBundle [38] | batch container which schedules similar tasks into the platform from different requestors | X | X | | | X | X | X |

The delay scheduling method [12] was specially designed for crowdsourced projects to maximize the probability of a worker receiving tasks from the same batch of tasks they were working on. Extension of this idea introduced a new method called "fair sharing schedule" [60]. In this method, heterogeneous resources would be shared among all tasks with different demands, which ensures that all tasks would receive the same amount of resources to be fair. For example, this method was used in Hadoop Yarn. Later, weighted fair sharing (WFS) [61] was presented as a method to schedule batches based on their priority. Tasks with higher priority are introduced first. Another proposed crowd scheduling method is QOS [59], a skill-based scheduling method with the purpose of minimizing scheduling while maximizing quality by assigning the task to the most available qualified worker. This method was created by extending standards of Web Service Level Agreement (WSLA) [64]. The third available method is a game with a purpose [62], in which a task will not be started unless a certain number of workers register for it. The most recent method is HIT-Bundle [38] a batch container which schedules heterogeneous tasks into the platform from different batches. This method makes for a higher outcome by applying different scheduling strategies at the same time. Table 2-2 summarizes different applied crowdsourcing scheduling methods.

## 2.4   Simulation Methods

Scheduling is an NP-hard problem since it is directly dependent on time. It is believed that NP-hard problems cannot be solved with their exact optimal solution, but with their near-optimal solution in a relatively short time. Simulation is one of the best methods to address NP-hard problems.

Also, there are many aspects that make CSD different and more challenging in evaluating crowd workers' performance than general crowdsourcing. First, tasks in CSD are typically more complex than general crowdsourcing tasks and also require much longer to complete [5] [8]. Tasks are uploaded as competitions in the platform, where crowd software workers would register for the challenges. Second, many workers may register for more than one task at a time and may drop the ones they can't complete [4]. A task with many untrustworthy workers is subject to a high risk of failure.

Crowd workers are individuals whose behavior affects task completion flow in the platform. Predicting workers' preference and performance helps with better task scheduling by managers. ABM is a good tool to simulate the trust factors among all stakeholders in CSD, while platform workflow will be simulated via SD methodology, and DES techniques will be used to simulate task scheduling and arrival to the platform.

### 2.4.1    Systems Dynamic Simulation

System dynamics (SD) was developed over 50 years ago by Jay Forrester at MIT to improve organizational structures and processes [65]. SD modeling provides understandings by investigating virtually any aspect of the software process at both macro and micro level. It can be used to evaluate and compare different life-cycle processes, defect detection techniques, business cases, interactions between interdisciplinary process activities, and so on. This simulation method provides the possibility of changing one or several factors (attributes) while the remaining ones are unchanged, thus supporting all possible scenarios to be checked in order to make decisions based on managerial policies [66].

The SD paradigm is appropriate for exploring the behavior of the systems or variables of interest over time. Its strength lies in its ability to accommodate the rich interrelationships between variables, particularly the effects of feedback [67]. In this study, we will use the SD model to simulate the platform level activities and understand the behavior of different stakeholders in the system due to task and worker arrival.

### 2.4.2    Discrete Event Simulation

Discrete event simulation (DES) roots to the 1960s when it was developed by Geoffrey Gordon [12]. DES is the process of simulating the behavior of a complex system to predict a specific event's result in which an event would contain a significant change in the system's state at any given time. DES is able to create a step by step simulation of the flow of the entities and their impact on the system. Any event happening will lead to a change in the state of the system. All the entities, attributes, events and relationships among stakeholders should be defined in the model of the system [68].

Since DES models present potential shocks, it is more suitable for analyzing specific problems in a system. However, this method has the limitation of not representing the dynamic aspect of the project variables such as human resources, productivity, defect injection and detection rates, etc. [68]. As of today, the DES model is commonly used in the different applications of stress testing, evaluating potential financial investments, and modeling procedures and processes in various industries, such as manufacturing and healthcare. In this study, we aim to use DES modeling in order to predict events occurring in the platform in task-level activities due to workers' performance and decision per task arrival. The main advantage of DES in this research would be the ability to find actual process levels as well as representing different steps of the development process by using attributes attached to each phase with regards to the potential complexity.

### 2.4.3    Agent-Based Modeling and Simulation

Since the 1990s agent-based modeling (ABM) and simulation has been used in different areas of science, which has led to new academic research. ABM simulates the actions of players in a specific design game [11]. While this method was used to replicate different elements of non-intelligent problems such as traffic management models, it is a unique tool for studying and replicating human behaviors.

Agent-based modeling simulation shows agents' behavior in 3 different levels: 1) macro level which is the overall system as a global view, 2) micro level which displays agent behaviors, and 3) interaction between macro and micro level through aggregation [69].

Macro-level techniques are related to the system when it is seen as the whole and agents are seen in different regions. The macro level agent behavior displays the complex system with many interdependent individual actors working, while agents have only a partial view of the system and no central controlling agent [70], hence the macro level is built based on the individual strategies and the interaction among individual agents. However, the micro-level is concerned only with individual agents, and due to the diversity of agents, it would be possible to implement different agents to sign up for the same tasks and make their own decisions based on their own utility and goals.

## 2.5   Challenges in CSD

Considering the highest rate for task completion and accepting submissions, software managers will be concerned about risks of adopting crowdsourcing. Therefore, there is a need for a better decision-making system to analyze and control the risk of insufficient competition and poor submissions due to the attraction of untrustworthy workers. A traditional method of addressing this problem in the software industry is task scheduling. Scheduling is helpful in prioritizing access to the resources. It can help managers optimize task execution in the platform in order to attract the most reliable and trustworthy workers. Normally, in traditional methods, task requirements and phases are fixed, while cost and time are flexible. In a time-boxed system, time and cost are fixed, while, task requirements and phases are flexible [71]. However, in CSD all three variables are flexible. This factor creates a huge advantage in crowdsourcing software projects.

Generally, improper scheduling could lead to task starvation [8], since users with high abilities tend to compete against users with low abilities in low skill required tasks [35]. Hence, users are more likely to choose tasks with fewer competitors [34]. Also, workers intentionally choosing less popular tasks to participate in could potentially enhance winning probabilities, even if workers share similar expertise. It brings some serious problems in the CSD trust system and causes plenty of dropped and non-completed tasks.

Moreover, tasks with relatively lower monetary prizes have a high probability to be chosen and be solved, which results in only 30% of problems in the platform being solved [55]. This may attract higher numbers of workers to compete and consequently causes a higher chance of starvation for more expensive tasks and project failure.

The above issues indicate the importance of task execution in the platform in order to attract a good amount of trustable and expert workers as well as shorten the release time. To tackle this problem this study focused on:

1. Workers' behavior patterns,
2. Team performance patterns,
3. Task competition patterns, and
4. Hybrid simulation model,

in open software development. The result of this research will empower managers to explore potential outcomes of open software development and impacts of different level of uncertainties and task configuration strategies.

# Chapter 3     Research Design

For software managers, utilizing external unknown and uncontrollable crowd workers is a great uncertainty and risk compared with in-house development [8]. By understanding crowd workers reaction to the task posting and opponents' trust factor, software managers will make tradeoffs among cost savings, the degree of competition, and the expected quality of the deliverables [60]. Most of the existing studies on software crowdsourcing are focused on the individual task level, providing limited insights on the practice as well as outcomes at the overall project level. There is an important need to understand different effective factors in crowdsourcing a project and task planning to avoid key crowdsourcing challenges. This fact shows the lack of research on appropriate task execution practice. Achieving lower failure rate and higher team elasticity can be a good measure for task success. Hence, it is important to attract the optimum number of trustable workers to take the tasks and receive the most qualified submissions in the shortest possible time.

## 3.1   Problem Statement

For a crowdsourcing software development manager, task completion rate is the primary factor to measure the effectiveness of the method. In competitive platforms, task submission ratio per project can be considered as a secondary factor to measure effectiveness. To develop a better understanding of crowdsourcing based software projects, this research focuses on a study on the crowdsourced platform and crowdsourced project success rate to answer the following research question:

*How can we provide a task execution model to improve understanding about the CSD environment in order to reduce task failure ratio?*

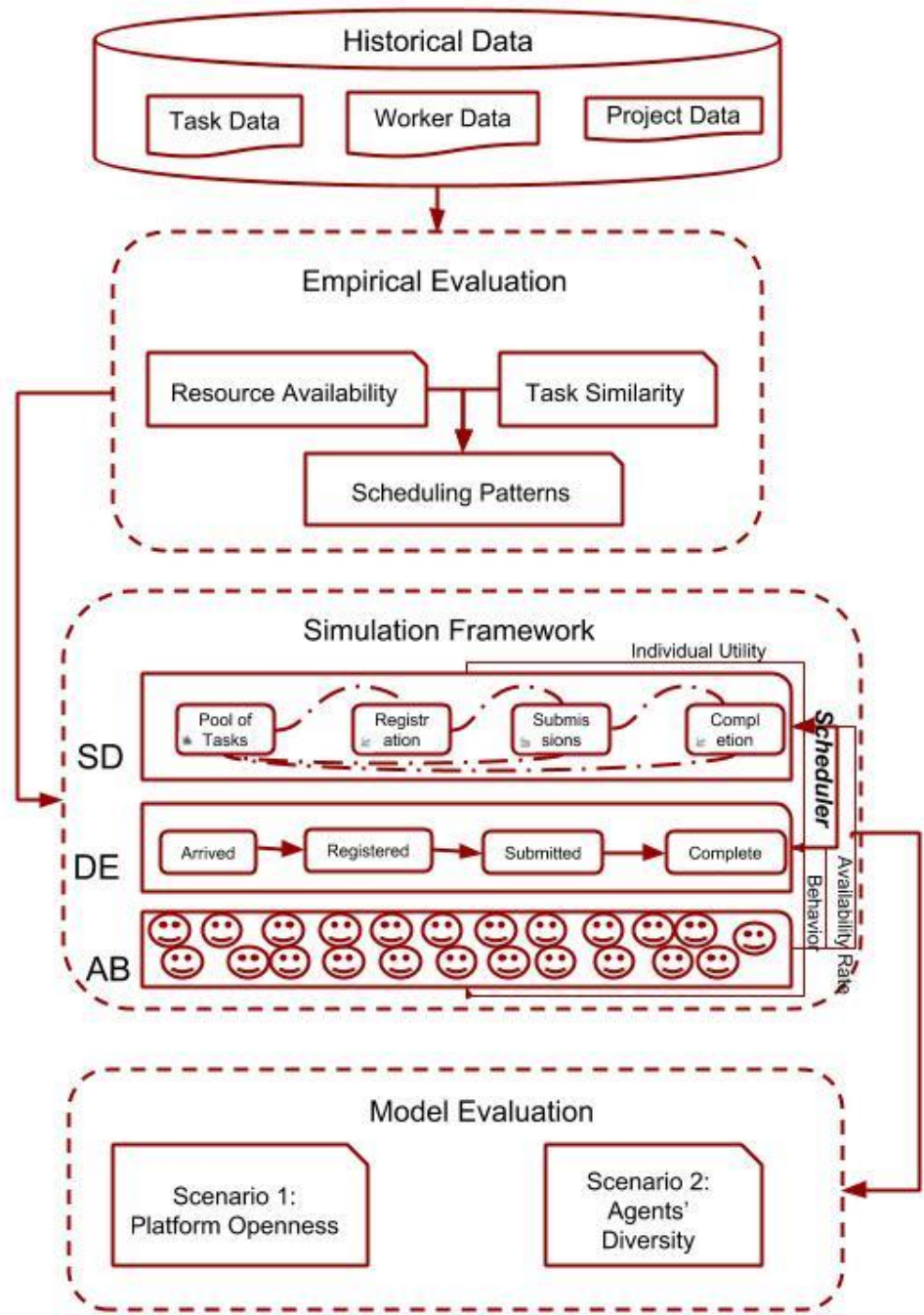

*Figure 3.1-1: Overview of Research Methodology*

There are four major factors to understand the key characteristics and challenges of the crowdsourced environment and task execution methods. First, it is important to understand the sequence of tasks per project. It is required to determine the impact of degree of task coupling on task success. Second, one should determine the effect of task similarity in the platform on task competition level. Third, it is required to investigate the available reliability and trust factors in the platform between both requestor and workers as well as workers themselves. Finally, we should analyze the relationship between task sequence, available similar tasks and trust factor on team elasticity and task execution strategies.

For further discussion on this question, it is vital to understand task completion, task similarity, resource availability and mini-tasks execution properties. This required information will lead us to the following sub-questions in sections 3.1-.1 and 3.1.2, which should be carefully analyzed and answered in order to come up with proper results of the main research question. Hypothesis associated to each sub-question will be presented afterwards.

Moreover, figure 3.1-1 shows the overview of the research methodology.

### 3.1.1    *Empirical Studies*

Based on the literature review, one of the best methods of utilizing the greater pool of workers in CSD is "parallelism in scheduling". To apply a good task execution approach, it is important to understand the most important factors in task failure ratio in the platform. To do so, the empirical sturdies divided to three groups of workers' behaviors, team performance and task completion patterns. Figure 3.1.2 illustrates the overview of the empirical studies.

#### 3.1.1.1    *Workers' Behaviors*

Considering the cost reduction in crowdsourcing, software mangers are more concerned about risks of starved tasks or poor quality of submissions. It is important for task owners to understand workers' behavior patterns in order to predict the best task execution plan. Therefore, answers to the following questions help task owners understand workers' behavior patterns better.

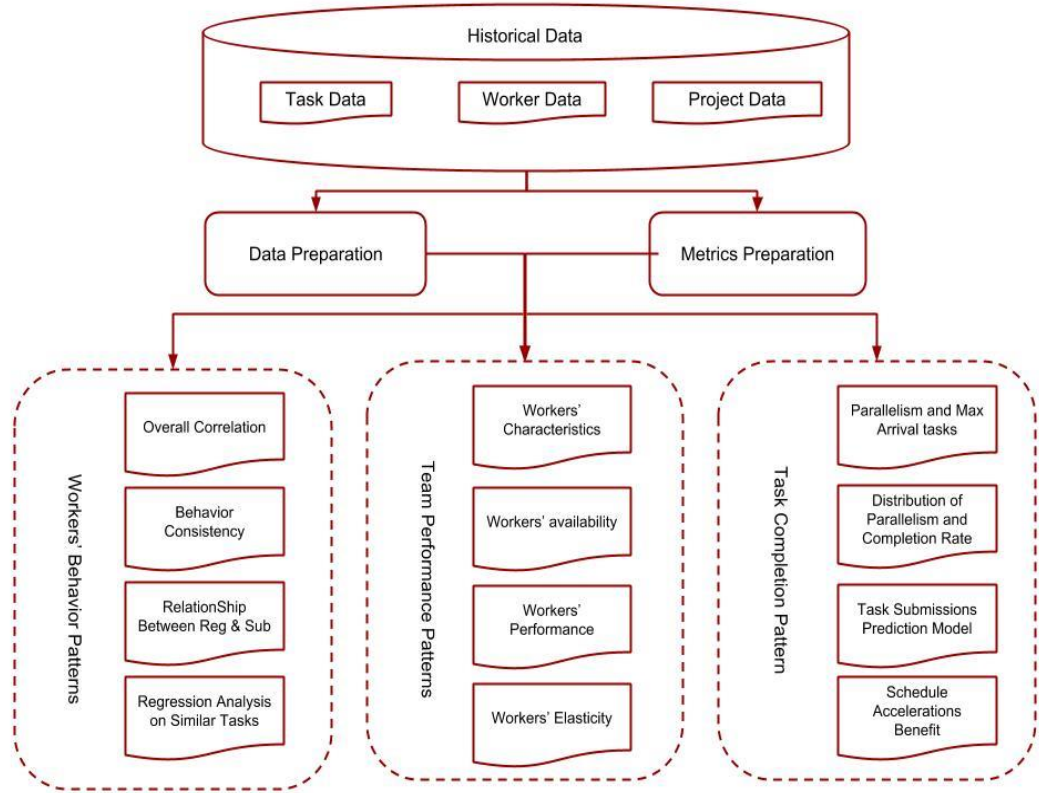

*Figure 3.1-2: Overview of Empirical Studies*

1) *How does the award correlate with worker's behavior in task selection and completion?*
2) *How consistent do workers behave from registering to submitting for tasks?*
3) *How does the number of registrants correlate to the quality of submission?*
4) *For similar tasks, will the number of registrants and submissions increase as award increase?*

**Research Approach:**

To design the appropriate research approaches, the following steps were undertaken:

Phase 1: Create a table of task summaries to identify the effective task attributes on workers' behavior.

Phase 2: Analyze the consistency of worker behavior based on workers' performance and qualified submissions.

Phase 3: Analyze the optimal monetary prize in similar set of tasks to attract higher level of competition.

This part will be fully discussed in chapter 4.

### 3.1.1.2  Team Performance

Software development is rarely done in an isolated environment; instead it increasingly depends on the collaboration among different groups of stakeholders. Such collaborative development processes frequently face challenges from rapidly evolving requirements, conflicting stakeholder needs and constraints, as well as unanticipated human factors. This fact requires a highly adaptive team to work on software tasks. For adaptive teams to leverage CSD to increase team elasticity, it is critical to understand crowd workers' sensitivity and performance and rate of task failure. To address this issue, the following questions were investigated:

1) *How diverse are crowd workers in terms of skill and experience?*
2) *How fast does crowd respond to a task call?*
3) *How reliable are the crowds in submitting tasks?*

**Research Approach:**

To design the appropriate research approaches, the following steps are undertaken:

Phase 1: Create a general overview of workers' characteristics in terms of membership age, reliability and skillsets.

Phase 2: Analyze workers' availability in response to a task, measured in terms of the number of registrants for tasks as soon as a task is uploaded in platform. Also, reliability of task submissions can be measured by the number submissions received and the highest score of the submissions.

Phase 3: Measure the consistency of worker performance by using relative velocity and submissions quality, indicating percentage of submissions duration usage by workers to submit a task and the quality of the submission for each task.

This part will be fully discussed in chapter 5.

### 3.1.1.3 Task Completion

Crowdsourcing has become a popular option for rapid acquisition, with reported benefits such as shortened schedule due to mass parallel development, innovative solutions based on the "wisdom of crowds", and reduced cost due to the pre-pricing and bidding effects. However, most of existing studies on software crowdsourcing are focusing on individual task level, providing limited insights on the practice as well as outcomes at overall project level. To develop better understanding of crowdsourcing-based software projects, following questions will be investigated:

1) *What are the task completion patterns in the CSD platform?*
2) *How does CSD benefit schedule reduction?*

**Research Approach:**

To design the appropriate research approaches, the following steps are undertaken:

Phase 1: Define the metrics of task attributes which describes the basic quantitative characteristics of a task including total associated award and total number of uploaded tasks in a limited period of time.

Phase 2: Analyze worker performance by measuring the total number of workers registering for the task and the total number of submitted work products for the task as well as stability.

Phase 3: Derive empirical evidence on comparing project schedule estimation of crowdsourced projects.

This part will be fully discussed in chapter 6.

### 3.1.2    Simulation Framework

Initially, we assume that workers' trust parameters are constants; however, it is possible that these parameters vary over time. Based on this assumption, we need to evaluate current task scheduling methods in order to come up with a more effective hybrid simulation task execution model to achieve lower task failure ratio.

1)  *How to provide a simulation model to improve the understanding about the CSD process in order to reduce task failure ratio?*

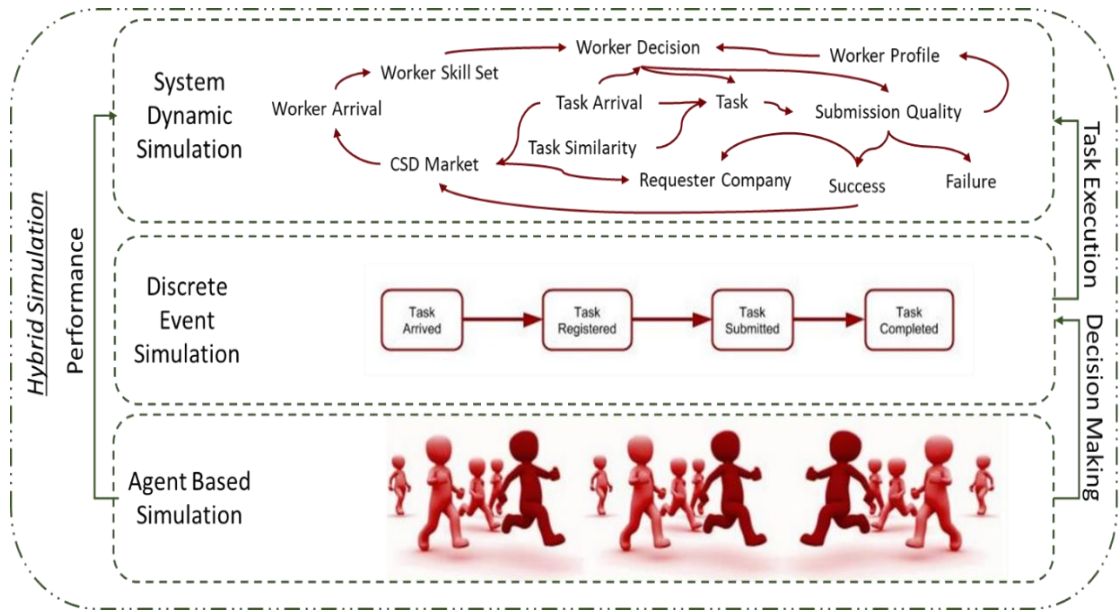

*Figure 3.1-3:  Overview of Hybrid Simulation Model*

**Research Approach:**

To develop the hybrid simulation model, it is important to understand the characteristics of crowd projects, workers' preference and performance as well as task situation and status in both project and platform. A combination of empirical analysis and simulation techniques provides the best tool to create a hybrid simulation model and address all of these factors in crowdsourcing task execution with the aim of minimizing task failure ratio. In this model, task completion will be modeled by discrete event simulation techniques, resource availability will be modeled by agent-based simulation techniques and the CSD platform will follow systems dynamic techniques. Figure 3.1-3 illustrates the overview of the hybrid simulation model.

This part will be fully discussed in chapter 7.

## 3.2   Dataset and Matric

As one of the most successful CSD platforms, Topcoder has over one million registered workers from over 190 countries, averaging 30 thousand logins every 90 days, 7 thousand challenges hosted per year and 70 million dollars in challenge payouts [88]. Software workers compete to design, develop, and deploy software, which, when combined, is launched to the market by the requestor companies, while a monetary award is granted to the two highest scored winners. Tasks are uploaded as competitions to the platform, where crowd software workers can register for the challenges. On average, most of the tasks have a life cycle of 2 to 4 weeks from first day of registration to the submissions deadline. When workers submit the final files, the work is by experts to check the final results and grant the scores. The granted score of a submission is dependent on the task's level of complexity and the time it took to code a solution. The final submission needs to gain a minimum score of 75 to be considered as successful in peer review.

At the project level, a total of 403 projects were further decomposed into a total of 4907 mini tasks, across 14 different challenge types, form Jan 2014 to Feb 2015. The average number of workers per task is about 18, the average task price is about $750, and the average task duration is about 16 days. As shown in Figure 2, 80% of the tasks have less

or equal to 14 registrants, with 3 or less submissions. 13% of all tasks in the database receive more than 20 registrants (650 tasks).

### 3.2.1     *Overall view of TopCoder*

Figures 3.2-1 illustrates task situation in TobCoder from Jan 2014 till Feb 2015. According to our database, on average, 13 tasks from nine different projects is uploaded to the platform, out of which two tasks failed and one was cancelled by the requestors. On average, 135 workers registered for the tasks per day but only 25 of them were submitted.

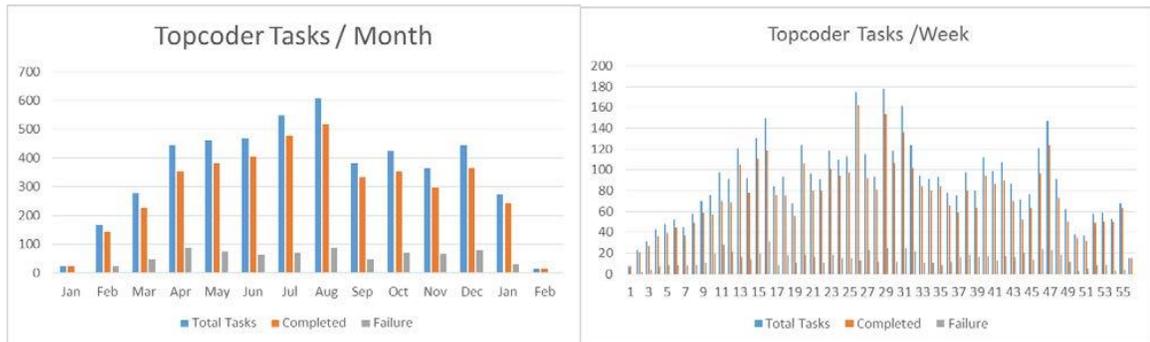

*Figure 3.2-1: Task Situation per Month and Week*

### 3.2.2     *Example of Worker Availability in TopCoder*

The Hercules Android App project is an example of task and resource scheduling in Topcoder. The project was decomposed into 277 tasks, which were uploaded within a duration of 39 weeks, while the whole project duration was 45 weeks. On average 18% of tasks in the entire project were failed. On a weekly basis, there were, on average, 7 tasks uploaded, which faced 13% failure which means that one task was failed per week. Figure 3.2-2 shows the distribution of task and workers in different weeks. As it is shown higher number of uploaded tasks will result in higher number of failure tasks per week; however, higher number of tasks does not guarantee higher number of workers taking tasks.

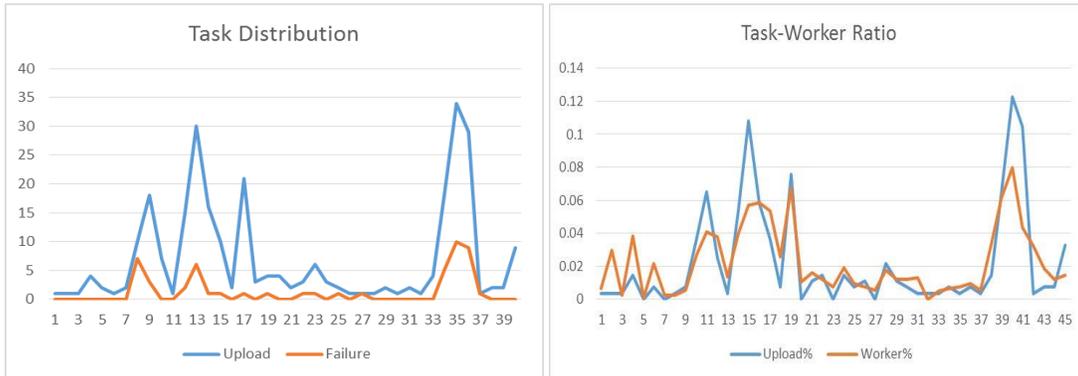

*Figure 3.2-2: Task Status per Week and Worker Arrival per Week*

### *3.2.3    Motivation Example*

The motivation example illustrates a real CSD project on the Topcoder platform. It comprises 19 tasks with a total duration of 110 days. The project experienced a 57% task failure ratio. Eleven of the 19 tasks failed. Two of the task failures are due to client requests and 1 based on failed requirements. The remaining 8 tasks failed due to zero submissions.

On average, the project's tasks were competing against 20 open tasks with similarity more than 60%. Also, the average reliability factor of workers participating in the competition was 0.12.

Deeper analysis showed that the failed tasks due to zero submissions are 3 tasks that were re-posted after each failure. As illustrated in figure 3.2-3, task 2 was cancelled and reposted as tasks 5 and 7. It was finally completed on the 3rd attempt with changes in the monetary prize and task type (i.e. task 7). Task 4 got cancelled and reposted as task 6. Task 6 was completed with no modification. Task 8 also failed at first and was reposted

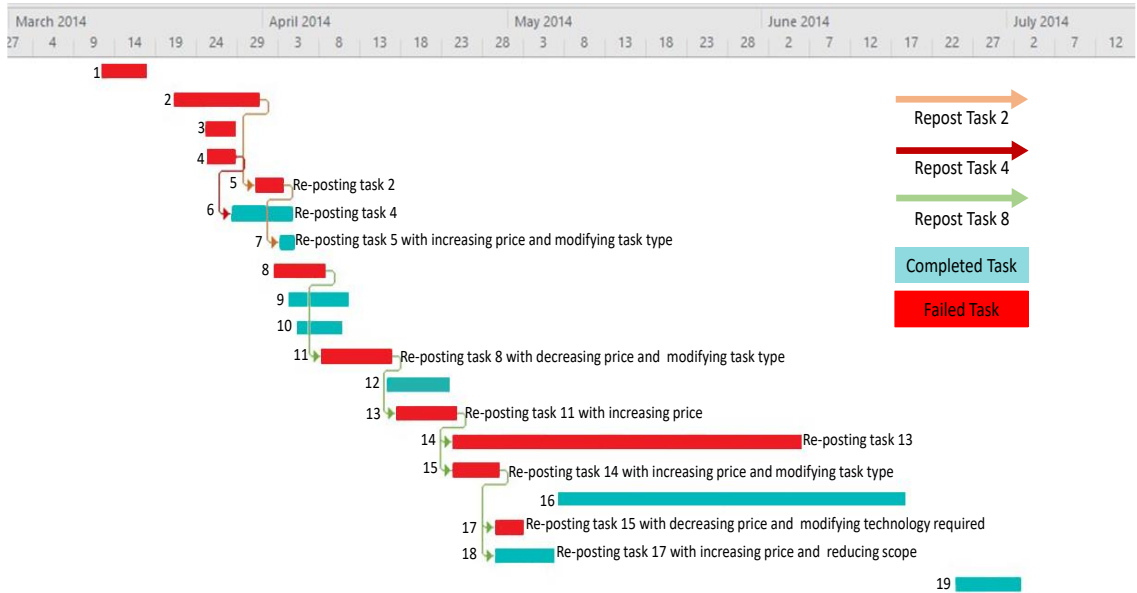

*Figure 3.2-3: Motivation Example Project*

6 times as tasks 11, 13,14,15,17, and 18. There were differences in the monetary prize, task type and required technology in each reposting task process for task 8. It was successfully completed as task 18.

Analysis on task 8, figure 3.2-4, shows that all the failed reposted tasks in the platform were competing on average with 27 similar open tasks with similarity of equal or greater than 80%, while the successful reposted tasks were only competing with 5 similar tasks. Interestingly the task with the lowest number of similar open tasks attracted workers with a higher average reliability factor of 0.19. Moreover, the duration of the reposted tasks of task 8 is 25 days, i.e. 23% of the project duration.

The journey of task 8 in the motivation example supports the fact that task similarity among open tasks impacts task success. Also, lower level of task similarity among open tasks positively impacts attracting workers with higher reliability factor (i.e. task 18 in figure2). Attracting workers with a higher level of reliability factor increases the chance of receiving qualified submissions and task success. Reported research reflects that crowd workers are more interested in working on tasks with similar contexts in terms of skillset requirements [3]. However, none of the available studies focused on the causal similarity among open tasks in the platform and its impact on workers' availability and task failure.

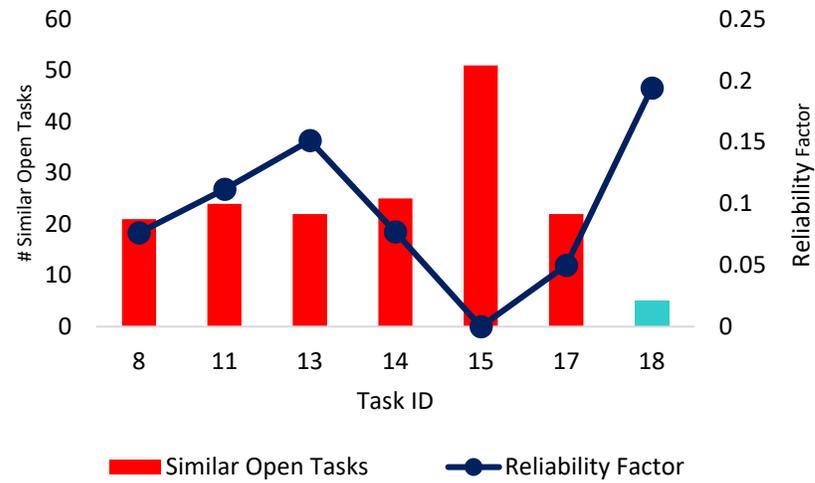

*Figure 3.2-4: Details of Task Similarity and Reliability Factor on Journey of Task 8*

Therefore, providing deeper insight on the relation between worker reliability and task in similarity is a missing factor that would help project managers' decision-making process in scheduling tasks.

# Chapter 4 Workers' Behavior Pattern

Compared with general crowdsourcing platforms with unpaid or low-paid, simple and labor-intensive micro-tasks, software development crowdsourcing platforms are mainly for crowdsourcing competitively paid, complex, knowledge intensive software development tasks. Therefore, it is assumed that there is a much stronger correlation between tasks' attributes and resources' behavior. To evaluate the overall correlation between task attributes and worker behavior we will investigate the following research questions in this chapter.

1) How does the award correlate with workers' behavior in task selection and completion?
2) How consistent do workers behave from registering for and submitting tasks?
3) How does the number of registrants correlate to the quality of submission?
4) For similar tasks, will the number of registrants and submissions increase as monetary award increases?

The details will be discussed in the next sections.

## 4.1 Monitory Prize and Workers' Behavior

In the basic conceptualization structure, three constructs, perceived value, perceived sacrifice, and willingness to buy are utilized to model the role of price and value perceptions [73]. Similarly, in this research, it is argued that in software crowdsourcing, the following constructs may be used to model the role of award and its association with workers' motivation and behaviors, as illustrated in Figure 4.1-1:

- The perceived value of an award drives the workers' motivation to win, i.e. whether the award is attractive enough to motivate potential workers to register to compete.

- The perceived sacrifice represents the required effort or skills to complete the tasks with high quality submissions that potentially leads to winning the award.

- The workers' willingness to compete, which is a trade-off between motivation to win against the perceived sacrifice, i.e. time and effort dedicated on the task.

-  Workers may change their mind due to various distracting or preventing factors. Such factors could be competition anxiety according to the Yerker-Dodson law, unavailable time and resources, etc.

- The chance of winning in the competition is mostly based of workers' skills, availability, and degree of competition. If distracting factors appear that decrease the chances of winning, workers may fail to complete and make submissions to registered tasks.

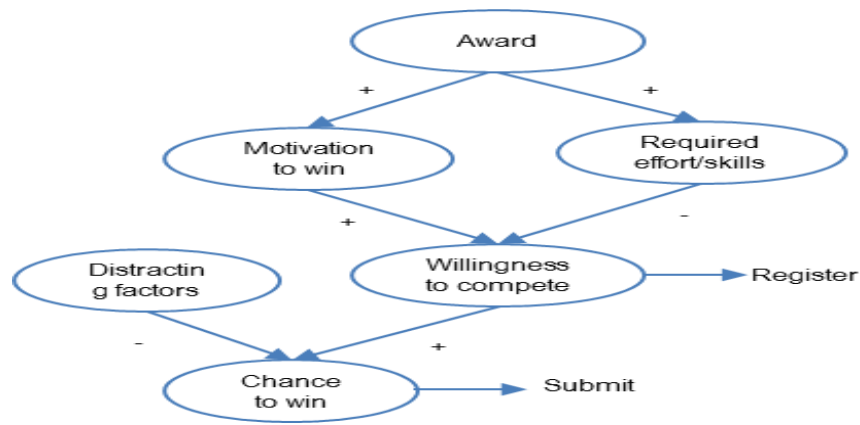

*Figure 4.1-1: Conceptual Award-Worker behavior model*

As shown in Figure 4.1-1, an award plays a dual role in a worker's trade-off in task selection and completion. On one hand, a higher award will lead to greater motivation, which drives a crowd worker to be more willing to register and compete. On the other hand, higher price frequently reflects large chunk of work, higher complexity, or specific requirements corresponding to a smaller pool of qualified workers [8] [34] [35] Consequently, the crowd worker will perceive a higher demand of skills and more effort to win the competition, which will negatively impact the worker's  willingness to compete. A worker's overall willingness to compete represents a trade-off between his motivations to win and effort/skill commitment.

*Table 4-1: Summary of metric data in the whole dataset*

| Metric | Min | Max | Median | Average | STDEV |
|--------|-----|-----|--------|---------|-------|
| Award | 112.50 | 3000 | 750 | 753.78 | 372.48 |
| Size | 310 | 21925 | 2290 | 2978.21 | 2267.99 |
| #Reg | 1 | 72 | 16 | 18.46 | 10.83 |
| #Sub | 1 | 44 | 4 | 5.15 | 4.84 |
| Score | 75 | 100 | 94.16 | 92.52 | 6.20 |

Similarly, the monetary award also plays a dual role in a worker's trade-off in task completion due to distracting or preventing factors. For example, a worker's belief on his strength of competition may change over time, esp. if more workers or strong competitors registered. It is also possible that the resource required for completing a task becomes unavailable, which prevents him/her from completing the task in time.

The available dataset implies that a typical task on Topcoder is priced, on average, as 750$ for 2290 lines of code, and the median numbers of registrations and submissions are 16 and 4 respectively. Also, the median score of the winning submission is 94.16 out of 100. Table 4-1 summarized the tasks attributes in the dataset gathered from Topcoder.

A current common impression is that crowdsourcing is more feasible for easy and simple tasks, the data shows that it is also feasible for complex component development at the scale of 21925 lines of code. The maximum number of registrants of 72 is surprising since the nature of the task is competitive considering that only top-2 winner gets paid. For the purpose of similarity analysis, dataset clustered tasks into different bins according to award size, each bin corresponding to a 100$ increase in award. This results in 30 bins from [100, 200) to [3000, 3100). After cleaning the dataset, it contains 9 application types categorized by TopCoder. The top-4 with the most data points are selected for further comparison analysis. These include "application management" (33 tasks), "communication" (34), "data management" (142), and "developer tools" (53).

Most of the comparison analysis is conducted on the top-4 subsets. Figure 4.1-2 illustrates the task characteristics across the selected 4 subsets. It can be seen that all four types are with the same median award, i.e. $750, as well the same median score, i.e. 94 out of 100. The median sizes are also very close, ranging between 2327 to 2541 lines of code. The "data analysis" type contains tasks with the largest variation in size, award, number of registrants and number of submissions. The "data analysis" subset involves the highest number of registrants and submissions, while "developer tool" subset involves the least number of registrants and submissions. Both "communication" and "developer tools"

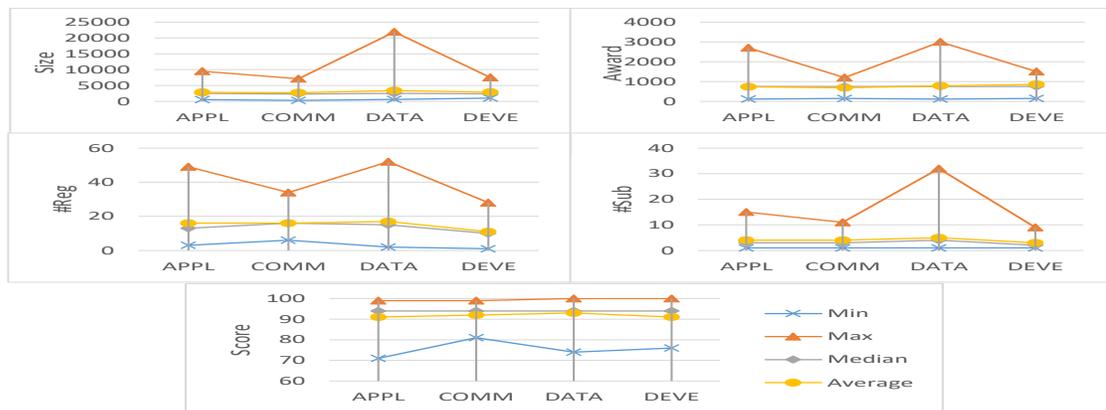

*Figure 4.1-2: Characteristics of 4 subsets*

types have smaller variation than the other two types for almost all the metrics compared.

### 4.1.1 Overall correlation of Monitory Prize and Task Competition Level

Figure 4.1-3 shows an overall profile of the scatter-plot between number of registrants and amount of award on main dataset. While there is no clear trend-line obtained from the data, the result shows a very weak negative correlation of -0.015 between them. There are certain dominant award patterns observed. One notable observation is that 311 out of the 494 (i.e. ~63%) tasks were priced with a typical award of $750. However, the number of registrants corresponding to these 311 tasks vary significantly from 2 to 72. As discussed earlier, about 63% of all tasks were priced the same as the median award of $750. Additionally, the top-7 award settings (i.e. $750, $450, $600, $900, $1050, $150, and $375) consist of 85% of all tasks. This is possibly due to the yesterday's weather effects when most similar tasks were priced at certain value, future tasks are more likely to follow the past experience and priced the same. This partly explains the very weak correlation between award and registrants.

The four regions in Fig. 4.1-3 are split by the vertical line of median award and the horizontal line of average #registrants. Table 4-2 lists additional rationales and statistics for each of the four regions. The results show that about 86.5% (i.e. 243 in region I and 185 in region II) of all tasks are priced under the median award (i.e. $750).

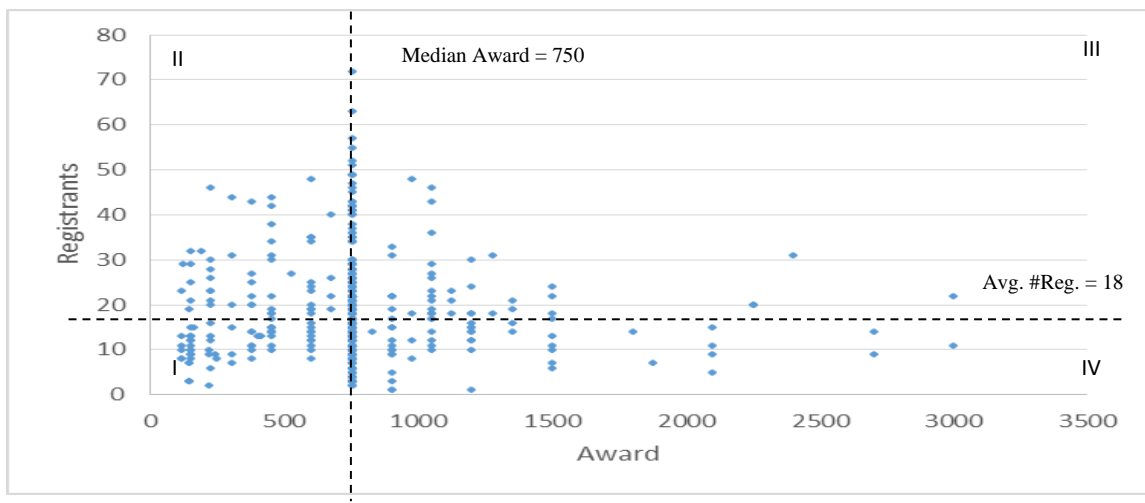

*Figure 4.1-3: Scatter-plot of #Registrants vs. Award*

For these cheaper tasks, 243 tasks involved fewer workers than the average (i.e. less competition), and 185 tasks involved more workers than the average (greater competition). The percentage of broader competition is about 43%. For more expensive tasks the percentage of broader competition is 33%, which is about 10% lower than that for the cheaper tasks. This indicates that lower priced tasks have a greater chance of attracting broader competition.

*Table 4-2: Rationales and statistics of tasks in four regions*

| Region | Award | #Registrants | #Tasks |
|--------|-------|--------------|--------|
| I | <=750; cheaper | <=18; less competition | 243 |
| II | <=750; cheaper | >18; broader competition | 185 |
| III | >750; more expensive | >18; broader competition | 22 |
| IV | >750; more expensive | <=18; less competition | 44 |

Pearson's r correlations were calculated between award and size, registrants, submissions, and score and listed in Table 4-3 The results show a strong correlation between award and score (i.e. -0.71), a medium correlation between award and submission

*Table 4-3: Summary of metric data in General dataset*

| Bin | #Tasks | Award | Size | #Reg | #Sub | Score |
|-----|--------|-------|------|------|------|-------|
| 1 | 32 | 142 | 2562 | 14 | 4 | 94 |
| 2 | 17 | 226 | 2582 | 18 | 4 | 95 |
| 3 | 19 | 353 | 2886 | 19 | 5 | 95 |
| 4 | 24 | 447 | 2766 | 20 | 8 | 96 |
| 5 | 25 | 612 | 2663 | 21 | 6 | 95 |
| 6 | 311 | 750 | 3129 | 19 | 5 | 92 |
| 7 | 23 | 913 | 3468 | 15 | 4 | 94 |
| 8 | 19 | 1050 | 2286 | 22 | 5 | 91 |
| 9 | 15 | 1210 | 2597 | 17 | 4 | 95 |
| 10 | 9 | 1500 | 2509 | 14 | 3 | 87 |
| Sum: | 494 | Correlation: | -0.09 | -0.13 | -0.40 | -0.71 |

(i.e. -0.40), and weak between award and registrants (i.e. -0.13). There is very minimum correlation between award and size.

Figure 4.1-4 illustrates the range of award in each bin, and the associated average registrants and submissions. For tasks priced between 100 and 500 (i.e. tasks in bin 1, 2, 3, and 4), both curves show an upward trend, which means as award increase, both registrants and submissions increase as well, which indicate positive association. However, for tasks

priced between 700 and 1000 (i.e. tasks in bin 6, and 7) and between 1200 and 1600 (i.e. tasks in bin 9 and 10), both curves show a downward trend, which indicate negative association. For tasks priced between 1000 and 1100 (i.e. tasks in bin 8), both curves also indicate a second peak effect.

One possible reason is that most cheap tasks (e.g. bin 1, 2, 3, and 4) are relatively easier and require lower experience and less skill sets, which corresponds to a much larger pool of potential workforce, considering the workforce supply pyramid. For example, on TopCoder platform, typical bug-fixing tasks are associated with $150 awards, compared with more complicated application development tasks priced around $750.

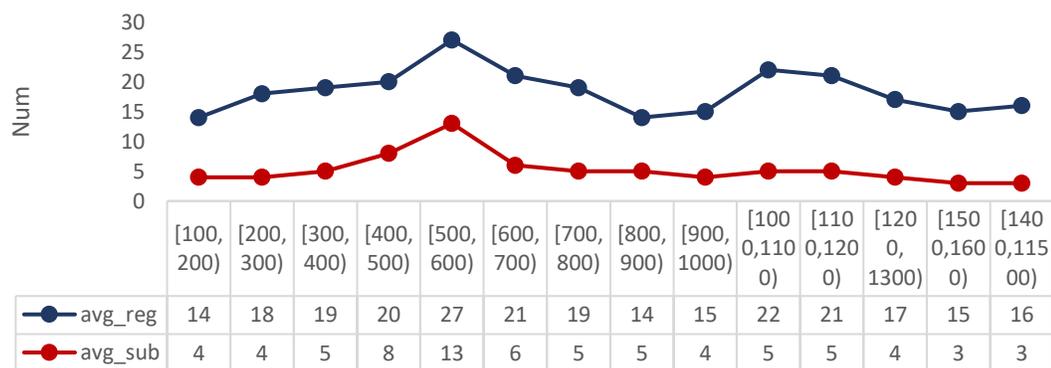

| | [100, 200) | [200, 300) | [300, 400) | [400, 500) | [500, 600) | [600, 700) | [700, 800) | [800, 900) | [900, 1000) | [100 0,110 0) | [110 0,120 0) | [120 0, 1300) | [150 0,160 0) | [140 0,115 00) |
|---|---|---|---|---|---|---|---|---|---|---|---|---|---|---|
| avg_reg | 14 | 18 | 19 | 20 | 27 | 21 | 19 | 14 | 15 | 22 | 21 | 17 | 15 | 16 |
| avg_sub | 4 | 4 | 5 | 8 | 13 | 6 | 5 | 5 | 4 | 5 | 5 | 4 | 3 | 3 |

*Figure 4.1-4: Relationship among award (X-axis), average registrants (blue curve, Y-axis), and average submissions (orange curve, Y-axis)*

Another interesting observation is that the average number of submissions for tasks over $500 remains relative stable, i.e. between 4 and 6. Considering the tasks in the dataset only award the top-2 winners, it is not practical and rationale for most workers to keep competing on a task with 4 or 5 submissions, esp. for generally effort-demanding expensive tasks.

Generally, the reported results confirm that, in task selection, the number of registrants will decrease as award increase; in task completion, the number of submissions and score will decrease as award increase.

It is interesting to observe how award negatively correlates to all four metrics in the General dataset. This indicates that overall, as award increase, the number of registrants, the number of submissions, and the quality of the final submission all decrease. This observation supports the negligible negative roles that award plays in worker behavior in the conceptual model, as shown in Figure 4.1-1. The negative coefficient between award and submission implies that the majority workers have a relatively low motivation to complete and submit for more expensive tasks, even though they are associated with high awards. The negative coefficient between award and score implies the average quality of submission is decreasing as award increase, which could be associated with considerably less competition for more expensive tasks.

This agrees the finding in [37], which reports that as the task becomes more complex, it will need registrants with higher skills compete on and consequently a smaller number of submissions. For example, the average number of submissions for general tasks above $500 remains quite stable, i.e. 4~6 according to results in Figure 4.1-4. Similar results about the negative impact of award on task selection were reported in [10] [8]. One related phenomena of "cheap talk" is reported in [36], referring for tough competitions, strong contestants strategically employ this strategy to deter additional competitor entry into the task.

One implication from these results is that it is not cost-effective to simply raise award in order to attract broader competition, esp. for competitive, 2-winner software crowdsourcing tasks. To encourage broader competition, it is recommended to decompose the task into smaller pieces for more accurate pricing and broader work involvement.

### 4.1.2    *Workers' Behavior Consistency Trend Analysis*

The behavior consistency analysis is mainly to investigate the number of submission and submission ratio data. A low submission ratio represents a low behavior consistency between registration and submission. Our analysis result shows that on the "Main" dataset, there is a strong positive correlation of 0.71 between number of submissions and number of registrants. However, the median and average submission ratio is 0.25 and 0.30, respectively. This implies that on average, only about 30% of registrants would submit work artifacts by given deadlines.

To understand whether different application types have different consistency patterns, we analyze the relationship between registration and submission data on four application subsets including the "APPL", "COMM", "DATA", and "DEVE". We compare different fitting models and the distributions used in the following discussion represent the better fits according to R-square values.

It is observed that the relationship between the number of registrants and submissions is better fitted in a linear distribution than the exponent distribution. As shown in Figure 4.2-5, the corresponding R-square values for the four trend lines are 0.48, 0.61, 0.43, and 0.43 respectively, which indicates a medium to good fit.

Though Figure 4-1.5 indicates that the number of submissions increases as the number of registrants increase, Figure 4.1-6 shows all decreasing trend in the submission ratio. In addition, the data shows that the relationship between number of registrants and submission ratio is better fitted in logarithm distribution than linear or exponential distributions. The model fitness may be improved through further outlier analysis, but for the purpose of our study, it is more important to observe evidence for such decreasing trend as the degree of competition goes up. More specifically, for all four application domains, there are decreasing trend lines between number of registrants and submission ratio. The decreasing trend is the most significant for tasks in "DEVE" domain.

Other interesting observations include the perfect submission ratio of 1 in both "DATA" and "DEVE" subsets. But the perfect submission ratio is only observed for tasks with fewer registrants, i.e. between 3 and 5. For tasks involving broader competition. For example, for tasks with more than 20 workers, about half (47%) has a submission ratio lower than 20%. "COMM" subset generally has lower submission ratio compared with the other three subsets. Moreover, there is a strong positive correlation of 0.71 between number of submissions and number of registrants. However, there is a decreasing tendency in making submission as the number of registrants increases.

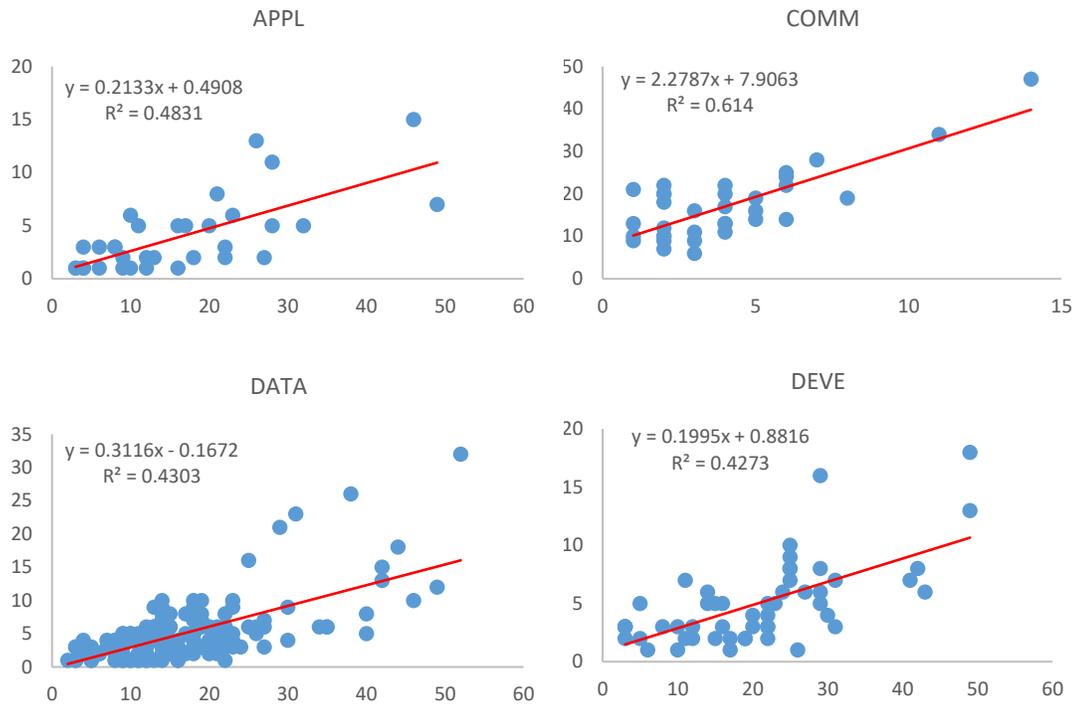

*Figure 4.1-5: Relationship between number of registrants (X-axis) and number of submissions (Y-axis)*

Reported results suggest that, the more registrants are attracted into the task, the more submissions are expected to receive (i.e. trend lines in Figure 4.1-5), but the lower the submissions ratio will be (i.e. trend lines in Figure 4.1-6). These results indicate that by attracting more registrants, there are higher chances of receiving satisfactory submissions, however the willingness to submit for each individual worker reduces. This reflects the behavior inconsistency from task registration to task completion, which supports the assumption of distracting factors in the conceptual model. Possible distracting factors include competition pressure, insufficient time to complete the task, etc.

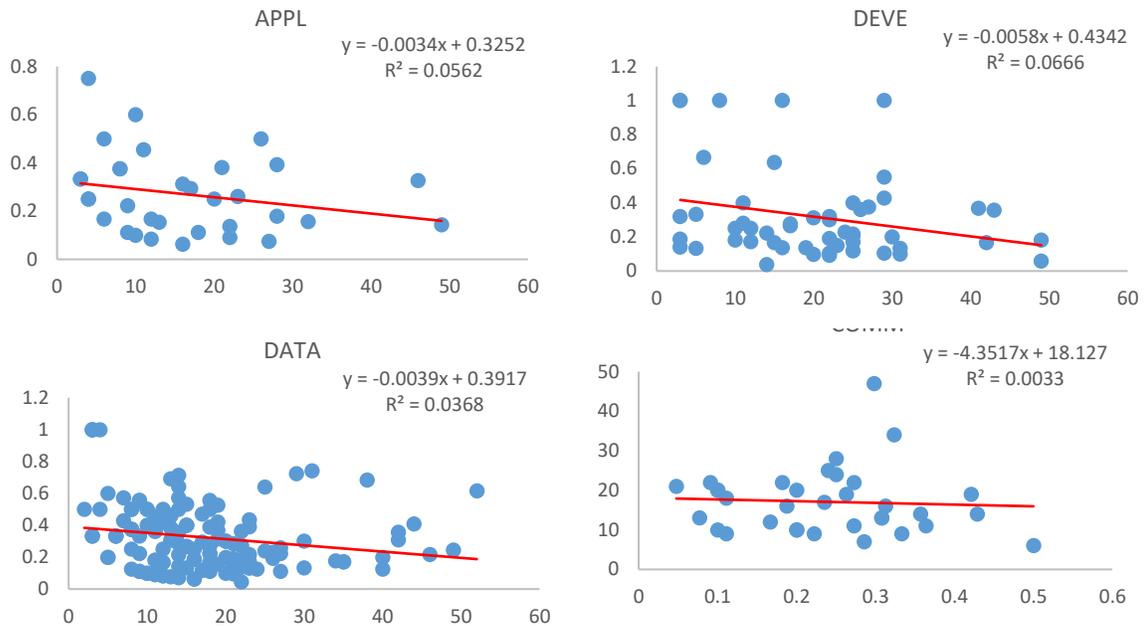

*Figure 4.1-6: Relationship between number of registrants (X-axis) and submission ratio (Y-axis)*

The submission ratio for past similar tasks would provide insights on deriving rules of thumb for crowdsourcing risk management. For example, the results show that for tasks involving broader competition of more than 20 workers, the majority only has a submission ratio lower than 20%. Task requesters may monitor the growth trend of registrants and assess the potential low submission risks with respect to different domains.

### 4.1.3    Relationship between Workers and Quality

Ideally, the submission quality could be assured through broader competition, i.e. greater number of registrants. The results, as shown in Figure 4.1-7, shows a positive correlation between that number of registrants and granted score. Due to the low correlation, the trend line equation and r-square details are not shown in the charts. The correlation coefficient in the Main dataset is 0.19, which indicates that higher number of registrants is associated with higher granted score.

It is noticeable that for tasks involving broader competition (i.e. more than 20 workers), there is a higher chance (71.5%) of receiving better quality submissions, i.e. scoring above average score of 92.5. It is clear that, there is a strong positive correlation of 0.71 between number of submissions and number of registrants. However, there is a decreasing tendency in making submission as the number of registrants increases.

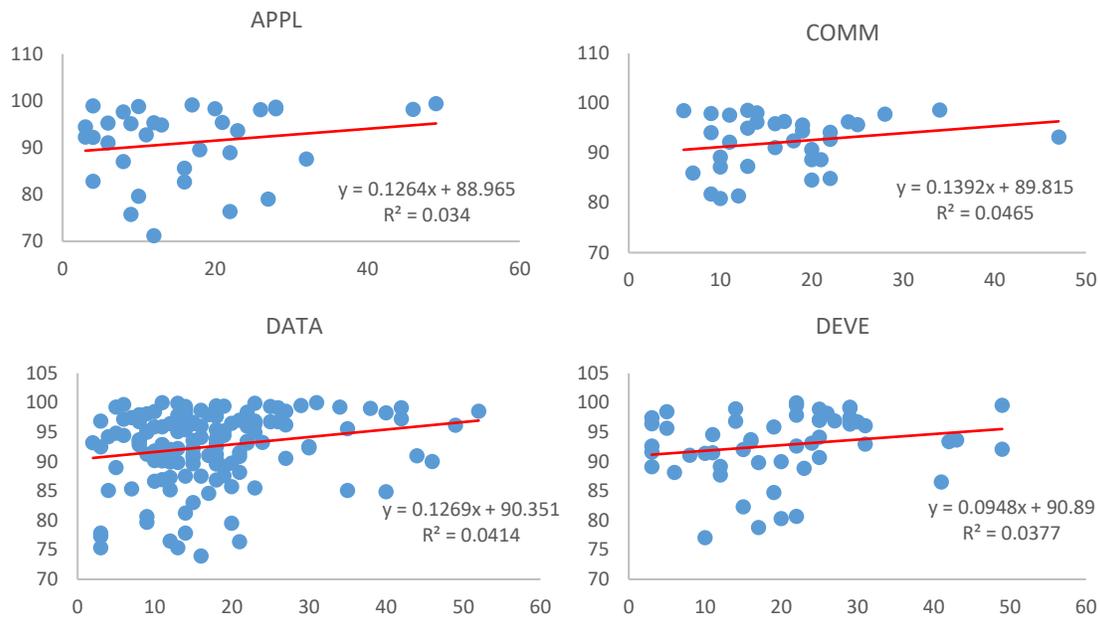

*Figure 4.1-7: Relationship between number of registrants (X-axis) and score (Y-axis)*

The positive correlation between number of registrants and submission scores confirms the improved quality due to leveraging on worker diversity through broader participation in crowdsourcing. However, the low correlation of 0.19 indicate that the previously reported great impact of team diversity [6] could be limited or weakened by many distracting factors in software crowdsourcing.

Similar viewpoint has been proposed in regarding a maximum point for this effect after which not only with the increase of diversity the performance does not improve but also it deteriorates due to problems that emerge because of coordination and communication [31]. Additionally, in our study, we believe that in competitive crowdsourcing, the increasing competition intensity is a greater problem than coordinating and communicating within large crowds. To control the risks of getting unsatisfactory submissions, one possible option would be to involve more than 20 workers into the competition.

### 4.1.4    Regression Analysis on Similar Tasks

For simplicity, task similarity is characterized based on application type, size, and award data in this analysis. The analysis is conducted on dataset "COMM", since it contains the most data points with the largest variation in almost all metrics. This provides us the most opportunity to observe how award change associates with other metrics, i.e. #registrants and #submissions.

The COMM dataset is further split into four subgroups corresponding to different size, as summarized in Table 4-4. Four abnormal cases are manually analyzed and identified as outliers, excluded from further analysis. For each group, the median #registrants and #submissions are calculated for each observed award value. Please note that this fourth sub-group as listed in Table 4-4 is not a good representation of similar tasks, since the size range in this subgroup is from 5kloc to almost 22kloc. It is included just for comparison purpose.

Comparison of different regression types including linear, power, exponent, and polynomial distributions on each subgroup was used, in order to fit the registrant data and the submission data with respect to the award. The results show that the polynomial distribution fits the best for all four groups, and the R-square values of the polynomial equations in each group are also listed in Table 4-4.

*Table 4-4: Model fitness details for the 4 subgroups*

| Sub-Group | # Tasks | #observed award | #outlier removed | Model Fitness - registrants | Model Fitness - Submissions |
|-----------|---------|-----------------|------------------|-----------------------------|-----------------------------|
| <2kloc | 56 | 13 | 1 | 0.69499 | 0.44283 |
| 2kloc~ 3kloc | 31 | 6 | 1 | 0.59802 | 0.85468 |
| 3kloc~ 5kloc | 27 | 9 | 1 | 0.76909 | 0.0619 |
| >5kloc | 26 | 11 | 1 | 0.46303 | 0.57195 |

Figure 4.1-8 illustrates the regression lines for #registrants and # submissions in all four subgroups. For the first three subgroups, the trend lines all follow the inverted U-shaped curve, which confirms the dual roles of award in the reported conceptual model. This implies that in this particular dataset, for smaller crowdsourcing tasks under 5KLOC, both the number of registrants and the number of submissions will first increase and then decrease as award increase. More specifically, the optimum award is observed to be around $1050 when the #registrants' curves reach the peak in the first two subgroup (i.e. <2kloc and 2~3kloc), and the optimum award in the third subgroup (i.e. 2~5kloc) is observed to be around $750.

Another interesting observation is that for smaller tasks, e.g. less than 2kloc, typical small differences between two award levels are $30 and $75, which indicates smaller tasks are easier to price more accurately than larger tasks, and workers will be more sensitive to award differences in smaller tasks.

As comparison, the data in the last sub-group (i.e. size greater than 5kloc) shows an opposite trend, with a valley at around $1500. One of the reasons is the large variation in size as introduced earlier, which limits its representativeness of similar tasks. The other reason is that the 6 large tasks with more than 10kloc in this subgroup are associated with a completion duration of less than 7 days, which implies that there is possibly substantial amount of code reuse. Finally, for similar tasks, the relationship between award and worker behavior follows a variety of inverted U-shape curves.

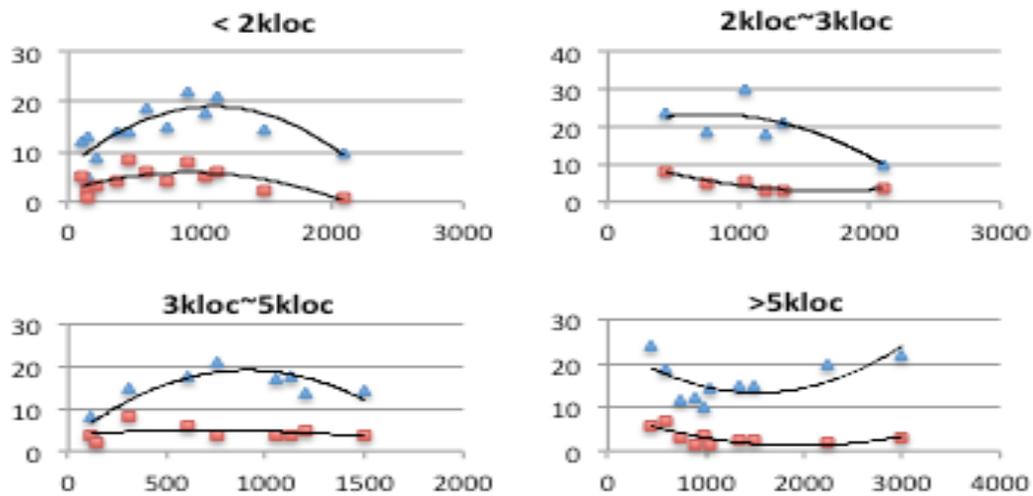

*Figure 4.1-8: Relationship between #registrants (blue triangle, Y-axis) and award (X-axis), as well as between #submissions (orange rectangle, Y-axis) and award (X-axis), for all four subgroups*

Results from regression analysis demonstrate some examples on optimal award within a similar set of tasks in terms of attracting the most registrants and submissions. Before reaching the optimal award, it is possible that the #registrants will increase as award increases; after the optimal award, the #registrants will decrease as award increases. According to the conceptual model, it is assumed that the general relationship between award and worker behavior, follow similar inverted U-shaped curve as Yerker-Dodson curve. On one hand, the upward phase of the U curve is mainly due to the positive effect of small increase in award that cause the workers to perceive the task to be high value and return, and hence more willing to compete. On the other hand, the downward phase of the U curve would be attributed to the negative effects of both required effort commitment and competition distracting factors. The observed U-shaped curves in the first three charts of Figure 4.1-8 supports our model assumption. The last chart of Figure 4.1-8 does not support or reject such assumption because it is not a representative set of similar tasks as discussed earlier in the study.

This further indicates that it is rather risky to attempt to incentivize broader competition via increasing the amount of award for competitive tasks, without considering potential negative effects. It is recommended for task requesters to design hybrid competition combining both collaborative and competitive tasks in order to not only involve diversified workers to contribute, but also mitigate negative competition effects, and eventually obtain higher quality of deliverables.

Another implication is that results from this analysis maybe leveraged to extend existing studies on award pricing for software crowdsourcing tasks [74]. Possible extensions include:

Incorporating typical pricing strategies such as broader competition or higher quality;

Providing sensitivity analysis for task requesters to explore different options with respect to their needs and preferences;

Deriving answers to questions like "what the strategic price should be?" in order to incentivize broader worker participation as well as higher quality of final submissions.

## 4.2    Task similarity and Workers' Behavior

It is important to understand the stability and failure rate of the parallel scheduling in both project level and platform level with respect to task similarity. To do so, it is required to understand tasks elements form different stakeholders' point of view. From project managers' perspective, a task in developing phase is representing required programing skill set, required platform, number of allowed using technology. While a crowd worker will put more attention to monetary prize, Task duration, Task type, Task size, Task context and even Requestor Company brand.

It seems applying TF-IDF is a great way to analyzed similarity analysis from project managers point of view, however for understanding the similarity from crowd workers perspective we need to apply a causal similarity analysis on top of TF-IDF.

We adopt the task model to only contain task attributes that is seen by workers in the platform. To answer task similarity in the platform, first we need to understand tasks local distance ($Dis_{i,j}$). The Task distance is represented as follows:

$$Dis_{i,j} = (max(A_i^j), max(ESD_i^j), max(LED_i^j),\ match(Tt_i^j),\ match(Tech_i^j),\ TF\text{-}IDF(Req_i^j))$$

In which award represents the maximum monetary prize of different uploading time of an individual task.

In which:

ESD$_i$          is the task (i) (T$_i$) registration start date in the platform,

LED$_i$          is the task (i) (T$_i$) submissions date in the platform,

In order to find distance of task requirements, the text converted to vector format, keeping only meaningful and descriptive tokens processed by tokenizing and stop word removal via TF-IDF method [81].

$$IDF(v) = \frac{M}{Y},$$

Where n is total number of task requirements, and Y is number of task requirements with term" in it.

$$TF(v,i) = \frac{Z}{W},$$

Where z is total number of term "X" repeated in task requirement (i), TReq$_i$, and W is maximum number term "x" repeated in any strings in TR$_i$.

$$TF\text{-}IDF(Req_i) = (log(IDF(v)))/(12+log(12TF(v,i)))$$

Therefor with respect to the introduce variables, task similarity (TS$_{i,j}$) would be:

$$TS_{i,j} = \frac{(DiSi * DjSj)}{|DiSi| * |DjSj|}$$

## 4.2.1   Task Availability

Now we should analyze task stability and failure in both project and platform level based on similar tasks uploading time frame. Also, task submissions trust-ability is another factor we need to analyze. Different studies focus on task submissions ratio but amount of qualified submissions.

Project level task stability will be calculated based on average submission ratio (SR$_i$) of a worker per task registration in a project:

$$Project\ level\ Stability = \frac{\sum_{i=1}^{n}(PRSR\ i)}{n}$$

In which submissions ratio is number of submission ($PRS_i$) based on number of registration ($PRR_i$) per worker in an individual project:

$$Project\ level\ Submissions\ ratio = \frac{\sum_{i=1}^{n}(PRS\ i)}{\sum_{i=1}^{n}(PRR\ i)}$$

While platform level task stability will be calculated based on average submission ratio ($SR_i$) per worker of similar tasks registration:

$$Platform\ level\ Stability = \frac{\sum_{i=1}^{n}(PLSR\ i)}{\sum_{i=1}^{n}(ST i)}$$

Where submissions ratio is number of worker i submissions ($PLS_i$) based on number of registration for similar tasks ($RST_i$) in the platform:

$$Platform\ level\ Submissions\ ratio = \frac{\sum_{i=1}^{n}(PLS\ i)}{\sum_{i=1}^{n}(RST\ i)}$$

Also, project failure rate defined as ratio of number of cancelled tasks (PRCTi) per total number of tasks per project ($PRT_i$) in same time frame:

$$Project\ Failure\ rate = \frac{\sum_{i=1}^{n}(PRCTi)}{\sum_{i=1}^{n}(PRTi)}$$

When, Platform failure rate defined as ratio of number of cancelled tasks (CTi) per total similar task in platform ($PLST_i$) in same time frame:

$$Platform\ Failure\ rate = \frac{\sum_{i=1}^{n}(CTi)}{\sum_{i=1}^{n}(PLSTi)}$$

Trust-ability ratio is number of accepted tasks ($AT_i$) based on number of submitted tasks in a project per worker.

$$Project\ Trust\text{-}ability\ Ratio = \frac{\sum_{i=1}^{n}(PRATi)}{\sum_{i=1}^{n}(PRSTi)}$$

When platform trust-ability ratio is number of accepted tasks ($PLA_i$) per total similar submitted tasks in platform ($PLS_i$) in the same frame time:

$$Platform\ Trust\text{-}ability\ Ratio = \frac{\sum_{i=1}^{n}(PLAi)}{\sum_{i=1}^{n}(PLSi)}$$

Also, project level task density is the ratio of arrival task per project per week by total number of tasks per project:

$$Project\ level\ Task\ Density\ Ratio = \frac{\sum_{i=1}^{n}(PRAT\ i)}{n}$$

And platform level task density is the ratio of similar arrival task (PLSA$_i$) per week by total number of open tasks (PLO$_i$) per week in platform:

$$Platform\ level\ Task\ Density\ Ratio = \frac{\sum_{i=1}^{n}(PLSAi)}{\sum_{i=1}^{n}(PLOi)}$$

Now we need to analyze total time development usage per project to be performed completely. Then we need to compare the result with available traditional schedule acceleration methods with the available result.

### *4.2.2   Project Level Overview*

Figure 4.2-1 presents success ratio elements for projects with different number of tasks in different time frame. As it is shown submissions ratio is around 20% for almost all the projects while the project stability is increasing by increasing number of decomposed tasks per project and it raised up to almost 40%. It is expected that by increasing number of task per project, task density in different time frame increase, however the average task density factor is decreasing by increasing number of decomposed tasks. One reason can be lower level of task dependency in project with higher number of decomposed tasks.

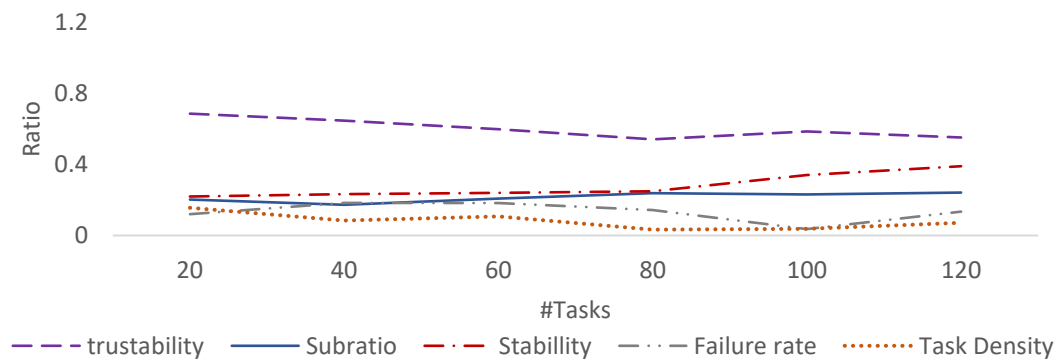

*Figure 4.2-1: Project level Status*

Another interesting factor in this chart is submissions trust-ability. It is expected that by decomposing a project to higher number of available tasks, competition level raised, and consequently higher number of qualified submission will be received. However, our result doesn't support such believe. Figure 4.2-1 clearly presents that by increasing the number of decomposed tasks in the same frame time, submissions trust-ability is decreasing. According to figure 4.1-1, a project with 20 decomposed tasks will receive almost 70% qualified submissions, i.e 3 qualified submissions out of 4 submitted tasks. While a project with 120 tasks will receive close to 55% qualified submissions, i.e 15 qualified submissions.

### 4.2.3   Platform Level Overview

Figure 4.2-2 shows the average distribution of task Submissions ratio and task similarity ratio per week in Topcoder platform. It is clear that higher task similarity in the platform does not guarantee higher submissions ratio. The average similarity ratio is 81% per week. And almost 63% of the weeks' task similarity is greater or equal to the average similarity ratio. Interestingly during this time task submissions ratio increased to on average 20% per week. However, the minimum similarity ratio of 67% occurred in week 55, the submissions ratio was 19% which is equal to average submissions ratio per week. Also, the maximum similarity ratio of 88% could lead to 21% submissions ratio. Also, the highest submissions ratio of 43% happened under similarity ratio of 87% which is very close to the maximum similarity ratio.

Interestingly, figure 4.2-2 shows that highest and lowest failure ratio per week is happened while the similarity ratio was 82% and 84% respectively, which is almost among top 5% of task similarities. Also, in almost half of the weeks, i.e 25 weeks, failure ratio is above average while the similarity ratio is above average as well.

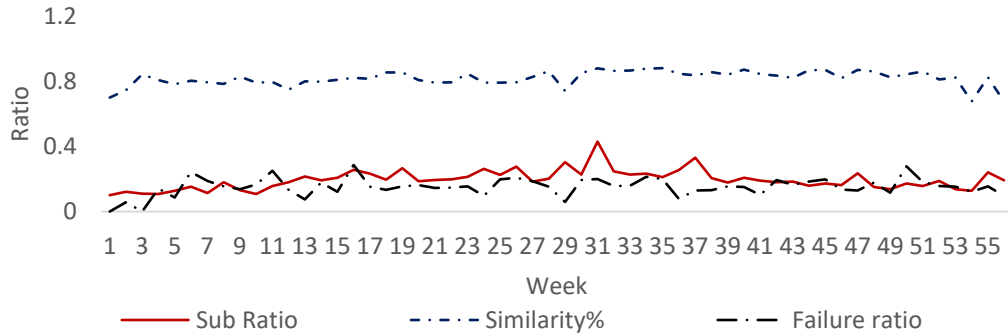

*Figure 4.2-2: Platform level Performance based on Task Similarity*

After applying Similarity analysis on the data base, we cluster the result to four different groups of tasks with 90% similarity, Tasks with 80% similarity, Tasks with 70% similarity and finally tasks with 60% similarity. We analyzed all these clusters at the same time to confirm the hypothesis regards to impact of level of task similarity on level of competition.

It is expected that accessing to higher number of similar available tasks will reduce the registration ratio. According to figure 4.2-3, with arriving higher number of similar tasks to the platform will reduce the attracting available workers, due to higher number of different task choice. However, our results confirm the fact that workers prefer to work on similar task contexts [61] [62] [12]. It is clear that tasks with more than 90% similarity successfully attract higher number of registrants in compare with other similarity cluster.

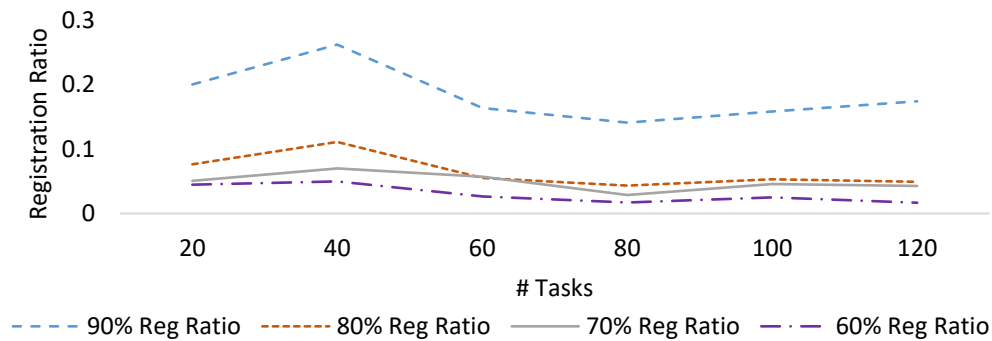

*Figure 4.2-3: Impact of task similarity on registration ratio*

Figure 4.2-4 presents impact of level of task similarity on submissions ratio. However higher number of available tasks will lead to higher submissions ratio, it seems that higher similarity level among tasks will cause lower submissions ratio. While 90% level similar tasks are providing higher submissions ratio for lower number of available tasks, interestingly it will provide one of the lowest submissions ratio for higher number of available tasks. Also 60% level of task similarity lead to almost highest task submission for higher available tasks. one reason can be higher number of available choices for workers which cause them register for a long queue of tasks and drop most of the due to short amount of available time to work on them.

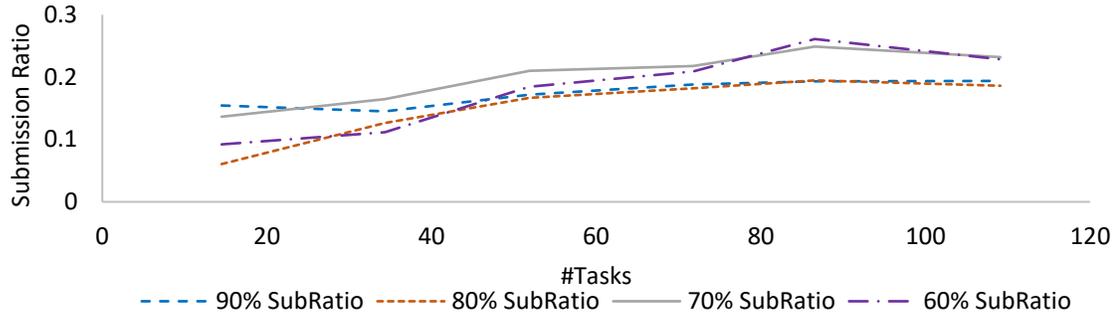

*Figure 4.2-4: Impact of task similarity on submissions ratio*

However, it is expected that higher number of available tasks in a project cause higher level of task density. Figure 4.2-5 shows that higher level of available task will negatively effect on task density in the platform. Interestingly, there is a higher level of task density among 90% similarity level, while the 60% similarity level represents lowest task density level. This shows that most of projects are decomposed to a higher level of task similarity.

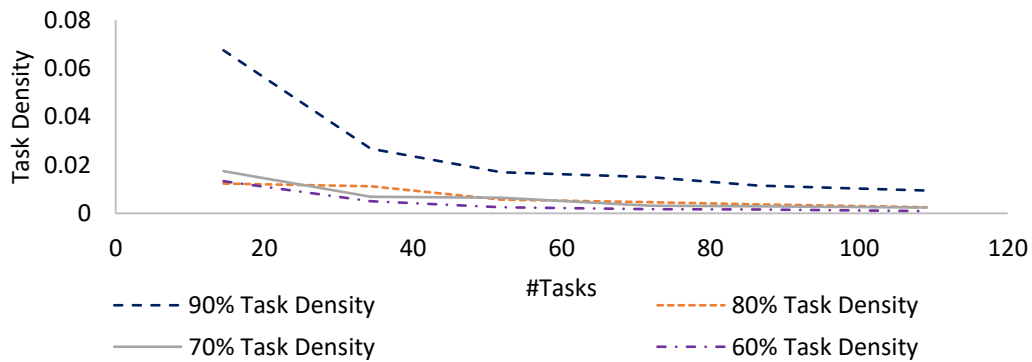

*Figure 4.2-5: Impact of task similarity on task density*

It is important to understand the impact of task similarity level on platform stability. Figure 4.2-6 illustrates the result of such analysis. Interestingly it seems that 70% of task similarity level guarantees higher level of average stability for different number of available tasks in same time frame. However, platform stability is almost following the same pattern of platform submission ratio, the difference between different clusters are very small. The highest average stability belongs to tow lowest task similarity level with 20%, and the lowest stability ratio belongs to 80% similarity level. The highest similarity level will bring 19% stability ratio. However, it seems stability ratio for similarity level greater than 80% is almost static.

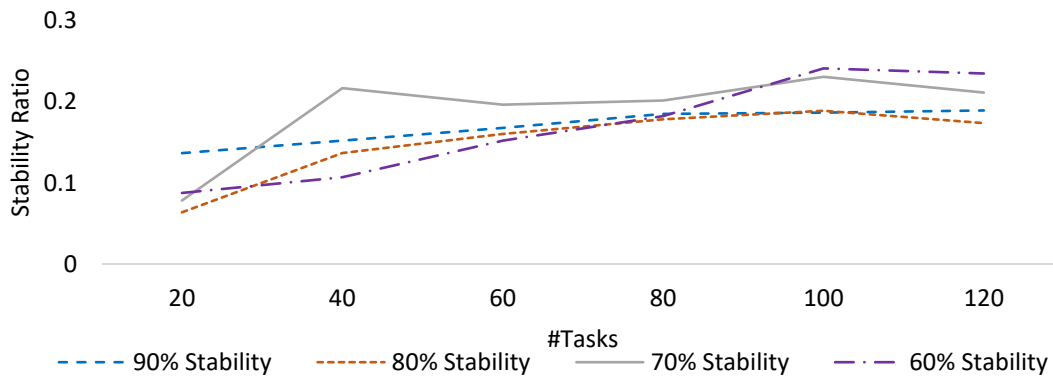

*Figure 4.2-6: Impact of task similarity on Platform Stability*

While it is expected that higher level of task similarity among available tasks correlates with higher task success level, due to similar skillset requirements, the initial result of similarity analysis confirms that higher number of available similar tasks will negatively impact on task failure. As figure 4.2-7 shows, with increasing the level of task similarity among available open tasks in the platform, failure rate will increase up to 17%. Interestingly 80% similarity level is leading to highest failure rate with average failure ratio of 21%, and 70% level of similarity causes the lowest failure ratio with average failure ratio of 7%. 90% similarity level and 60% similarity level are causing 15% and 11% failure rate reprehensively.

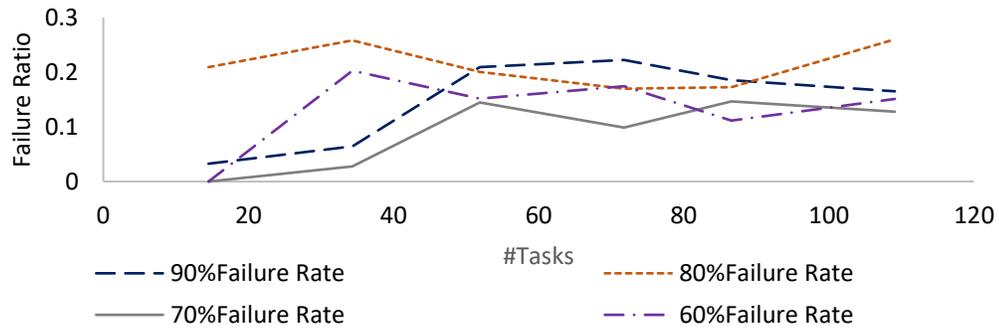

*Figure 4.2-7: Impact of task similarity on failure ratio*

## 4.3 Conclusion

One prerequisite for the success of crowdsourced software development is that enough participation level should be achieved and it's ideal for most registered workers to behave consistently and reliably towards completion. Existing studies have shown that setting the Monitory prize for software crowdsourcing tasks can be accurately predictable to reflect the size and complexity of the tasks [74]. In order to incentivize broader worker participation as well as higher quality of final submissions, it is very important to recognize the dual roles that award plays in crowd workers behavior in task selection and completion. This chapter reported an empirical study on worker's behavior in task selection and completion, based on data extracted from TopCoder platform. Major findings of this research showed that:

1. In task selection, the number of registrants will decrease as award increase; in task completion, the number of submissions and score will decrease as award increase.

2. There is a strong positive correlation between number of submissions and number of registrants. However, there is a decreasing tendency in submission as the number of registrants increases. While there is a weak positive correlation between number of registrants and score of the winning submission.

3. For similar tasks, the impact of award change on registrants follows a variety of inverted U-shape curves. Small award increase may be helpful in attracting additional registrants and submissions. However, big award increase may associate with decrease in both number of registrants and submissions.

The findings of this chapter will be used as the initial setting of systems dynamic model as CSD platform in the hybrid simulation model in chapter 7.

# *Chapter 5      Team Performance Pattern*

Leveraging crowd work force in CSD has a great potential to increase rapid delivery. To create an adaptive team to the changes it is required to increase the team elasticity. Therefore, it is crucial to understand crowd workers sensitivity and performance pattern to tasks and rate of task failure. To develop better understanding of worker performance in software crowdsourcing, this chapter reports an empirical study at Top Coder. The aim of this chapter is to investigate the following questions:

How diverse are crowd workers in terms of skill and experience?

How fast do crowd workers respond to a task call?

How reliable are crowd workers in submitting tasks?

To perform the empirical analysis after data cleaning, workers were sorted based on registration order per task. Then we clustered workers based on 5 different belts according to Top coder's definition of worker rating, i.e. Red, Yellow, Blue, Green and Gray, which are representing the highest skillful worker to the lowest one. Rating belt is showing the skill level and success rate of workers in platform. Crowd worker's reliability of competing on the tasks is measured based on last 15 competition workers register and submit for. For example, if a worker submitted 14 tasks out of 15 last registered tasks, his reliability is 93% (14/15). Top coder considers workers with a minimum reliability rate of 80% to be eligible to get a bonus (12 submissions out of 15 registrations).

We defined Team Elasticity (TE) as the ratio of the maximum number of registrants for project tasks per week divided by the minimum number of registrants per week, across the total project duration.

The 5 worker groups are defined into 5 belts of Red, Yellow, Blue, Green and Gray, which corresponds to the highest skillful workers to the lowest ones [75]. Table 5-1 shows the statistics of worker belts for the TopCoder dataset used in this study. It is shown that almost 90% of the workers are in Gray belt, which is non-experienced group. We will analyze different elements of a worker tuple in different five worker belts. To perform the empirical analysis after data cleaning, workers were sorted based on registration order per task. Workers are clustered based on 5 different belts per Top coder's definition of worker rating, i.e. Red, Yellow, Blue, Green and Gray, which are representing the highest skillful worker to the lowest one [17]. Rating belt is showing the skill level and success rate of workers in platform. Crowd worker's reliability of competing on the tasks is measured based on last 15 competition workers register and submit for. For example, if a worker submitted 14 tasks out of 15 last registered tasks, his reliability is 93% (14/15). Top coder considers workers with a minimum reliability rate of 80% to be eligible to get a bonus (12 submissions out of 15 registrations).

*Table 5-1: Summary of different worker belt*

| Belt | Rating Range (X) | # Workers | % Workers |
|---|---|---|---|
| Gray | X<900 | 4557 | 90.02% |
| Green | 900<X<1200 | 146 | 2.88% |
| Blue | 1200<X<1500 | 273 | 5.39% |
| Yellow | 1500<X<2200 | 78 | 1.54% |
| Red | X>2200 | 8 | 0.16% |

Worker availability in response to a task call in CSD is derived using two measures: worker's registration order and average submissions rate per registration order in the platform. Since the average number of registrants per task is 18 [76], the availability of workers with registration order less than or equal to 20 will be analyzed. Average response time (day) to new task call per registering order per belt as well as average submissions ratio per registering order per belt is a good measure to analyze worker's availability.

## 5.1   Research Design

   Tasks are uploaded as competitions in the platform, where crowd software workers would register and complete the challenges. On average, most of the tasks have a life cycle of one and half months from first day of registration to the submission deadline. When workers submit the final files, it will be reviewed by experts to check the results and grant the Scores. In order to analyze uploaded mini tasks in a project , we categorized the available data and defined the following metrics, as summarized in table 5-2.

Task attributes describe the basic quantitative characteristics of a task including total associated award and total number of uploaded tasks in a limited period of time;

- Worker attributes measure workers' overall submission ratio, reliability, year's membership, and rating;
- Worker-task pair level attributes include all worker' registration and submission dates, final scores, worker's registration order for each task.

The first step is to remove workers who never registered or submitted any task in their membership history. This reduced number of active workers to 5062 workers.

Second, because it is our interest to study worker's behaviors and performance in responding to task calls, we calculate, for each worker-task pair, the worker's registration order on that task. As shown in Table 5-2, this is derived from the rank of worker arrival

*Table 5-2: Summary of different worker belt*

| Metric | Definition |
|---|---|
| Task attributes | |
| Duration (D) | Total available time from registration date to submissions deadline. Range: $(0, \infty)$ |
| Task registration start date (TR) | Time when task is available on line for workers to register |
| Task submission end date (TS) | Deadline that all workers who registered for task must submit their final results |
| Award (P) | Monitory prize (Dollars) in task description. Range: $(0, \infty)$ |
| Worker attributes | |
| # Registration (R) | Number of registrants that are willing to compete on total number of tasks in specific period of time. Range: $(0, \infty)$ |
| # Submissions (S) | Number of submissions that a task receives by its submission deadline in specific period of time. Range: (0, #registrants] |
| Submissions Ratio (SR) | The ratio between the number of tasks a worker submitted and the total number of tasks that a worker registered. |
| Reliability (RE) | The percentage of successful task submissions in a worker's most recent 15 task registrations. Range: (0, 1) |
| Years of Membership | Year since a worker become a member. Range: $(0, \infty)$ |
| Rating (R) | Platform reputation rating of a specific worker. Range: $(0, \infty)$ |
| Worker-task pair level attributes | |
| Worker registration date (WR) | Date and time that a worker registered for a task. |
| Worker Submissions date (WS) | Date and time a worker submitted for a task. |
| Registration order (RO) | Rank of a worker's arrival time at a task by his registration date |
| Score (SO) | Point value of submission evaluated through peer review. Range: (0, 100] |

### *5.1.1 Worker's Characteristics*

In this part, we will be analyzing the impact of worker's experience level and reliability on worker's behavior per rating belt.

### *5.1.2 Worker's Availability*

We investigate two aspects of worker availability: 1) how fast they respond to task calls? 2) how reliable are they in completing tasks? Worker availability in response to a task call in CSD is derived using two measures: worker's registration order and average submissions rate per registration order in the platform.

Since the average number of registrants per task is 18 [76], we will analyze the availability of workers with registration order less than or equal to 20. Average response time (day) to new task call per registering order per belt as well as average submissions ratio per registering order per belt is a good measure to analyze worker's availability. To analyze worker's availability, we will use following definitions:

***Def. 1:*** Response time, RTi, k, measure the speed of a worker i arrival on task k, derived from the difference in the number of days between the task's registration starting date, i.e. TRk and the worker's registration date on the task, i.e. WRi, k:

$$RTi,k = WRi,k - TRk \qquad\qquad Eq.5\text{-}1$$

***Def. 2:*** Average Response Time of workers in the same registration order i, i.e. ARTi, is the average respond time of n worker with the same registration order i.

$$ARTi = \sum_{k=1}^{n}(c * RTi,k)/n \qquad\qquad Eq.5\text{-}2$$

***Def. 3:*** Average Submissions Ratio, ASR, is the average submissions ratio of workers from the same rating group.

$$\text{ROS(i)} = \sum_{i=1}^{n}(SRi)/n$$

Eq.5-3

### 5.1.3   Worker's Performance

Workers Relative Velocity and Quality would be used to analyze worker performance and responsiveness in task taking and their reliability in submit tasks in the platform and task completion rate in this analysis. The related measures are defined at below.

*Def. 4:* Relative Velocity, i.e. RVi,k, measures the ratio between a worker i's actual duration on completing a task k and the allowed task duration of task k.

$$\text{RVi, k} = (WSi,k - WRi,k)/(TSk - TRk$$

Eq.5- 4

*Def. 5:* Average Relative Velocity ARV(i) measures the average of RV ratio between actual duration and allowed task duration of workers from the same rating group.

$$\text{ARV(i)} = (\sum_{i=1}^{n}(WSi - WRi)/ \sum_{i=1}^{n}(TSi - TRi))/n$$

Eq.5-5

Also, the quality of task submissions is the second important factor in analyzing worker's performance.

*Def. 6:* Quality Q(i) of worker i is defined as the average score SO(i) of worker i per registration order per belt.

$$\text{Q(i)} = \sum_{i=1}^{n}(SOi)/n$$

Eq.5-6

*Def. 7:* Average Quality AQ(i) in the same registration order, average quality Q(i) of worker per registration order per belt.

$$\text{AQ(i)} = \sum_{i=1}^{n}(Qi)/n$$

Eq.5-7

### 5.2   Empirical Results

### 5.2.1    *Overall Worker Characteristics*

At project level, a total of 403 projects were further decomposed into a total of 4907 mini tasks, across 14 different challenge types [17]. The average number of workers per task is about 18, the average task price is about $750, and the average task duration is about 16 days. As shown in Figure 5.2-1, 80% of the tasks have less or equal to 14 registrants, with 3 or less submissions. 13% of all tasks in our data base receive more than 20 registrants (650 tasks).

We further look at the worker distribution on the top 5 task types, including Code, First2Finish, Assembly, Content Creation, and UI Prototyping. Figure 3 shows the worker distribution results, in five different rating belts, for these top 5 task types, which accounts for 94% of all tasks. The left vertical axis shows the percentage of workers in each rating belt, and the right vertical axis reflects the unit price for tasks. The dotted line shows the average unit prices of the top 5 task types.

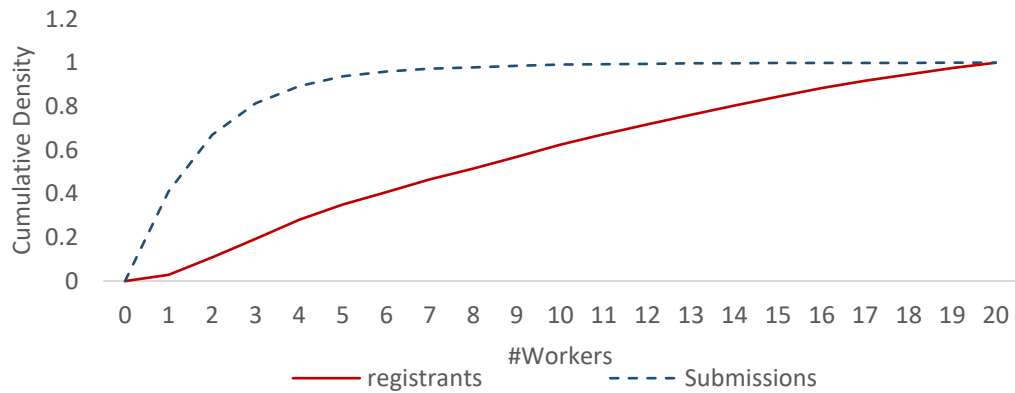

*Figure 5.2-1: Cumulative density plot of #Submissions and #Registration per task*

As illustrated in Figure 5.2-2, Gray workers, with the lowest rating among all groups, are mostly (i.e., 48%) interested in participating in coding, while Red workers as highest rated workers prefer to apply for first2finish challenges. Almost 35% of Red workers are concentrated in First2finish task types, none of them applied for UI prototype challenges. Yellow workers are mostly interested in assembly tasks with 23% of workers and less interested in content creation. It is also observed that the Green and Blue categories have very similar task preference patterns.

Another important aspect in software crowdsourcing is worker's reliability. According to Figure 4, Green and Blue category workers contain more than 50% members with reliability above 0.5 (which means 50% of the time they successfully deliver jobs.). Additionally, there are about 20% of members with reliability above 0.8. Yellow and Red groups, though with relative higher rating, have only 20% of their members with reliability above 0.5 and 0.3, respectively.

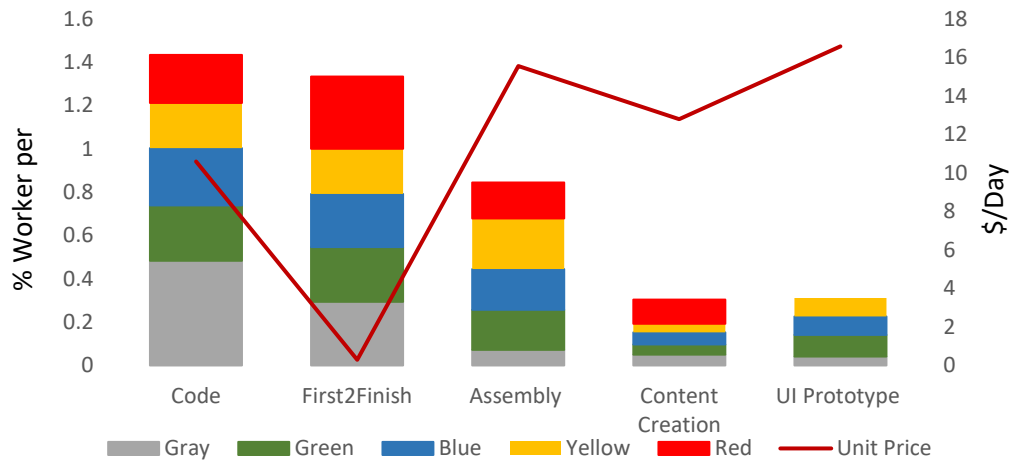

*Figure 5.2-2: Workers belt contribution in top 5 task types*

Another important aspect in software crowdsourcing is worker's reliability. According to Figure 5.2-3, Green and Blue category workers contain more than 50% of members with reliability above 0.5 (which means 50% of the time they successfully deliver jobs.). Additionally, there are about 20% of members with reliability above 0.8. Yellow and Red groups, though with relative higher rating, have only 20% of their members with reliability above 0.5 and 0.3, respectively.

We run repeated measure one-way ANOVA test on the worker's reliability data from the five groups. Based on ANOVA test results, the worker's reliability is significantly different across the five groups (i.e. p-value is 0.005).

***Finding 1.1***: Workers from different rating groups (denoted in different color columns) have somewhat different preferences in task selection in terms of task types, prices, and durations.

***Finding 1.2***: Workers from different rating groups have absolutely different experience level and reliability distribution. Workers joining after year 2010 are mostly experienced workers with strong interest and motivations; however, workers with higher ratings are more skillful, not necessarily more reliable than workers with lower ratings.

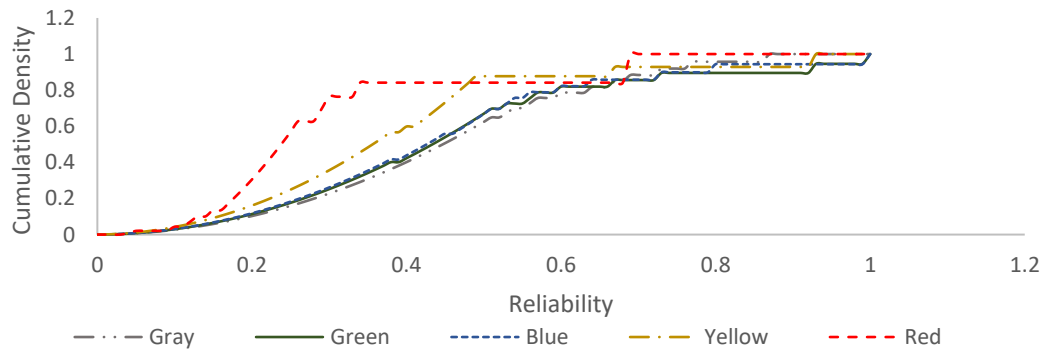

*Figure 5.2-3: Cumulative density of worker's reliability across five rating groups*

### 5.2.2 Worker Availability

To derive empirical evidence on worker availability in CSD, we focus on the first 20 clusters, i.e. worker-task pairs with registration orders below 20. In each rating group, we calculated the average response time (in days) for each cluster. Figure 5.2-4 shows a line chart of average response time and registration order for all five rating groups. The results show that the average response time for the top-20 registrants is not exceeding 1 day (i.e., 23.57 hours).

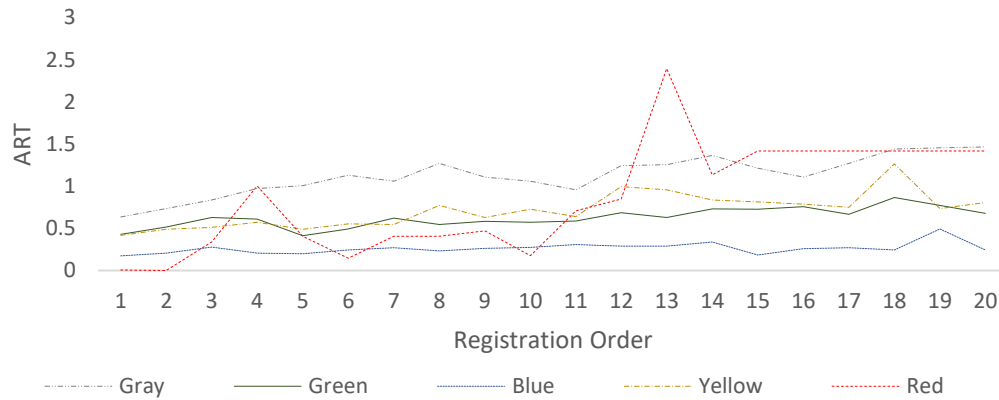

*Figure 5.2-4: Average Response Time (ART) per day in dependence of the order of registration (up to 20 workers per task)*

On average, 59% of workers respond to a task call within the first 24 hours after task being uploaded; 48% of the first 5 workers in Green, Blue and, yellow belts are registering within the first 12 hours. Gray category is relatively slow in responding to task calls, and 65% are responding in the first days, 18% are still registering in the second day. It is observed from Figure 6 that for workers from Green and Blue group, the response time is not very sensitive to registering orders, which means the workers are very motivated to register for new uploaded tasks, even though they are aware of potentially dozens of competitors. Another interesting observation is that, first and second registrants in the Red group tend to respond to tasks immediately after it is uploaded. The reason for the fluctuation in Red group is because of the small sample size of this group, which only contains 8 workers.

The result of the ANOVA test showed that worker registration order is significantly different across all 5 groups (i.e. p-value is 0.000).

Figure 5.2-5 shows the relationship between average submissions ratio w.r.t. registration orders. There is a general trend of decrease in submissions ratio as registration order goes up from 1 to 20. This is a clear evidence of the strong motivation in completing

tasks and winning prizes among early registrants. On average, more than half of early registrant workers (first and second registrants) have submitted tasks. As it is illustrated in Figure 6 that workers in higher belt are not actively submitting for tasks where they are not among the early registrants, i.e. with a registration order greater than 12. Workers in Gray, Green, Blue, and Yellow categories are following the same pattern; however, workers in Blue and Yellow categories are facing more fluctuated trend.

More specifically, first registrants in the Yellow category have an average submissions ratio of 60%, by going up in registering order the rate is decreased to 20% for the 20th registered worker. The submission ratio for the first register order in Blue category is about 39% and it gradually drops to 10% for the 20th registrant.

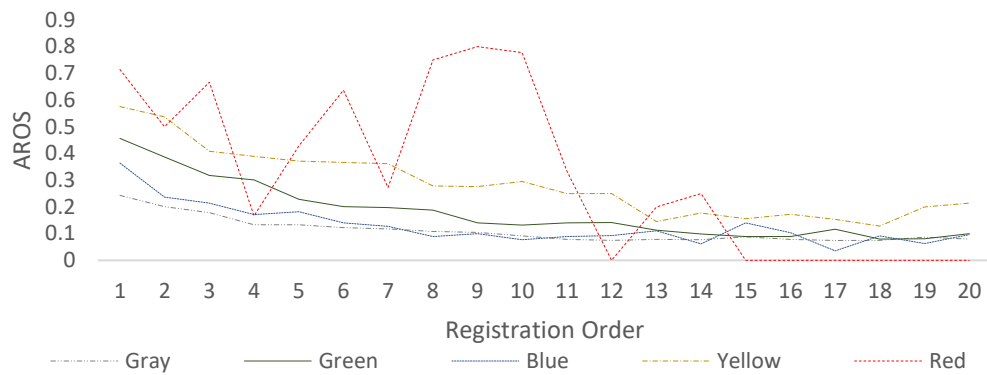

*Figure 5.2-5: Average submission ratio per registration order (AROS) per belt*

Among all the belts, Gray and Green members seem to have smoother patterns. First registration order for these clusters has 25% and 45% submissions ratio, respectively. Submissions ratio in both categories is decreasing to 10% for the 20th registrant. In the Red category, the average submissions ratio for top 3 registrants is 60% which drops to 20% for the 4th registrants. Statistical analysis supports that average submission ratios for the five group are significantly different (i.e. p-value at 0.000067).

**_Finding 2.1:_** The average response time for the top-20 registrants is not exceeding 1 day (i.e. 23.57 hours).

**_Finding 2.2:_** 59% of workers are registering for a new task within 24 hours of uploading; and 48% of the first 5 workers in Green, Blue and, yellow belts are registering within the first 12 hours.

**_Finding 2.3:_** There is a decreasing trend of submissions ratio in reverse to registration order.

### 5.2.3    *Workers performance in task submissions*

Figure 5.2-6 shows the average relative velocity of workers for different rating belts, which measures the percentage of acceleration in task completion. Interestingly, lower rated workers are using less time; while higher rated workers, required higher amount of time to submit. This can be due to different task choices, different worker motivation patterns, or individual skills.

On average 34% of the workers can complete the requested tasks within only 10% of allowed time for submissions, and 60% of workers are submitting the final files by using 70% of total submission duration. More specifically, 20% of Red group workers are using less than 40% of duration to submit, while this rises to around 40% of Blue and Green

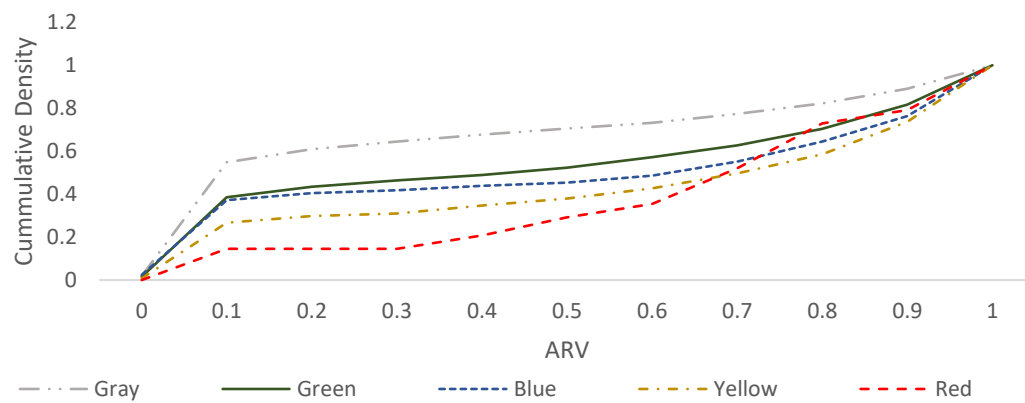

*Figure 5.2-6: Average Relative Velocity (ARV) in different belt categories*

groups. It takes almost 35% of Yellow group workers and increase to 70% of Gray group workers.

However, results of ANOVA test do not support that rating clusters are significantly influencing workers' velocity, i.e. p-value is 0.14, which is greater than 0.05.

Moreover, the quality of worker's submissions is another important factor in task success. We can measure quality based on final score that workers granted per submission.

Figure 5.2-7 shows the cumulative distribution of average score per belt for workers with registration order less than 20. As it is illustrated higher rated belts are granted higher scores.

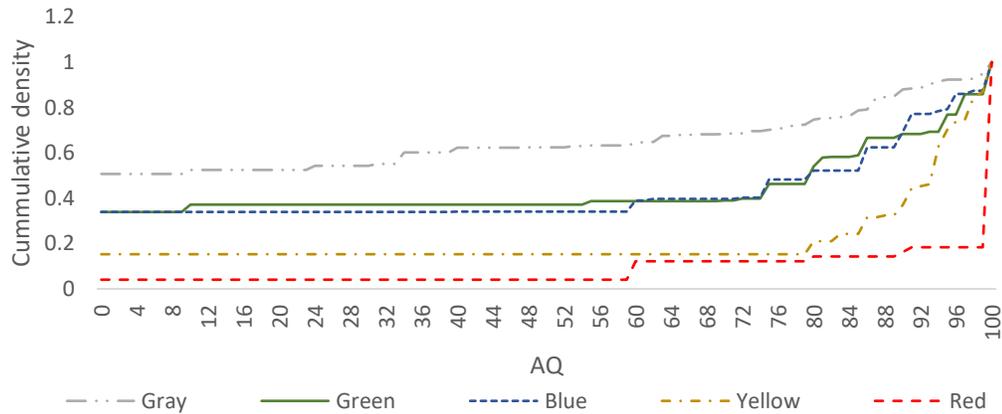

*Figure 5.2-7: Evolution of Average Quality (AQ) in different belts*

Almost 50% of Gray members are granted score of 0, which means they never have any submissions or they have granted 0 as their score, there is a slow increasing trend up to score of 35 for 65% of members, and only 25% of gray members granting score higher than 85. In Red category, around 18% of members are either receiving scores less than 69 or dropped the task out. The granted score increased to 90 for next 2% of the population and there is a sharp rise for net 80% of members with score of 100. In Yellow category, almost 20% of workers are granting scores less than 80, then scores gradually increase to 100 for next 80% of population. Almost the same patterns happen for members of Green and Blue workers, however, they experience a slower rising trend than the Yellow ones.

ANOVA test showed that worker quality of submission is significant different across all 5 groups (i.e. p-value is 0.00).

**_Finding 3.1:_** Average relative velocity to complete tasks within 80% of allowed duration. 34% of the workers can complete the requested tasks within only 10% of allowed time.

**_Finding 3.2:_** Higher rated workers are more reliable in task submissions quality, on average 86% of higher rated workers (Red and Yellow) successfully deliver products passing reviews (i.e. granted scores above 75 out of 100), while the number is less than 50% for the workers from the other lower rating groups.

*Table 5-3: Team Elasticity*

| Project | Project I | Project II | Project III | Project IV |
|---|---|---|---|---|
| #tasks | 156 | 306 | 177 | 277 |
| # Reg / #Sub | 2174 / 428 | 4136 / 807 | 4222 / 494 | 2412 / 867 |
| Team Elasticity | 187 | 14.42 | 8.5 | 33.3 |

### *5.2.4  Team Elasticity*

According to one of the author's experience with TopCoder projects, this number is reasonable yet conservative. In a recent report [76], it is reported with a total schedule acceleration rate of 3 through CSD, based on an interview study conducted in one TopCoder's client organization. One of the possible reason for this difference is that in our study, the corresponding management overhead within the requesting company is not included as we don't have access to such data. Considering the additional in-house management and coordination effort, the estimated schedule of the corresponding nominal in-house development would most likely be longer, and the resultant schedule acceleration ratio would be even greater.

Based on the findings in the previous steps, we are interested to investigate the relationship between team elasticity and schedule acceleration effects. We define the measure of team elasticity as follows:

Def. 8: Team Elasticity (TE) is the ratio of the maximum number of registrants for project tasks per week divided by the minimum number of registrants per week, across the total project duration.

Pearson correlation analysis result shows there is a 0.76 coefficient between team elasticity and average SAR, which indicates a high correlation between degree of team elasticity and schedule acceleration potential. Table 5-3 summarized the team elasticity.

This indicates CSD may help organizations with integrating elastic using external human resource to reduce cost from internal employment and exploring the distributed production model to speed up the development process [78].

### 5.2.5  Discussion and Evaluation

The empirical findings reported above provides insights for software managers, platform providers, and researchers that might help them to better understand worker performance and propose solutions to better governance in CSD.

Finding 1.1 and 1.2 confirm the intuition that workers from different rating groups have very different preferences in task selection and reliability distribution. Obviously, different rating groups are associated with different expertise and experience levels. As summarized in Finding 2.1, the average response time for the top-20 registrants is less than 1 day (i.e. 23.57 hours), and this implies the great potential for organizations to immediately tap into talents beyond traditional boundaries. The evidence in Finding 2.3 shows the strong motivation in completing tasks and winning prizes among early registrants, with more than 50% of submission rate among the first two registrants. This result does not support Top coder assumption of 90%-96% success rate with having more than 1 registrant [79].

However, it is still promising for organizations to take advantage of software expertise in the crowd in successfully delivering requested software solutions.

In terms of crowd worker performance, finding 3.1 provides additional support for the rapid delivery effect attributed to CSD. On average 34% of the workers can complete the requested tasks within only 10% of allowed time; and 80% of workers can complete within 70% of allowed time. When considering the quality of delivery from the crowd workers, as anticipated, higher rated workers are more guaranteed to make satisfactory submissions (Finding 3.2). Though, the data also show that workers in lower rated belts are more reliable in adopting to change and following a more stable submission pattern. Hence, it suggests additional insights for CSD tasks decomposition in a way that can better attract early registrants and leverage various patterns of different rating groups.

Understanding how workers with different experience respond to different types of tasks, different prices, and different durations will help managers in planning for crowdsourcing.

## 5.3   Conclusion

For adaptive teams to leverage CSD to increase team elasticity, it is critical to understand crowd worker's sensitivity and performance to the tasks and rate of task elasticity and success. This chapter reports an empirical study to address that end. Based on the available empirical data and related research, we designed and developed a study about the impact of worker performance with different skill and experience level. The main findings of this study showed that on average:

1.  59% of workers respond to a task call in the first 24 hours,

2.  24% of the workers who registered early will make submissions to tasks, and 76% of them exceeding the acceptance criteria,

3.  On average 60% of submissions by the high ranked workers, 40% of the submission by middle level workers and 25% of submissions by low experienced workers are qualified.

The findings of this chapter will be used as workers' personal knowledge to create the agent-based model in chapter 7.

# Chapter 6    Task Completion Pattern

Crowdsourcing has become a popular option for rapid acquisition, with reported benefits such as shortened schedule due to mass parallel development, innovative solutions based on the "wisdom of crowds", and reduced cost due to the pre-pricing and bidding effects. However, most of existing studies on software crowdsourcing are focusing on individual task level, providing limited insights on the practice as well as outcomes at overall project level. To develop better understanding of crowdsourcing-based software projects, this chapter reports an empirical study on analyzing Topcoder platform that intensively leverage crowdsourcing throughout the product implementation, testing, and assembly phases.

## 6.1   Task Execution overview in CSD

It is reported that CSD platforms are following waterfall development model [4]. Therefore, all tasks in the platform will follow development phases of Requirements, Design, Implementation, Testing, and Maintenance one after the other. Topcoder divided different phases of the task life cycle to seven different clusters of tasks:

- First2Finish: The first person to submit passing entry wins
- Assembly Competition: Assemble previous tasks
- Bug Hunt: Find and fix available Bugs
- Code: Programming specific task
- UI Prototype: User Interface prototyping is an analysis technique in which users are actively involved in the mocking-up of the UI for a system.
- Architecture: This contest asks competitors to define the technical approach to implement the requirements. The output is a technical architecture document and finalized a plan for assembly contests.

- Test Suit: Competitors produce automated test cases to validate the quality, accuracy, and performance of applications. The output is a suite of automated test cases.

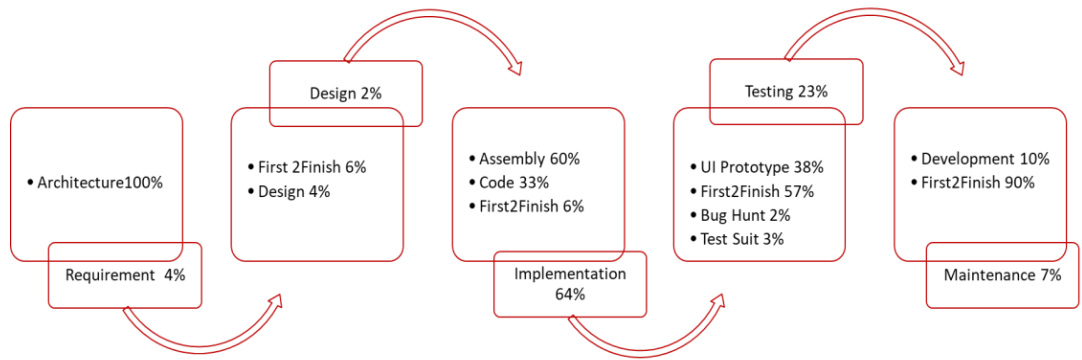

*Figure 6.1-1: Distribution of task failure ratio among different development phase in task life cycle*

Figure 6.1-1 presents the distribution of failure ratio for different task types per task phase in the task life cycle. As it is shown, First2Finish cluster contains tasks from all different development phase. Highest task failure takes place in the implementation and testing phase with 64% and 23% respectively, while design and requirement are sharing less than 5% of task failure each. One reason can be a lower number of the available task in these phases. Interestingly maintenance is holding 7% of task failure in the platform.

First2Finish tasks can be assigned to all different phases. As it is shown in Table 4-1, 90% of task failure in maintenance and 57% of task failure in testing phase belong to this task type. While in implementation, only 6% of task failure is under First2Finish task type, and interestingly 60% of task failure happens in assembly tasks.

In addition, task failure ratio in CSD is based on one of the following situations:

- Failed Request: arrival task failed to be uploaded

- Client Request: The initial client decide to cancel the tasks

*Table 6-1: Failure distribution in different development phases*

|                | Failed R | Client R | Zero Sub | Requirement | Failed S |
|----------------|----------|----------|----------|-------------|----------|
| Architecture   | 15%      | 4%       | 62%      | 12%         | 8%       |
| Design         | 33%      | 27%      | 7%       | 33%         | 0%       |
| Implementation | 11%      | 14%      | 65%      | 5%          | 5%       |
| Testing        | 20%      | 31%      | 35%      | 12%         | 3%       |
| Maintenance    | 20%      | 30%      | 30%      | 20%         | 0%       |

- Requirement Failed: Task requirement is not complete
- Zero Submissions: none of the registered crowd workers submit the submissions file
- Failed submissions: none of the final submissions can pass the peer review.

Our empirical analysis supports that on average 48% of the tasks failures happen due to zero submissions per tasks. This fact is a direct impact of lack of skillful or motivated workers to take the tasks which cause resource discrepancy in the platform. Table 4-6 shows the summary of task failure type per development phase.

Moreover, following waterfall development model in CSD makes each batch of tasks be always from a prior batch and a fresh batch. This sequence of task arrival will provide four cluster of task cycle patterns per project in CSD. In this study, the patterns are grouped into four clusters of Prior Cycle, Current Cycle, Orbit Cycle and fresh Cycle.

- Current Cycle is the batch of tasks that are scheduled to complete following the initial task life cycle.
- Prior Cycle is the batch of tasks that went into task life cycle before the batch of current cycle arrives at the platform.
- Fresh cycle is the batch of tasks that will start their life cycle after the current cycle arrives at the platform.
- Orbit Cycle is the batch of tasks that all following the same development phase in the task life cycle.

| Phase | Requirement | Design | Implementation | Testing | Maintenance | Pattern |
|---|---|---|---|---|---|---|
| Requirement | RR | *RD* | RI | RT | RM | |
| Design | DR | DD | *DI* | DT | DM | Prior Cycle |
| Implementation | IR | ID | II | *IT* | IM | |
| Testing | TR | TD | IT | TT | *TM* | Current Cycle |
| Maintenance | *MR* | MD | MI | MT | MM | Orbit Cycle |
| Pattern | Current Cycle | | Fresh Cycle | | Orbit Cycle | |

*Figure 6.1-2: Summary of Task Cycle in CSD Projects*

Figure 6.1-2 shows the summary of task cycles in a CSD platform.

Our analysis shows that on average 44% of tasks in Topcoder are in Fresh Cycle while only 4% of tasks are in the Prior cycle. Also, only 15% of tasks are in Current Cycle and 37% of tasks are in Orbit Cycle. 18% and 20% of task failure is respectively associated with Orbit Cycle and Fresh Cycle. Figure 6.1-3 illustrates the details of task failure pattern in different task cycles.

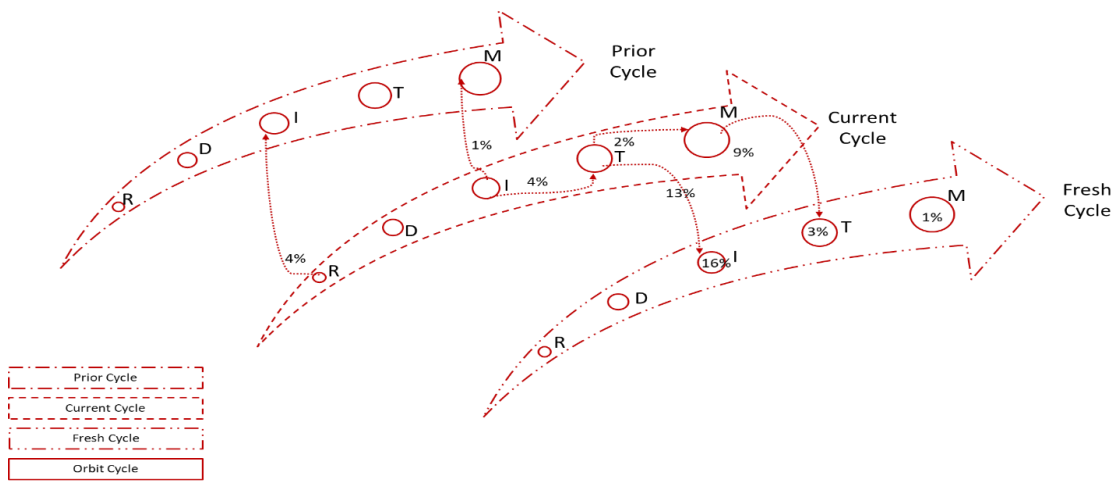

*Figure 6.1-3: Task Failure Pattern in different Task Cycle*

## 6.2    Parallelism in Task Scheduling

Ideally, mass parallel production through Crowdsourcing could be an option to rapid acquisition in software engineering by leveraging infinite worker resource on the internet. It is important to understand the patterns and strategies of decomposing and uploading parallel tasks in order to maintain stable worker supply as well as satisfactory task completion rate. According to the finding in part 4.1.1, projects with more than 100 decomposed mini-tasks are large enough to perform parallel task competition. Therefore, for the purpose of this study, we conduct a comparative analysis on the four largest projects of the available dataset in different periods of time in terms of uploading number of tasks. On average, most of the tasks of these projects have a life cycle of one and half months from the first day of registration to the submission deadline. In this part we aim to investigate following questions:

1) What are the task completion patterns in the CSD platform?
2) How does CSD benefit schedule reduction?

### 6.2.1   Parallelism and Maximum Arrival Tasks

Figure 6.2-1 shows an overall profile of the distribution pattern of task uploaded and associated award rate per month per sample projects. As it is shown the number of uploaded tasks rises with time to reach a maximum number, which includes the greatest number of small decomposed tasks per project, then it smoothly starts to decrease. It again, increases in September which leads us to the second peak in December, this can be due to new product launch, or a lot of debugging tasks uploading in order to complete the projects.

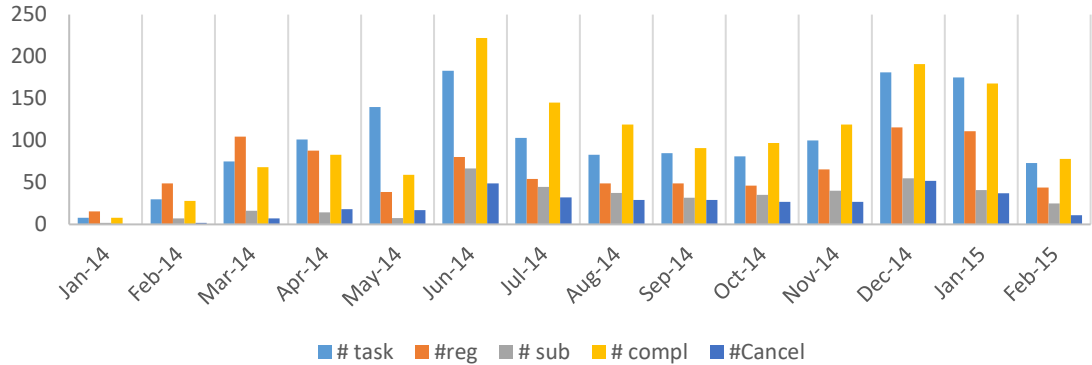

*Figure 6.2-1: Trend of uploaded tasks and associated award in the four biggest projects in the dataset*

As it is illustrated in Fig.6.2-2, total uploaded task rate and total associated award rate do not follow the same trend as above. The maximum uploaded task does not represent the maximum award associated with the projects since the tasks size and award have direct influence while number of decomposed tasks and total amount of associated award have negative influence [22] [66]. This fact makes the opportunity of better resource allocation since by decomposing the project to a greater number of tasks and following the parallel task uploading method not only the requestor would use minimum budget but also the pool of indefinite crowd workers is available to compete on the task.

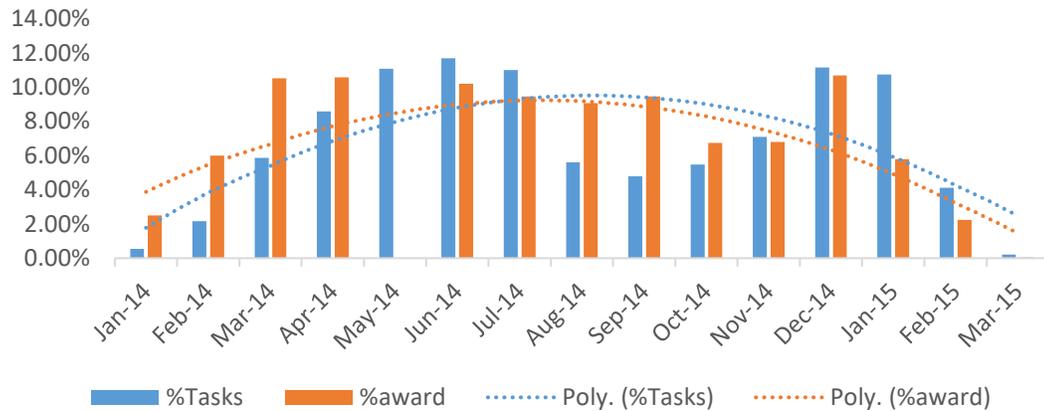

*Figure 6.2-2: Relationship between Award and Tasks uploading trend*

One possible reason can be that, based on project scheduling methods, at first, most of arrival tasks are in the design category which has a relatively large size, then it turns to implementation and development which undergoes the project decomposition into more number of mini-tasks; next step would be test and deployment that develop bigger task size to work on, consequently cause reduction in the total number of uploaded tasks and total award associated with them. This results in an increase in the complexity of tasks or required skills to perform them at this step [30]. We focus on four biggest projects in the platform to analyze the result, Table 6-2 summarized the statistical analytics of these projects.

*Table 6-2: Statistical analytics of sample projects per month*

|  |  | # Task | Award | # Reg | # Sub | # Comp | #Cancel |
|---|---|---|---|---|---|---|---|
| Sample Project 1 | Min | 1 | 30 | 6 | 0 | 0 | 0 |
|  | Median | 5.5 | 3850 | 69 | 12 | 5.5 | 0 |
|  | Max | 57 | 43225 | 763 | 117 | 51 | 6 |
|  | Average | 16.35 | 8233.21 | 155.28 | 30.57 | 15.21 | 1.14 |
| Sample Project 2 | Min | 3 | 280 | 18 | 4 | 3 | 0 |
|  | Median | 26 | 20910 | 274 | 37 | 23 | 3 |
|  | Max | 95 | 34671 | 752 | 159 | 89 | 10 |
|  | Average | 33.92 | 20853.84 | 318.15 | 62.07 | 30.84 | 3.076 |
| Sample Project 3 | Min | 2 | 2300 | 30 | 5 | 2 | 0 |
|  | Median | 25 | 10632 | 235 | 49 | 16 | 8 |
|  | Max | 47 | 22426 | 410 | 97 | 38 | 10 |
|  | Average | 23.81 | 12842.09 | 219.72 | 49.18 | 17.45 | 6.36 |
| Sample Project 4 | Min | 2 | 355 | 43 | 5 | 2 | 0 |
|  | Median | 26.5 | 8667.5 | 147 | 43.5 | 22 | 3 |
|  | Max | 93 | 21045 | 464 | 173 | 82 | 25 |
|  | Average | 35.64 | 10180.92 | 172.28 | 61.92 | 29.21 | 6.57 |

Base on arrival task rate in available empirical data, about 75% of all tasks are priced under the 67% of the total award and maximum percentage of arrival tasks does not represent the maximum sum of award rate associated with the project, which can be due to the task complexity or required skills to compete on the task. However, increasing the number of arrival tasks at the same time may represent higher number of registrants and chance of acceptable submissions [22] [25]. While total number of arrival tasks and associated award are not following almost the same pattern, it still supports individual task size and the specific associated award follow negative influence [8] [53]. We further focus on the four biggest projects individually based on the total award rate and the uploading task rate to predict the task completion rate.

According to the empirical analysis and assumption in pervious section, although less than 25% of the total registrants will submit the final files, yet almost 87% of the uploaded tasks will be completed. This result does not support Topcoder assumption of 90%-96% success rate with having more than one registrants [79]. Moreover, the analysis suggests that uploading more number of parallel tasks lead the different tasks to be chosen by a greater number of crowd workers at the same time hence, a shorter project time and higher chance of success will be achieved.

## 6.2.2   Distribution pattern of Maximum Parallel Arrival Task

As it is illustrated in Fig 6.2-3, a higher number of uploaded tasks per period do not ensure a higher total amount of associated award in four different projects, however according to the available empirical data, there is an average positive correlation of 0.44 between total number of uploaded tasks and total prize associated with them, yet task type and project domain would influence this fact.

When analyzing individual projects, we observed that the number of uploaded tasks depends on the project type and domain of the project. The number of uploaded tasks increase to a maximum number of tasks per project and then decreases based on the different phase and planning. Among all the projects, project II took longer period of time to decompose the project to more number of uploaded tasks, which based on available data, as maximum tasks type is bug hunt, the time making sense. However unexpectedly, in project IV we are facing two peaks in number of uploaded tasks, which may happen due to different version of the software production and also, the specific production phase that project started to be crowdsourced. In rest of sample project, task uploading almost following the total platform trend, meaning, in the beginning there are smaller number of tasks for compete on and by improving the project to further phases, project can be decomposed to smaller task size, and a greater number of uploaded tasks available to be chosen from. In late production phase, again number of decomposed tasks would drop, due to the nature of deployment projects. Figure 6.2-4 clearly shows that higher number of uploaded tasks in specific period of time do not significantly mean higher associated award rate and more budget to spend. For example, in project III, higher uploading task rate happened in July which is not represents higher associated award rate.

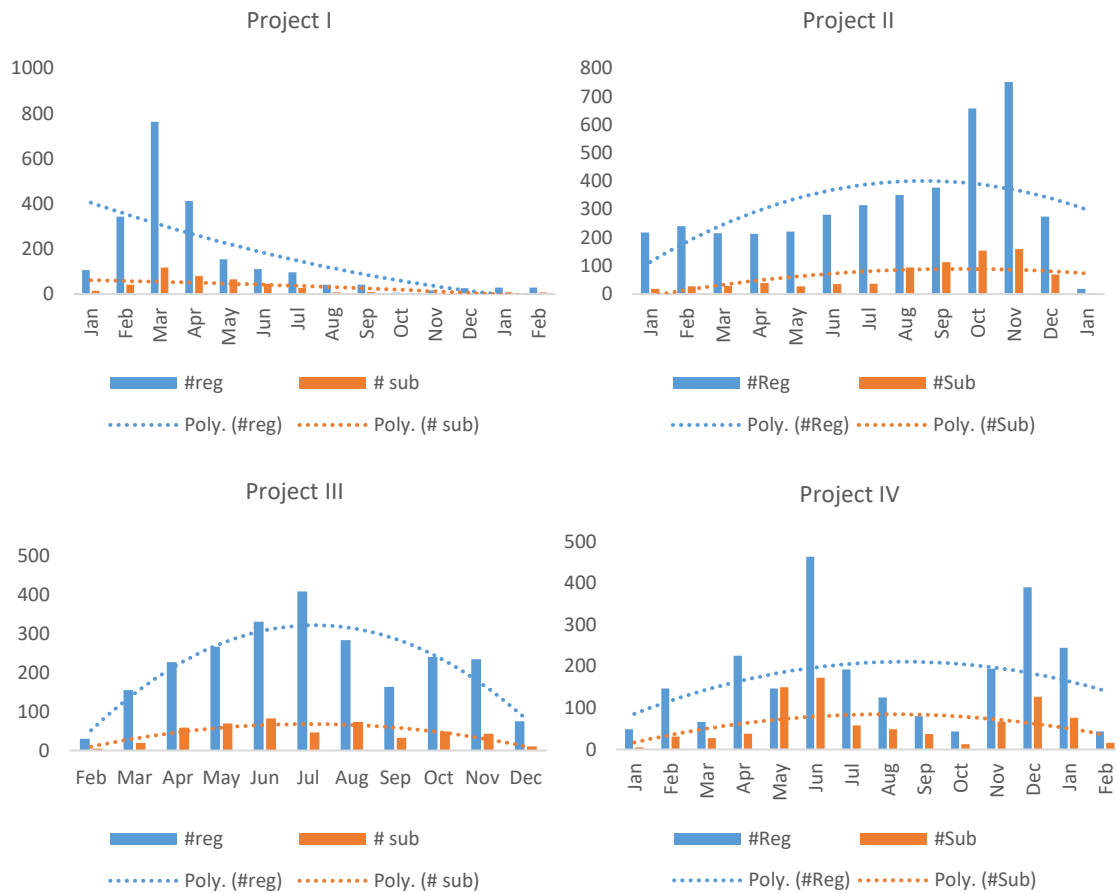

*Figure 6.2-3 : Relationship between Award and Tasks and Registration*

Decomposing project to a greater number of tasks and parallel uploading them cause more choice for crowd workers to utilize tasks and compete on, and consequently better rate of resource allocation for project managers in terms of budget and project scheduling. When further analysis is conducted on the four projects, we also observed that the relationship between number of uploaded tasks and associated award rate generally follows exponential distribution, Figure 6.2-5, still, it does not guarantee that by increasing number of uploading task, total associated award would increase, as it depends on task size and complexity [22]. Expectedly, it is possible that as number of uploaded tasks increases, some previously registered workers might perceive distracting factors such as competition pressure, insufficient time to complete the task or misunderstanding the task requirement while registering [8] [22]. Hence by raising the number of uploaded tasks in parallel flow at the same time, number of registrants and consequently submissions would increase [90].

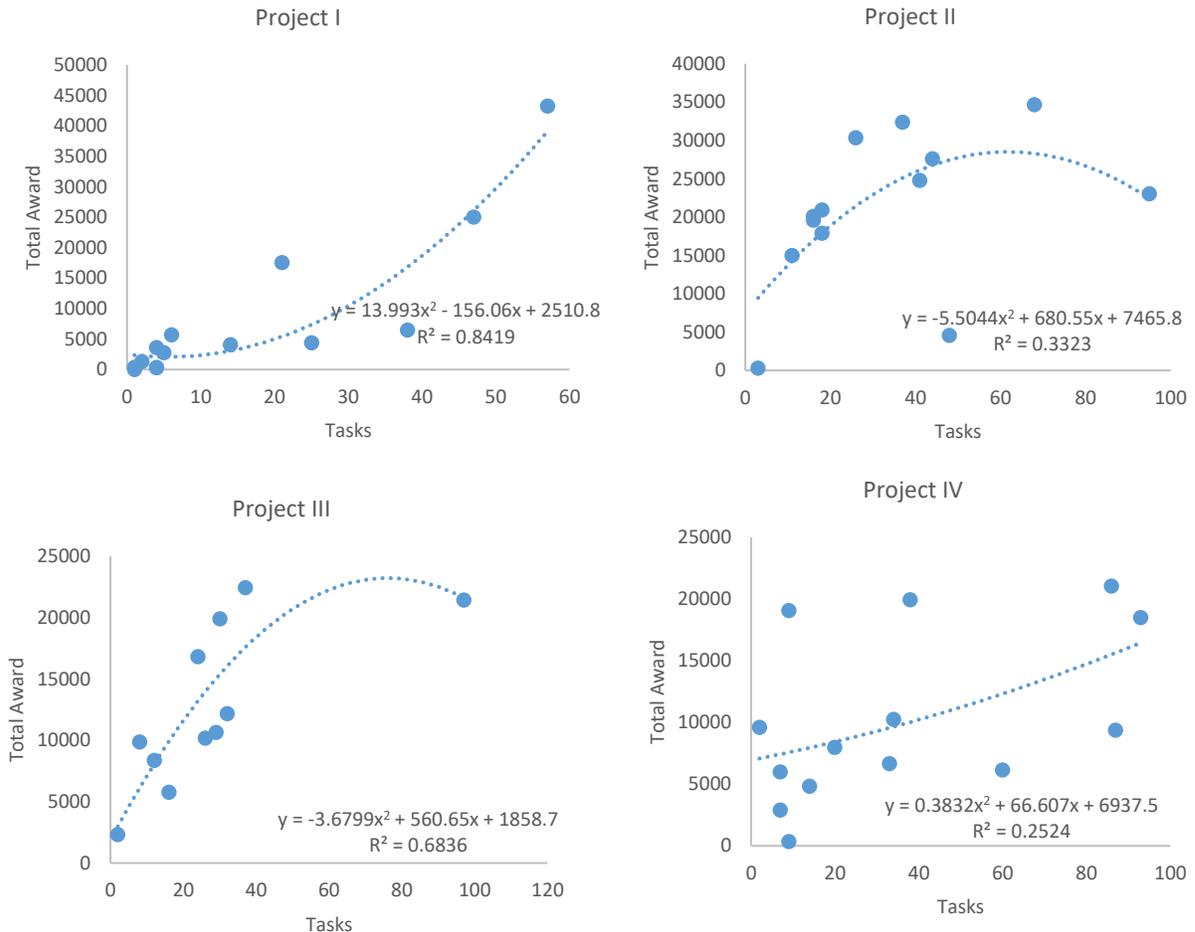

*Figure 6.2-4: Relationship between total number of uploading tasks and total associated award rate*

Based on the available empirical data, for large projects arriving on average 90 parallel tasks at the same period of time, may lead us to average of 86% success. Since in comparison with the in-house production there would be a greater number of crowd software workers available with wild variety of skills, there is no need to wait to finish one task and start the next on. Therefore, in shorter time line higher rate of the tasks would be done parallel and consequently higher rate of resource allocation would happen.

### 6.2.3    Prediction model of number of submissions

Since the result of empirical analysis of parallelism in uploading tasks shows positive effect on the tasks success and workers performance, it is important to present an empirical model to predict the number of submissions per tasks as the main factor of performance based on main projects attributes.

In order to build a linear regression model, four main drivers in uploading parallel tasks was used: number of registrants, associated award, lead time for submissions, and number of parallel tasks in the same period.

The result of the linear regression model is described in table 6-3. Further, to evaluate the model we compare the results via six different Machine Learning methods which will be explained in the next part.

The aim of the validation models is to understand A) which predictive model gives the best overall predictive performances assessed by performed analysis in pervious section? B) What actionable insight can Figure 6.2-6 illustrates the summary of different performance measures.

*Table 6-3 : Regression Model Parameters*

| Source | Value | Standard error | t | Pr > \|t\| | Lower bound (95%) | Upper bound (95%) |
|---|---|---|---|---|---|---|
| Intercept | 2.768 | 0.494 | 5.601 | < 0.0001 | 1.791 | 3.745 |
| #Parallel Tasks | 1.000 | 0.386 | 2.593 | 0.025 | 0.151 | 1.850 |
| # Reg | -0.001 | 0.001 | -1.435 | 0.179 | -0.003 | 0.001 |
| Award | 0.151 | 0.084 | 1.805 | 0.099 | -0.033 | 0.335 |
| Duration | 0.294 | 0.799 | 0.368 | 0.720 | -1.464 | 2.052 |

Considering the highest rate for task completion and acceptable submissions, software mangers will be more concerned about risks for adopting crowdsourcing and need to better decision support on analyzing and controlling the risk of insufficient competition and poor submissions. This paper reports an empirical study to address that end. The analysis results conclude that:

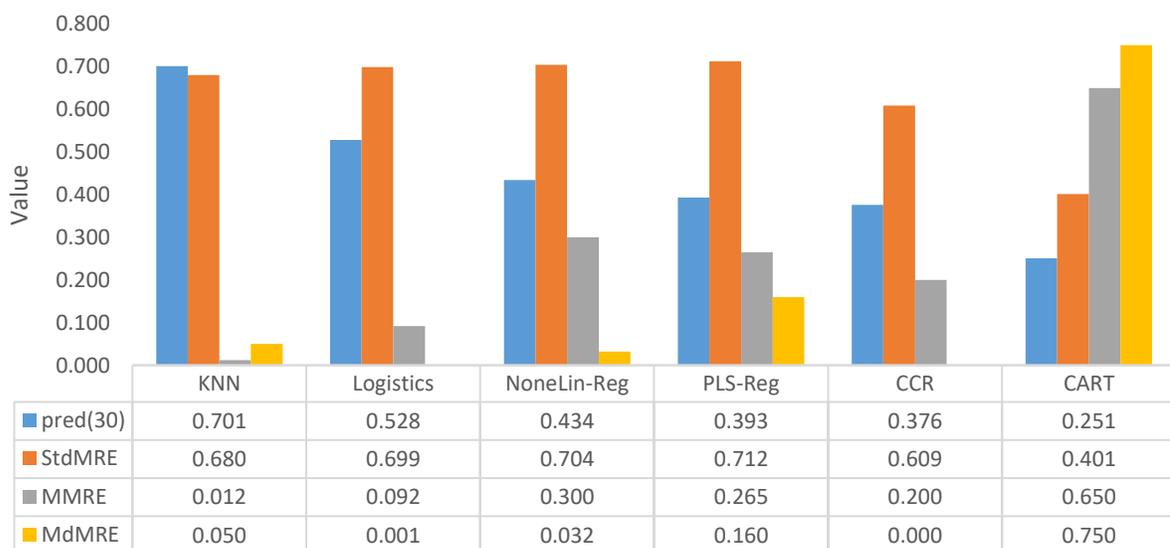

| | KNN | Logistics | NoneLin-Reg | PLS-Reg | CCR | CART |
|---|---|---|---|---|---|---|
| pred(30) | 0.701 | 0.528 | 0.434 | 0.393 | 0.376 | 0.251 |
| StdMRE | 0.680 | 0.699 | 0.704 | 0.712 | 0.609 | 0.401 |
| MMRE | 0.012 | 0.092 | 0.300 | 0.265 | 0.200 | 0.650 |
| MdMRE | 0.050 | 0.001 | 0.032 | 0.160 | 0.000 | 0.750 |

*Figure 6.2-5: Performance of number of submissions by each approach*

1) Crowdsourcing task scheduling follows typical patterns including prototyping, component development, bug hunt, and assembly and coding;

2) Budget phase distribution patterns does not follow traditional patterns, and uploading task rate not representing same budget rate associated with them as about 75% of uploaded tasks would price under 67% of total project budget;

3) Higher number of uploading parallel tasks would make greater Stability and lower Failure rate for the project, however stability and failure rate are not following the same pattern.

4) Higher degree of parallelism would lead to higher demand for competing on tasks and shorter planning schedule to complete the project consequently better resource allocation and shorter project schedule planning.

## 6.3   Schedule Acceleration in CSD

In order to study the schedule acceleration effects of crowdsourcing, we choose three baseline software schedule estimation models to derive representative nominal schedule from the effort data of the 4 CSD projects. It is our assumption that the derived nominal schedule will reflect the corresponding in-house development schedule to complete a project with similar amounts of effort. Finally, we compare the results with the actual crowdsourcing schedule.

To calculate the total CSD effort, we use the combination of the effort from the top-2 workers with higher submissions scores to represent the effort for each task. More specifically, the effort for a CSD task is defined as:

***Def 6.1***: Effort spent on each task is measured by the weighted sum of the top-2 workers' effort spent on the task, with corresponding weights of 80% and 20%, respectively. Worker's effort is measured by the number of days between his registration date and his submission date. If there is only one active worker, the effort would be calculated based on that worker's actual effort only.

Next, we select three baseline schedule estimation models which are commonly used in industry, for deriving corresponding nominal schedules to be compared with CSD project duration. Most software schedule estimation methods adopt an underlying arithmetic model for predicting duration from development effort, which contains some model constants C and exponent D as follows:

$$Duration = C*(Effort)^D$$

The first is schedule estimation model in COCOMO II [80], which follows the basics model shown t bellow:

$$Duration\_I = 3.67*(Effort^{0.28})*(SCED\%)$$

Since we are deriving a nominal schedule, the SCED% is assumed to be 100%, meaning no schedule compression or stretch-out will be considered.

The second model is CORADMO [81], which is a sub-model in the COCOMO II family, specialized for agile software development projects. To reflect the schedule acceleration effects in agile development, CORADMO assumes a square root relationship between duration and effort, instead a cubic root relationship as that in the COCOMO II model:

$$Duration\_II = Effort0.5$$

The third model is proposed by McConnel [78], similar to COCOMO II but with different model constant parameters:

$$Duration\_III = 3.0*(Effort0.33)$$

Finally, we analyze schedule acceleration rate per project in CSD, based on the following definition:

**_Def 6.2_**: Schedule Acceleration Rate, SARi, is the ratio of estimated nominal duration if following traditional in-house development, i.e. Duration_I, II, and III, and the actual CSD duration for each project.

Table 6-4 summarizes results from the analysis on schedule acceleration effect in four CSD projects. The "Effort (worker-day)" row shows the effort aggregated from individual submitter's effort on all tasks within a CSD project. The "Effort (worker-month)" row converts the previous effort into a unit of worker-month by dividing 22 (i.e. usually 22

workdays per month as defined in COCOMO II [94]). The "Actual CSD Duration" row shows the duration from the beginning of the first task to the end of the last task within a project. The next three rows provide the estimated representative durations following the three baseline models as discussed earlier assuming the same project being developed in traditional methodologies other than CSD. Then the next three rows are the calculated SAR by comparing the estimated duration with the actual CSD duration.

*Table 6-4: Derived schedule and schedule acceleration ratio*

| Project | I | II | III | IV |
|---|---|---|---|---|
| Effort (man-day) | 6005.7 | 11037.2 | 8552.7 | 13841 |
| Effort (man-month) | 273 | 501.7 | 388.8 | 629.1 |
| Actual Duration (months) | 9 | 13 | 11 | 14 |
| Duration_I | 17.7 | 20.9 | 19.5 | 22.3 |
| Duration_II | 16.5 | 22.4 | 19.7 | 25.1 |
| Duration_III | 19.1 | 23.3 | 21.5 | 25.2 |
| SAR_I | 2 | 1.6 | 1.8 | 1.6 |
| SAR_II | 1.8 | 1.7 | 1.8 | 1.8 |
| SAR_III | 2.1 | 1.8 | 2 | 1.8 |
| Average SAR | 1.97 | 1.7 | 1.87 | 1.73 |

The last row shows the average SAR for each project, ranging from 1.7 to 1.97. This indicates an overall average SAR of 1.82 (the mean of the four-project level average SAR), with standard deviation of 0.1258. These results reflect that the duration of crowdsourcing software projects is significantly reduced, i.e. by a factor of 1.82 compared with the nominal schedule if following with traditional in-house development methodologies. This indicates a huge potential of schedule shortening in crowdsourced development if with properly design and governance.

## *6.4   Conclusion*

Considering the highest rate for task completion and acceptable submissions, software mangers will be more concerned about risks for adopting crowdsourcing and need to better decision support on analyzing and controlling the risk of insufficient competition and poor submissions. This chapter reports an empirical study to address that end.

In this chapter we introduced a research design to understand the available task completion patterns as well as tasks situation and workers' performance. The findings of this study conclude that:

1. Crowdsourcing task scheduling follows typical patterns including prototyping, component development, bug hunt, and assembly and coding,

2. Budget phase distribution patterns does not follow traditional patterns, and uploading task rate is not representing same budget rate associated with them as about 75% of uploaded tasks would price under 67% of total project budget,

3. Higher degree of parallelism would lead to higher demand for competing on tasks and shorter planning schedule to complete the project consequently better resource allocation,

4. An overall average of 1.82 schedule acceleration rate is observed through organizing mass parallel development in 4 software crowdsourcing projects.

The findings of this chapter will be used as initial setting of task posting pattern to create the discrete event model in chapter 7.

# *Chapter 7*      *Hybrid Simulation*

In many cases, simulation is a tool for decision making that can help in risk deduction at a tactical or operation level [82]. Simulating task scheduling helps with operational level plans. In the presented model, we extended the proposed model in [10] and created a hybrid simulation model based on SDS, DES and ABS using Anylogic [83]. Driven by the resource related challenges in software development, the model includes a meso level, micro level, and macro level to create a hybrid simulation model. This model addresses decision making under planning and understanding purposes [82].

The consistency of agents' availability to respond to tasks and level of qualified performance will be measured based on agents' experience level in the competition.

## *7.1 Working Definition*

Driven by the resource related challenges in software development, process models in CSD contain four main components: tasks, project, crowd-worker and the platform. Details of each component are below:

**Task;** Sequence of time dependent tasks which are complex with usually high-level interdependencies. In this research, task is a tuple of different characteristics of task Id ($ID_i$), task duration ($D_i$), associated award ($AW_i$), task requirement ($Req_i$,), task type ($Tt_i$,), required technology ($Tech_i$), task sequence to other tasks as series of parallel ($P_i$) or sequential tasks ($S_i$) and task status as completed tasks ($C_i$) or failed task ($F_i$). Combination of above characteristics creates a crowdsourced task ($t_i$) in a CSD platform.

$$t_i = (ID_i, D_i, Aw_i, Tt_i, Tech_i, Req_i, P_i, S_i, C_i, F_i)$$
$$\text{where } i = 1, 2, 3, \ldots, n$$

**Project;** Sequence of time dependent tasks make a project ($P_k$). In this study, time sequence of arriving task is considered as a measurement of defining the series of parallel or sequential tasks followed finished to start task dependency. Task (n), $t_n$, is parallel with task (n-1), $t_{n-1}$, if it starts before $t_{n-1}$. And Tn is a sequential task for tn-1, if it starts after $t_n$ is complete. The total duration of sequential tasks per project makes the project duration. Also, the ratio of failed task per project is a measure of project success. Therefore, in this research, a project is defined as tuple of different characteristics of decomposed tasks $\{t_i\}$, project duration ($PD_k$), project failure ($PF_k$), and the assigned crowd workers to the task $\{A_r\}$.

$$P_k = (\{t_i\}, PD_k, PF_k, \{A_r\}), \text{ where k} = 1, 2, 3, \ldots, m$$

**Crowd Workers;** According to Howe [1], crowd-workers ($A_z$), are large and undefined group of skillset workers who have access to the task via internet globally. In this research, workers are a tuple of different characteristics of workers' Id ($WID_r$), reliability factor ($Re_r$), rating ($Ra_r$), Skill-Set ($Sk_r$), Score ($So_i$), number of winning ($Wi_r$), Location ($L_r$), and membership age ($MA_r$).

$$A_r = (WID_r, Re_r, Ra_r, Sk_r, So_r, Wi_r, L_r, MA_r)$$
$$\text{where z} = 1, 2, 3, \ldots, r$$

**Platform;** The customer organization that owns the tasks and has to specify and prepare the task. The platform is where there is an open call for specified software development tasks to be accomplished on behalf of an organization by a large pool of undefined groups of external software workers [4]. The general purpose of a crowdsourcing platform is to provide a market in which:

- Requestor companies can post tasks to be completed,
- Specify prices paid for completing the tasks,

Allows crowd workers perform tasks which are difficult for computers to perform [14].

Therefore platform (Plat) is a set combination of 'm' different projects (P$_m$) which provides 'i' number of tasks (t$_i$) and will attract 'r' number of workers (A$_r$) to perform them. Platform is defined as follow:

$$\text{Plat} = \{\{ti\}, \{P_k\}, \{A_r\}\} \text{, where} \begin{cases} k = 1, 2, 3, \dots, m \\ z = 1, 2, 3, \dots, r \\ i = 1, 2, 3, \dots, n \end{cases}$$

While Platform outcome is not included as aCSD component it is important to measure it in order to measure platform success. Clearly, both tasks' arrival time as well as crowd workers' arrival time and competing strategies have a significant impact on platform failure rate. In order to analyze elements of different components in the model, it is very important to identify the characteristics of all the components.

## 7.2   Conceptual Model

Crowdsourcing is a dynamic market which contains three direct stakeholders: software workers as agents, demand companies as requestors and crowdsourced platforms [3]. Its success is based on the individual stakeholder's behavior and their interactions with each other. Requestors are providing the tasks to the platform. Such platforms need to be designed to ease software workers' understanding of crowdsourcing tasks, as well as form the relationships and practical communication between software workers and requestors [30]. While the ABS model is responsible for individual agent's behavior and DES manages task sequence, SDS shows the interactions among system parameters and feedbacks within the platform. In the hybrid model, DES is responsible for executing tasks in the platform. Once a task arrives in the platform, agents' decision-making process starts via an ABS model. Different decision-making strategies by agents creates differing agents' performance in the platform. Agents' performance directly impacts task failure rate in the platform and platform trust-ability rate in the market. Figure 7.2-1 illustrates the overview of the hybrid simulation model. The proposed model is created based on TopCoder [17] workflow and measures the task failure ratio in the platform.

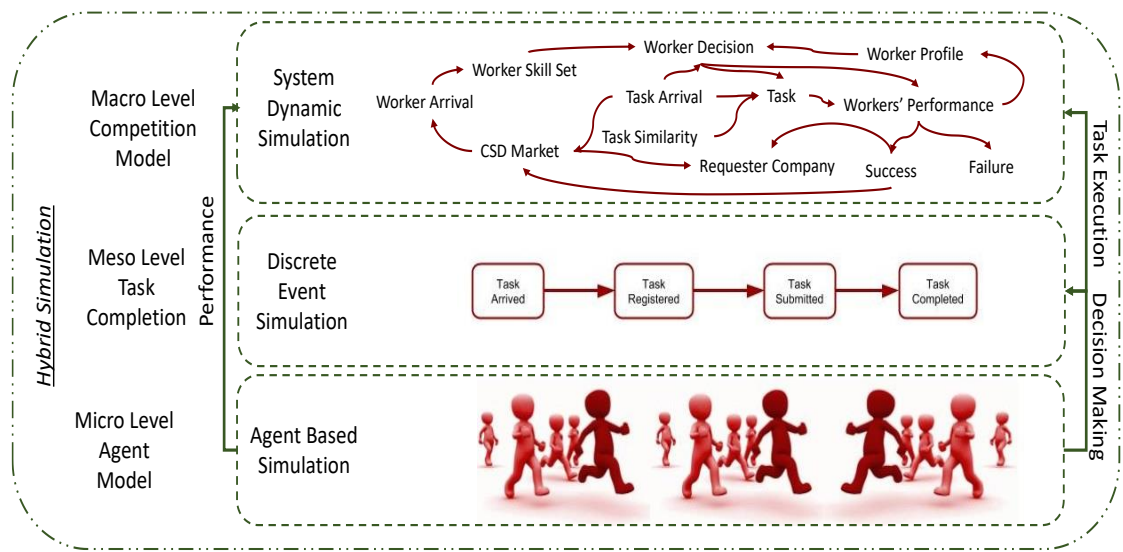

*Figure 7.2-1: Overview of the Hybrid Simulation Model*

*Table 7-1: Summary of Metrics Definition*

| Metric | Definition |
| --- | --- |
| *Task Completion* | |
| # Registration (R) | Number of registrants that are willing to compete on total number of tasks in a specific period of time. Range: $(0, \infty)$ |
| # Submissions (S) | Number of submissions that a task receives by its submission deadline in specific period of time. Range: $(0, \#registrants]$ |
| Peer Review (PR) | Process of reviewing a submitted task to check the quality of submissions. Range: $(0, \#registrants]$ |
| *Task Similarity* | |
| Award (P) | Monetary prize (Dollars) in the task description. Range: $(0, \infty)$ |
| Registration Date (TRD) | Earliest time that task (i) is available on line for workers to register |
| Submission Date (TSD) | Deadline that all workers who registered for task have to submit their final results |
| Technology (Tech) | Required programing language to perform the task |
| Task Type | Type of challenge depends on development phase |
| Task Requirement (TReq) | Detailed requirement describes in task description cosseted with uploaded tasks. |
| *Resource Reliability* | |
| Registered Worker (RW) | Number of tasks a worker registered for in the specific period of time. Range: $(0, \infty)$ |
| Submitted Worker (SW) | Number of tasks a worker submits in a specific period of time. Range: $(0, \#registrants]$ |
| Submissions Ratio (SR) | Percentage of number of tasks a worker submits for of total number of tasks that a worker registers for |
| Reliability (Re) | The percentage of successful task submissions in a worker's most recent 15 task registrations. Range: $(0, 1)$ |
| Metric | Definition |
| Accepted Task Ratio (AcTR) | Number of qualified submitted task per worker to total registered tasks by the same worker |
| Submitted Task (ST) | Number of submissions that a worker does in the history of performance |
| Trust Ratio (TR) | Percentage of number of qualified submissions a worker submits of the total number of submissions that a worker does. |

| Platform | |
|---|---|
| Failure Ratio (FR) | The ratio of number of canceled tasks per total number of tasks per platform. |
| Arrived Task (ArT) | Task uploaded in the platform to be work on |
| Failed Task (FT) | Task that is not receiving any qualified submissions |
| Task Duration (TD) | Total available time from registration date to submissions deadline. Range: $(0, \infty)$ |
| Task Submissions Ratio (TSR) | Percentage of number of tasks that submitted to total number of registered tasks at any given time. |
| Task Completion Ratio (TCR) | Percentage of number of tasks that can pass the peer review to total number of registered tasks at any given time. |
| Task Drop Ratio (TDR) | Percentage of number of tasks that are not submitted while is registered to the total number of registered tasks at any given time. |
| Worker Registration Date (WRD) | Date and time that a worker registered for a task. |
| Worker Submission Date (WSD) | When a worker submits their final results after registering for a task |
| Worker Earliest Registration Time (WERT) | First time that a worker registered for a task in their history of membership in the platform |

### 7.2.1    Macro Level: Platform Competition

Crowdsourcing platform as a dynamic market that contains both newly arrived tasks and available agents. Therefore, a systems dynamic (SDS) model is more appropriate for modeling the effect of agents' error in the reliability of returning qualified tasks in the system with showing a broader behavior of agents [84] [85].

Proposed SDS model for the platform, figure 7.2.2, contains 14 variables including Task, Agent Decision, Workers' Performance, Task Similarity, Worker Profile, Worker Skill-set and different available Crowdsourced markets. The SDS model represents the causal loops among different levels of the platform.

"*Tasks arrival*" in the platform is an event following Poisson distribution. Arrival task executed by DES model join the pool of open tasks in the platform. Each new task is associated to a similarity factor upon arrival. The task similarity factor is a causal similarity analysis based on arrival tasks. According to empirical analysis, higher degree of "*Task Similarity*" among the pool of open tasks in the platform, leads to higher competition level.

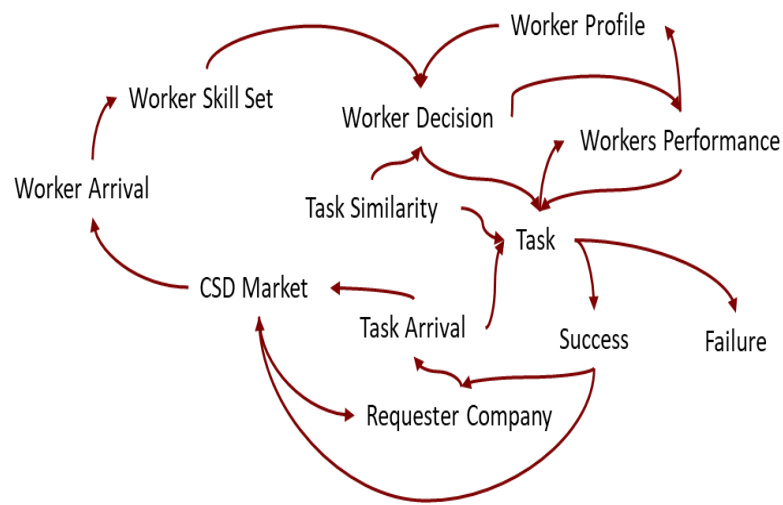

*Figure 7.2-2: Overview of Platform Model (SD)*

*Table 7-2: Variables Used in Systems Dynamic Model*

| Metric | Definition | Model Variables |
|---|---|---|
| Systems Dynamic Simulation | | |
| Task Similarity (TS) | Tuple of monitory prize, task duration, task complexity, required technology and task type. Range [0,1] | ts(double) |
| Task Duration (D) | Total available from task registration start date to submissions deadline. Range: $(0, \infty)$ | duration(time) |
| Failure Ratio (FR) | Ratio of number of cancelled tasks per total number of tasks per platform. | tFR(double) |
| Task Arrival (TAr) | Task uploaded in the platform to be worked on | tTAr(int) |
| Task Failure (TF) | Task that is not receiving any qualified submissions | tTF(int) |
| Task Submissions Ratio (TSR) | Percentage of number of tasks that are submitted to total number of registered tasks at any given time. | tTSR(int) |
| Task Completion Ratio (TCR) | Percentage of number of tasks that can pass the peer review to total number of registered tasks at any given time. | tTCR(double) |
| Task Drop Ratio (TDR) | Percentage of number of tasks that are not submitted while registered to total number of registered tasks at any given time. | tTDR(double) |
| Worker Decision (WD) | Worker decides to register for a task and/or make a submission for the same task | tWD(Register, Submit) |
| Workers Performance (WP) | Number of qualified submissions that a worker makes Range: (0, i) | tWP(double) |
| Worker Skill Set (WSS) | Workers' list of skillsets in their profile. | tWSS(Java, C) |

Average task similarity of 90% leads to average competition level of 55%. However, degree of task similarity does not impact submissions level and failure ratio increased by increasing task similarity in the pool of open tasks. Average task similarity greater than 80% provides an average failure ratio of 20%, while average task similarity of 60% provides an average failure ratio of 13%.

As illustrated in figure 7.2-2, other available CSD platforms in the market impact "*workers' arrival*" on platform as well as requestors' company decision to work with a specific platform at any given time. Also, "*agents' skillset*" and task similarity are the main factors to impact agents' Decision-Making process to take a specific task in the platform. Agents' decision to "*Register*" and "*Submit*" a task provides information for "*workers' performance*" and consequently updates "*workers' profile*".

Workers'd performance in the platform lead to different task status of success or fail. If the task is successfully complete it counts as "*Complete*" and is reported to the "*Requestor Company*" for further process. In this model, Task completion ratio ($TCR_k$) is calculated based on ratio of passing review tasks to the total number of registered tasks, equation 6.1.

$$TCR_k = \frac{\sum_{i=1}^{n} C_i}{\sum_{i=1}^{n} R_i} \qquad \text{Eq.-6.1}$$

However, if the task is not qualified, it will count as "*Failure*". Task failure ratio ($TFR_k$) in this model is the ratio of tasks that are registered but are dropped or not passed peer review to the total number of registered tasks. This model is reported as third failure prediction metrics, equation 6.2.

$$TFR_k = 1 - \frac{\sum_{i=1}^{n} C_i}{\sum_{i=1}^{n} R_i} \qquad \text{Eq.-6.2}$$

Table 7-2 summarized the structured attributes used in the SD model.

### 7.2.2    Meso Level: Task Completion

Task lifecycle is one of the most important factors in workers' decision-making process in crowdsourcing since it is a representative of task priority and complexity. In this model, tasks are defined as a set of discrete events (DE) that have start and end dates.

Arrival Task ($T_i$) in this model is defined as a list of 'w' number of tasks from the same project arriving at the same time in the platform.

$$T_j = \{t_i\}, \text{ where } \begin{cases} j = i, 2, 3, \dots, w \\ Dj \leq Di \leq D(j+1) \end{cases}$$

Task arrival in the model is an event that follows Poisson distribution of population of tasks with lambda equal to 87. Figure 7.2-3 illustrates the overview of the DSE model based on TopCoder work flow [76].

When a task arrives ("*Arrived*") in the platform, it can be "*Registered*" by available agents to start the process. Empirical analysis shows that, arrival tasks will attract agents to work on them at a rate of 70%. If the task cannot attract any agents and faces zero registration, it will be *"Starved"*. Based on different decision-making scenario by competing agents, the proposed model will provide 2 different failure prediction ratios in different phases of a task lifecycle.

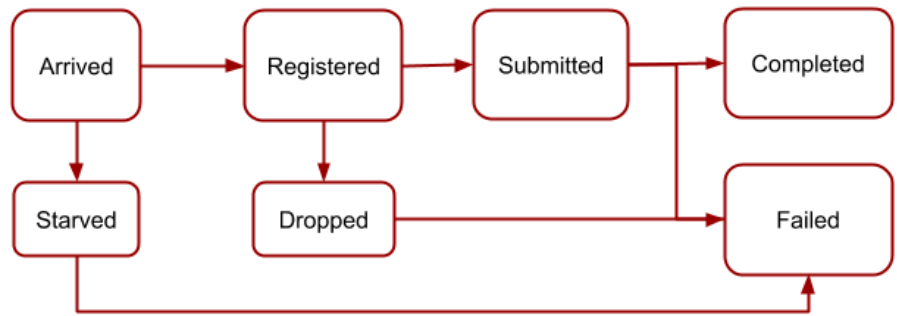

*Figure 7.2-3: Overview of the Task Completion Model (DE)*

TopCoder currently applies a heuristic-based color coded (RED, YELLOW, and GREEN) predictor for a task out come in the registration phase. The model involves three simple rules with respect to the sum of the reliability ratings of all registered agents for a task [8]. If the sum of reliability ratings is greater than or equal to 2, the predictor produces a GREEN label, as a symbol of success for the task to the requester. If it is less than 1, the predictor produces a RED color, which represents a task likely to fail. YELLOW is the result when the sum of reliability ratings is between of 1 and 2, which is an uncertain situation for a task status. In this model we extend the heuristic model used by TopCoder to predict the first failure ratio in the registration phase, $FPR_i$, equation 6.3.

$$FPR_k = \begin{cases} \frac{\sum_i^n Re_j * P_j}{3} & \sum_i^n Re_j > 2 \\ \frac{\sum_i^n Re_j * P_j}{2} & 1 < \sum_i^n Re_j < 2 \\ \frac{\sum_i^n Re_j * P_j}{1} & \sum_i^n Re_j < 1 \end{cases} \qquad \text{Eq.-6.3}$$

Where, $P_j$ is the probability of making no qualified submissions by the registered agents as followed in table 1.

By the end of the task duration, registered tasks may be "*Submitted*" by agents. If registered tasks face zero submissions, they will be *"Dropped"*.

*Table 7-3: Variable used in Discrete Event Simulation Model*

| Metric | Definition | Model Variables |
|---|---|---|
| Discrete Event Simulation | | |
| Registered (R) | Number of registrants that are willing to compete on total number of tasks in specific period of time. Range: (0, r) | tregister(int) |
| Task Duration (D) | Total available time from task registration start date to submissions deadline. Range: (0, 30) | duration(time) |
| Submitted (S) | Number of submissions that a task receives by its submission deadline in specific period of time. Range: (0, #registrants] | tsubmit(int) |
| Peer Reviewed (PR) | Process of reviewing a submitted task to check the quality of submissions. Range: (0, #registrants] | tpeer(int) |
| Completed (C) | Qualified task that has successfully passed the peer review | tcomp(int) |
| Reworked (RW) | Qualified tasks that need some adjustment in order to pass peer review | trework(int) |
| Failed(F) | Non-qualified task that has not passed peer review | tfail(int) |
| State | Each task has a state that declares the task situation on the process | state (Arrived, Registered, Submitted, Reviewed, Complete) |

In this model, as soon as the first submissions happen, task submission ratio ($TSR_k$) will be dynamically calculated based on the ratio of submitted tasks to the registered tasks in the platform ($TSR_k$), equation 6.4.

$$TSR_k = \frac{\sum_{i=1}^{n} S_i}{\sum_{i=1}^{n} R_i} \qquad \text{Eq.6.4}$$

Statistical correlation analysis on different drivers in a crowdsourced task presents that task failure and task submissions ratio in the platform are directly related. Therefore, second task failure prediction ratio ($FPS_k$) will be analyzed based on a linear regression model of dynamic task submission ratio, equation 6.5:

$$FPS_k = 0.0473(TSR_k) + 0.014 \qquad \text{Eq.-6.5}$$

It is reported that the average CSD task duration is 16 days [92]. Also, according to our empirical analysis, task duration is follows triangle distribution with the maximum 30 days, minimum 1 days and a mode of 16 days. Submitted tasks should be checked to ensure the quality of submissions. In this phase, if the quality of task is greater than or equal to 75, it is completed, if not it is failed.

If the submitted task is qualified, it is recorded as "*Completed*", otherwise, the task will be recorded as a "*Failure*". Moreover, all the starved and dropped tasks will be reported as failures. Table 7-3 summarizes the structured attributes used in the DES model.

### 7.2.3    *Micro Level: Agent Model*

Crowdsourced projects integrate online and unknown workers elements into the design. It is reported that crowd workers often overestimate their productivity [38] and register for more tasks than they can complete. Therefore, simulating crowd-workers with various characteristics, decision making process and performance ratio is difficult. Applying agent based (ABS) method to simulate crowd-workers' behavior individually provides the option of observing the diversity of attributes among them. Crowd-workers are represented as agents who have one or more of the following characteristics:

- Identifiable with a set of rules that directs their behavior,

- Autonomous agents that can act independently in the environment and has control over their actions,

- Situated workers that work in the same environment and interact with each other,

- Flexible agents that can adapt their behavior to be a better fit to the environment [86].

Agents arrival to the platform follows non-homogenous Poisson distribution [8]. Agents are assigned unique IDs upon creation in the simulation. As shown in the figure 7.2-4, the model contains:

- The agent environment,

- Set of agents' attributes, and

- The agent decision making process [87].

In any crowdsourcing platform, an individual agent has an "*Agent's Knowledge*" which is based on their skill-set, background and the society s/he is coming from. Also, the agent is a member of the "*Pool of Agents*". This membership makes the agent interact and communicate with other agents in the pool. TopCoder divides "*Pool of Agents*" into 5 groups. The 5 agent groups are defined into 5 belts of Red, Yellow, Blue, Green and Gray, which correspond to the highest skillful agents to the lowest ones respectively [17]. Agents update their "*Personal Knowledge*" based on other agents' performance in the platform. The combination of interaction and communication with other agents in the pool and the agent's personal knowledge creates the agent's "*Strategy*" in the system.

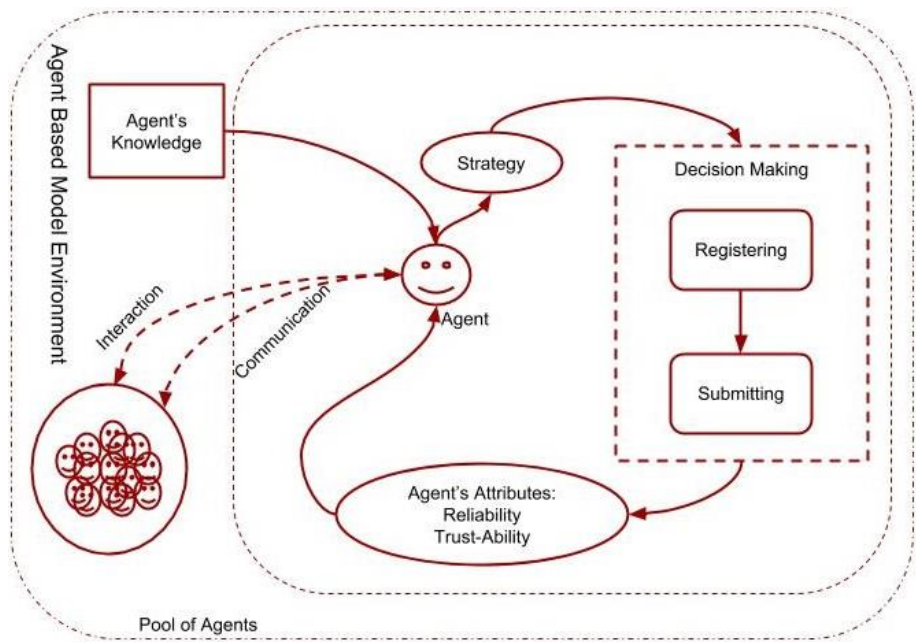

*Figure 7.2-4: Overview of Agent Model (AB)*

To address the rest of the agent's characteristics, the agent's strategy provides their utility factor [8] that defines their individual behavior. Therefore, each agent has an internal "*Decision Making*" process consisting of two components: *"Registering*" for a task and "*Submitting*" the task. Agent decision making is related to the information that the agent receives from the agent's community and social environment, as well as another agent's choice of competing on a task [88].

Once a task generates in the system, agents decide to register for the task. Agents' registration for a task categorizes them as active agents. Agents' registration in the model occurs at the rate of one registration per day per agent if the associated random number to the agents was greater than 0.8. Agents perform and eventually submit a task based on a variety of factors, including number of open tasks in their list of tasks, task complexity, task competition level and probability of winning in the competition. Agents analyze their probability of winning based on the task competition level and number of higher ranked opponents [35].

It is reported that some agents are using their history of victory as a policy to win the registered tasks and assure an easy competition [35]. Also, agent may drop some of the registered tasks due to lack of time to finish them or a better suited task arrives in the platform. Agent's final decision to submit a task and level of submission impacts the agent's profile. The agent may decide to submit the registered task. Task submission decision follows an event with rate of 0.51 per day per agent if the random number associated to the agent is greater than 0.86. By submitting a new task, agents' attributes of reliability factor will be updated.

*Table 7-4: Variable Used in Agent Based Simulation Model*

| Metric | Definition | Model Variables |
|---|---|---|
| Agent Based Simulation | | |
| Registering (RW) | Number of tasks a worker registered for in a specific period of time. Range: (0, i) | worker_registering (ArrayList<Task>) |
| Submitting (SW) | Number of tasks a worker submits in a specific period of time. Range: (0, #registrants] | worker_submitting (ArrayList<Task>) |
| Quality (Q) | The quality of submitted task based on associated score. Range: (0, 1) | tQ(int) |
| Reliability (Re) | The percentage of successful task submissions in a worker's most recent 15 task registrations. Range: (0, 1) | tRe(int) |
| Trust-ability (TA) | Percentage of number of qualified submissions among total number of submissions that a worker makes | tT-A(int) |

According to previous research, 59% of the agents are responding to the task call in the first day while only 24% of them who register early for a task will make a submission [88]. 76% of the agents were exceeding the submissions deadline [88]. 48% of registrants are among average rated agents (i.e. green, blue and yellow belt). 86% of higher rated agents (i.e. yellow and red) submit qualified submissions. The personal knowledge in the proposed ABS model reflects the reported empirical analysis.

According to empirical data, reviewed tasks with scores greater than 75% are considered qualified and the submitter agent is reported as winner or runner-up. Therefore, qualified submissions are determined by assigning a random number greater than 0.75 as quality score. When the submission passes peer review, the agent's attribute of winning will be updated. Moreover, the new score is reported in the agent's profile.

Decisions from simulated agents will determine task progress. The details of the model integration are explained in part 7.3.3-algorithm 1. Table 7-4 summarizes the variables used in the agent-based model.

## 7.3 Integration of the Hybrid Simulation Model

### 7.3.1 *Macro Level: Platform Competition*

Arrived tasks and agents created by daily events enter to the pool of open tasks and available agents in the platform as the dynamic crowdsourced market place. Task execution follows the project schedule provided by DES. Each task associates to a similarity rate which creates an event following uniform distribution between 30% and 98%. In this model, agent arrival follows Poisson distribution. Arriving tasks in the platform will impact an agent's decision making and consequently an agent's experience. Agent's experience in the model follows beta distribution with minimum = 0, Maximum = 3000, $\alpha$=1, and $\beta$ = 5. The result of the SDS model provides the platform failure ratio.

### 7.3.2 *Meso level: Task Completion*

Task arrival in the DES model follows specific project schedule defined based on task requirements and sequence. Each task must pass 3 states to be successfully completed upon arrival.

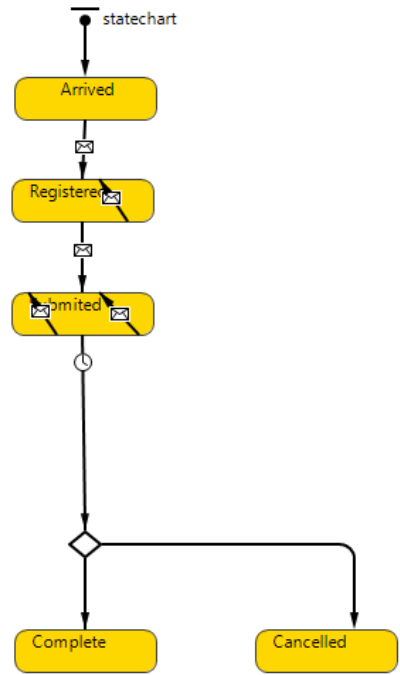

*Figure 7.3-1: State Chart, DE model*

As figure 7.5-1 illustrates, as soon as the task is executed, it is added to the state of arrived tasks. Once the first agent takes the task to register for, a message will be released as a trigger and the task is moved to the state of being "*Registered*". In this state, the model will provide failure prediction ratio based on the reliability and experience of agents who take on the task. Registered tasks are associated with a duration following triangle distribution reported in part 7.2-2 Registered tasks move to a "*Submitted*" state as soon as one of the registered agents makes a submission and the model releases a submission message.

Once a task moves to the submissions state, the failure prediction model switches to the submissions phase and continues the prediction based on the task submission ratio at any given time. When the submissions deadline is over, the task moves to the peer review phase. If the quality score was greater than 75, the task is reported as complete, otherwise it is reported as a failure. The completed task is reported to the requester company and its sequential task will be posted in the system to be performed.

### 7.3.3    Micro level:  Agent Model

Agent arrival in the AB model is an event which follows a Poisson distribution with $\lambda$ = 800. Once the model runs, agents start arriving in the model.

When an agent decides to register for a task, the agent's personal profile gets updated with new registered tasks. In this model, agent's registration decision follows an event with rate of 1 registration per day. Also, none of the agents can work on more than 5 open tasks at the same time. If the agent meets the above criteria and s/he has the required skillset to perform the task, s/he can decide to register for the task. To make the registration decision, agents are assigned to a random number in the range of (0,1). If the random number is greater or equal to 0.8, s/he will register for the task. As soon as the agent registers for the task a message will be sent to the task to change its state to registered. Moreover, according to the empirical findings the average competition level per task is 18 [92]. Therefore, in this model, the probability of an agent registering for a task with more than 18 assigned registrants follows a Bernoulli distribution (P= 0.3), and the probability of the agent competing on a task with assigned registrants less than or equal 18 is 1.

```
// Procedure of Agents Decision Making (ADMi)
1: i = 1 to n
2: for (int i = 0; i < n; i++)
// Agent Registering for a Task
3: Agent Registering = create_ArrivalEvent(1, days);
4: for (Task t: Model.Tasks) {
5:         if ((t.isRegistering) && (this.OpenList < 5) && (this. Expertise == t.ExpertiseRequirements) && ( Rating > 0)){
6:                 if ((uniform_pos ()>0.80) ){
7:                         if(t.registers < 18){
8:                                 send("Worker Register", t);
9:                                 this.WRlist.add(t);
10:                                t.TRList(this);
11:                                t.Rating = this.Rating;
12:                        }else{
13:                                double X= bernoulli(0.3);
14:                                if( X==1){
15:                                send("Worker Register", t);
16:                                this.WRlist.add(t);
17:                                t.TRList(this);
18:                        }}}}
// Agent Submitting for a Task
19: Agent Registering = create_ArrivalEvent(0.51, days);
20: for (Task t : this.WRlist) {
21:        if (( t.isSubmitting = true) && (this.OpenList >1)){
22:                double X= uniform_pos();
23:                double Y = 0;
24:                if(this.Belt == 4){
25:                        Y= 0.4;                        }
26:                if(this.Belt == 3){
27:                        Y= 0.4;                        }
28:                if(this.Belt == 2){
29:                        Y= 0.61;   }
30:                if(this.Belt == 1){
31:                        Y= 0.55;   }
32:                if(this.Belt == 0){
33:                        Y= 0.75;   }
34:                if(X*Y<0.051){
35:                        send("Worker Submit", t);
36:                        this.WSlist.add(t);
37:                        t.TSList(this);
38:                        this.Score= 100*uniform();
39:                        if(this.Score > t.Score){
40:                                t.Score = this.Score;
41:                                Score = this.Score;
42:                                t.Winner = this;
43:                                this.WWlist.add(t);
44:                        }}}}
```

*Figure 7.3-2: AB Integration Algorithm*

If an agent registers for a task, s/he can decide to make a submission for the same task. Agents' submissions decision for the registering tasks follows an event with rate of 0.51 submissions per day. For deciding to make a submission, each agent is assigned to a random number in the range of (0,1). If the product of the random number and the agent's registration probability based on the agent experience belt is less than 0.051, the agent will make a submission for the task. Once the agent makes a submission, a message will be sent to the task as a trigger to change the state of task from registered to submitted.

By sending the submission message, the number of submissions, reliability, and rating factor will be updated in the agent's profile. The reliability of agents who make a submission follows pert distribution with an average of 10%. In this step, if the random score assigned to the submitted task is greater than 75, the agent will be reported as the winner.

AB integration algorithm as shown in figure 7.3-2, presents the associated pseudo code with the AB model in the simulation.

## 7.4    Evaluation

### 7.4.1    Overview of the Example Project with in the Hybrid Model

In the first 100 days, 15 tasks have arrived in the platform, of which 5 have been reposted and 3 are under process. As is presented in figure 7.4-1, Task 1 has attracted registrants on the very first day and received submission on week 5. Unfortunately, the review phase shows that the submission was not qualified, and it is reported as a failed task. Therefore, task 1 will be reposted in the platform as task 3 with 1 day added to the task duration. Task 3 got registered on the second day of arrival and received the qualified submission. Meanwhile, task 2 was posted in the platform on day 7. It got registered in the same day and received the first submission on day 10. Peer review marked this task as successful and reported it as complete to the platform.

Task 4 arrives in the platform on day 23, while task 3 was in the peer review phase. It got registered on day 25 and received a qualified submission on day 27. On day 30, the 5th task arrived in the platform; it took 5 days before task 5 could finally attract an agent to register for it. However, the agent dropped the task and the task was reposted as Task 7 in week 39. It took 6 days for task 7 to receive a registration, but on day 50, it received a qualified submission and reported as a complete task. Task 6 posted on day 37 with duration of 5 days and received a registrant in the same day.

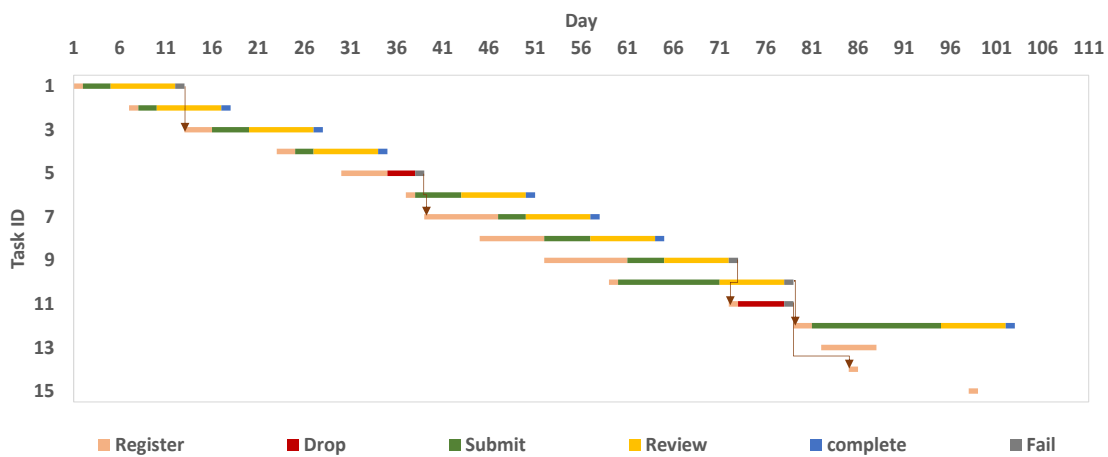

*Figure 7.4-1: Overview of a project with in hybrid simulation model*

Task 8 was posted on day 45 with duration of 5 days, when task 7 was still open. It took a week to get registered and received the qualified submission on day 57 and reported as complete on day 64. Task 9 was posted on day 52, got registered on day 61, and received a not qualified submission on day 65 and was reposted as task 11 on day 72. Interestingly, task 11 was dropped on day 82 and reposted as task 14 again.

Task 10 was posted on day 59 and received registration on the same day. After 11 days, task 10 received a submission which was not qualified; therefore, it was reposted as task 12 on day 76 with 3 days of duration increase. Task 12 got completed following a qualified submission on day 95. Tasks 14 and 15 arrived on days 85 and 98 respectively.

### 7.4.2 Performance of Failure Prediction Simulation in the Platform

#### 7.4.2.1 Platform Evaluation Design

The success of a CSD platform is based on the level of task failure in the platform. Since task failure is the result of agents' performance in the platform, it is very important to understand agents' utilization in CSD. To address that end, the simulation was updated with associated schedule of the example project and ran under the basic setting of simulation design for 30 times, 60 days per time. During the simulation time, if a task failed due to not receiving a qualified submission or no submissions at all, it will be reposted.

The last step was to understand the accuracy of failure prediction in different task state. To do so, mean relative error (MRE) of each task failure prediction model ($FP_{zy}$) was calculated based on the available actual failed task ($AF_{zy}$) data gathered from TopCoder in the same day was calculated, where z is time and y is task state, as shown in equation 7-6.

$$\text{MRE}_{\text{FP}} = \frac{\sum_{z=1}^{L} AF_{zy} - \sum_{z=1}^{L} FP_{zy}}{\sum_{z=1}^{L} AF_{zy}} \qquad \text{Eq.7-6}$$

The t-test was also applied to the prediction results in each state to confirm the accuracy of the models.

### 7.4.2.2 Platform Performance

As presented in figure 7.4-2, the average success ratio in the platform while running the project is 71%. The average none-qualified submissions are 19% and the average zero submission is 7%. Therefore, the average task failure ratio is 13%. Also, the mean relative error (MRE) of the failure prediction models is only 1.1% in the registration phase and 2% in the submissions phase.

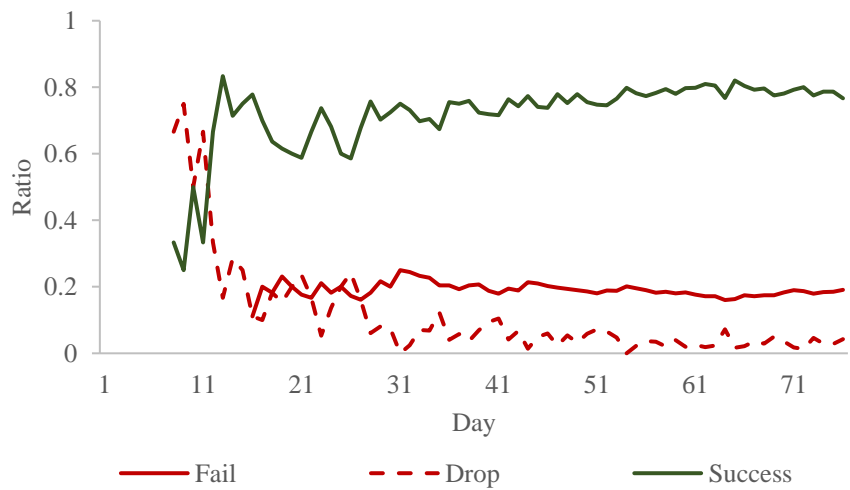

*Figure 7.4-2: Simulation Platform Status*

The result of the t-test on 60 observations per failure prediction model showed that the probability of error in failure prediction in registration phase is almost 0 and the error in the submissions phase is almost 1% with 0 hypothesized mean difference in both states. The Pearson Correlation provides very close correlation between the actual failure for the project based on the data set and failure simulation results. According to this test, the correlation is 0.21 in the registration phase and 0.42 in the submissions phase.

Additionally, it is reported that agents' utilization in the platform impacts the platform success ratio. a deeper look at the agents' utilization in the platform at the time of sample project, as shown in figure 7.4-3, reveals that the average agents' utilization factor is 43% with a standard deviation of 0.15. This information can be used to create a control chart to study the agents' utilization in the platform. Figure 7.4-3 shows the control chart of agent utilization in the platform. Unlike industrial control charts where the goal is to keep the quality as close as possible to the average, in CSD the goal is to have as high a utilization as possible and closer to the upper level Therefore, we took a closer look at the days with utilization factor close to or more than the upper level.

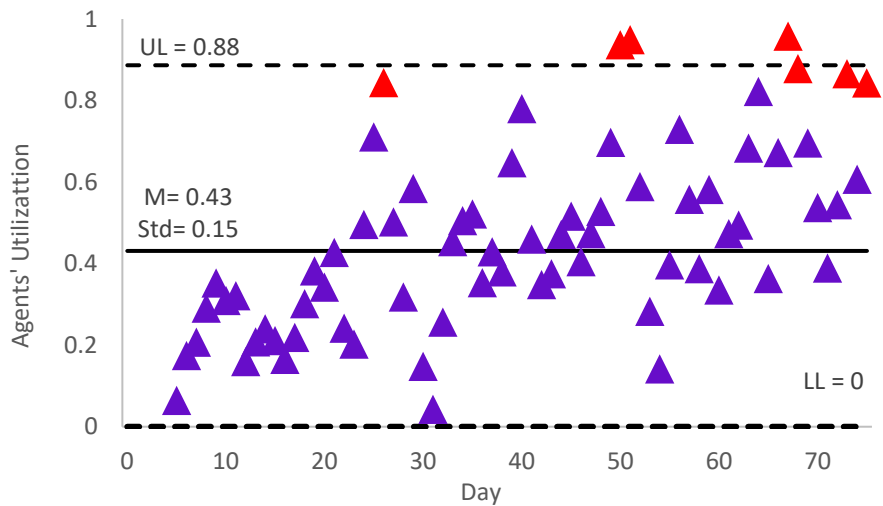

*Figure 7.4-3: Agent Utilization*

Analyzing days with higher level of agents' utilization factor showed that on days with highest utility factor, the average similarity of the open pool of tasks is less than 60%. Interestingly, in those days, the availability of agents in the middle ranking clusters (i.e. green and blue) are higher and almost close to each other, figure 7.4-4.

These results lead us to design and test two different scenarios to answer our research questions.

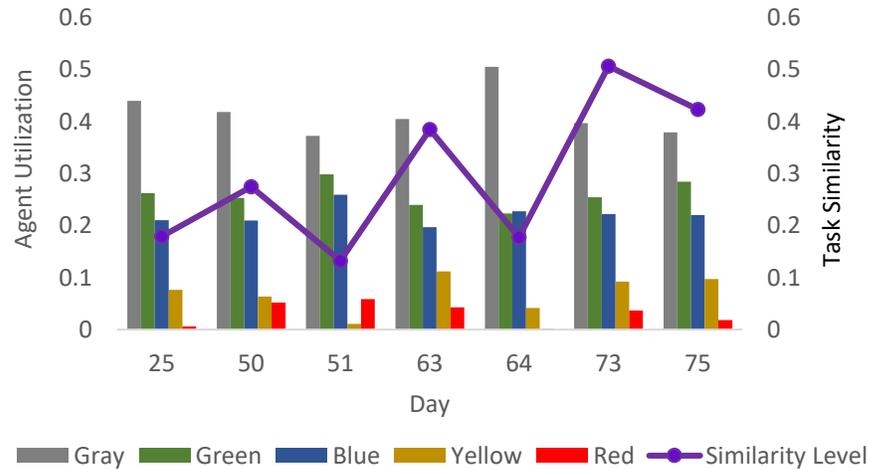

*Figure 7.4-4: Agents Availability*

## 7.5  Evaluation Case Study

The aim of the proposed hybrid simulation model (HSM) in this research is to understand the optimal level of openness in the pool of available tasks and diversity in the competition level in order to minimize task failure ratio in any CSD platform. The results of the evaluation model lead us to design and test two different scenarios to achieve the research goal.

*Scenario1 (Task Openness):* To successfully crowdsource a software project in a crowdsourcing platform not only is it important to fully understand task dependency in a project, but it is also vital to know the impact of available open tasks on each other in the platform. Different task similarity level may be representative of task difficulty, task size and task priority [76]. Also, degree of task similarity represents task openness in the pool of open tasks in the platform. Therefore, understanding workers' behavior and performance based on the degree of task similarity in the pool of open tasks is helpful to present a more effective task planning.

Different researchers discuss that shorter task duration is associated with less complex work, which may impact workers' decision making of task taking based on higher number of tasks getting done. Also, it is reported that there is a general negative correlation between monetary prize, as one of the elements of task similarity, and workers' behavior [76] besides the fact that workers are more attracted to tasks with a certain range of associated monetary prize (i.e. $750) [76]. Therefore, task similarity can be an important factor in task failure rate.

Empirical studies show that though higher similarity level among tasks leads to higher registration, it also leads to lower submissions and higher failure ratio. Scenario 1 aims to investigate the impact of task similarity level on agents' performance in the CSD platform. In this scenario, we will test 4 different testing policies based on previous empirical findings. Each testing policy will provide arriving tasks with limited similarity rate into the platform. The result of different testing policies will be compared to address the findings of scenario1.

*Scenario2 (Agents' Diversity):* According to TopCoder, higher rating level among agents represents higher level of experience. It is also reported that higher reliability factor among agents presents highest chance of submitting qualified tasks by agents [40] [46]. In general, it is expected that higher rated agents are associated with higher reliability factor due to their experience. However, the reliability factor measure is based on number of agents' submissions and is not related to the quality of the submission. It is reported that higher experience agents may take advantage of their reputation and apply cheap talk to the task [35]. Scenario2 aims to investigate the impact of diversity among agents in terms of agents' experience level on agents' performance.

In this scenario, we will test 4 different testing policies based on different agents' experience level introduced by TopCoder [17]. Each testing policy will attract agents with specific experience level to register for the chosen task, and agents' behavior and performance level will be reported. The result of different testing policies will be compared to address the findings of scenario 2.

To apply evaluation scenarios, we will simulate the motivation example. In the motivation example, task 8 was reposted 7 times to be completed and we aim to analyze this task under different presented scenarios.

### 7.5.1    Scenario 1, Task Openness

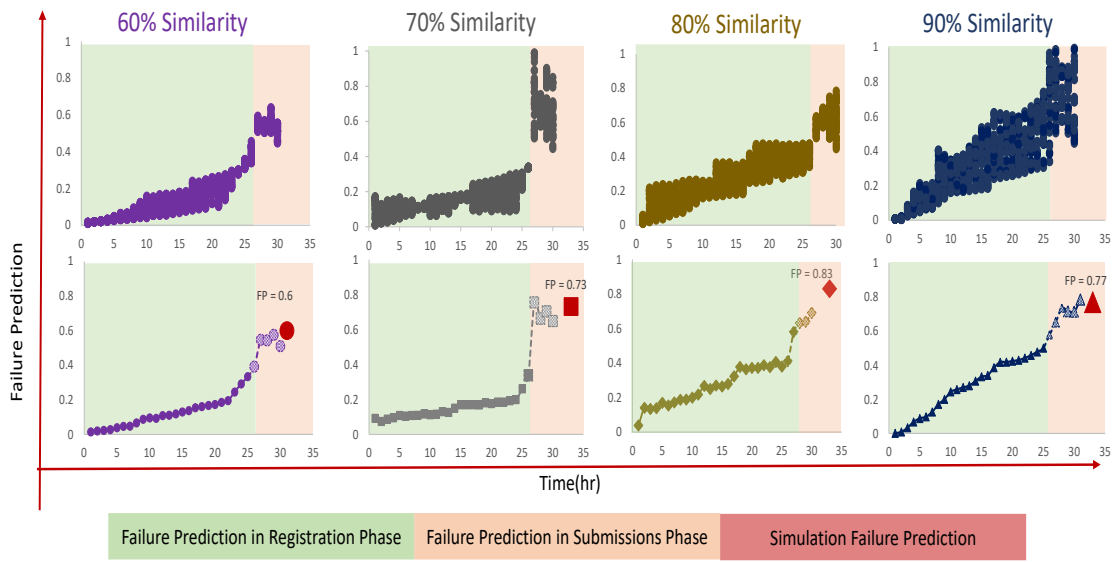

*Figure 7.5-1: Scenario 1, Impact of Task Similarity on Task Failure*

New arrival tasks in the platform will join the pool of open tasks and wait in the queue to be registered by agents. It is reported that agents are often more attracted to similar tasks in terms of monetary prize [76], task complexity and context [38] [59]. This fact creates a demand market with an associated similarity factor to each task. Workers are using this factor as one of the inputs to their decision-making process. Therefore, it is expected that higher number of available similar tasks directly impact the level of competition per task. In this scenario, we aim to study such an impact.

According to empirical analysis, new arrival tasks are competing with tasks with average similarity of 69% per week. In scenario 1, it is assumed that the project manager can control the openness of the pool of open tasks in terms of task similarity when they post their tasks. The simulation ran 30 times and the average of all the runs is used in this scenario. Four different testing policies based on degree of task similarity will be analyzed. The result of this study reports task failure prediction ratio in different states of a task in the platform. This scenario ran 120 times, 30 times under 4 different testing scenarios. Figure 5.7-1 illustrates failure prediction per testing policies and table 7-5 summarized agents' participation per testing policy in scenario 1.

*Table 7-5: Summery of Scenario 1*

| Scenario 1 | | TP1 | TP2 | TP3 | TP4 |
|---|---|---|---|---|---|
| Task Status | Fail | 18 | 22 | 25 | 23 |
| | Success | 12 | 8 | 5 | 7 |
| | Failure Prediction | 60% | 73% | 83% | 77% |

### 7.5.1.1  Testing Policy 1: 60% Degree of Task Similarity

The project manager chose to post the task in the task pool with degree of similarity equal to 60%. The result of this policy lead to an average 1% failure prediction on the early registration stage; by time 26, which is the moment the first submission was made, the failure prediction was increased up to 35%. This fact shows that as time passes the chance of receiving high reliable agents to complete the task is low even though the average failure ratio in this state is 8%. At time 26, failure prediction shifts to the submission state and increases to 45%. Unfortunately, the task received only 4 submissions out of an average 48 registrations which increased the failure prediction to 55% on task submission deadline. The results of the simulation show that out of 30 runs, 18 failed which provides actual failure ratio of 60%.

 Applying this policy would attract a good diversity of agents. The result shows that competing with 60% similarity attracts on average 44% of Gray agents, 22% of Green agents, 11% Yellow agents and almost 6% Red agents.

On average 67% of submissions were made by Gray agents and 33% by Blue agents.

### 7.5.1.2  Testing Policy 2: 70% Degree of Task Similarity

Under testing policy 2, the project manager will keep the similarity level of pool of open tasks at 70%. At the first registration the failure predictor predicted a 9% chance of task failure. As time passed and the task received a higher level of competition, task failure prediction increased to 33% in the registration phase. The average failure prediction during the registration state is 14%. At time 26, the first submission was made and increased the failure prediction to 75%. Finally, at time 30, the task received its third and last submission with failure prediction of 67%. This task attracted 36 registrants and received only 3 submissions.

The average result of 30 runs of simulation under this policy showed that out of 30 runs, 22 failed, indicating 73% task failure.

Moreover, this task could attract competitors among Gray, Green, Blue and Red ranking belts with participation ratio of 37.5%, 25%, 25% and 13% respectively. Among all the participants, Gray and Green agents made submissions.

It seems that under this policy, the task cannot attract high reliable agents to make a submission; however, it successfully attracts higher ranked agents to compete.

### 7.5.1.3  Testing Policy 3: 80% Degree of Task Similarity

Under this policy, the project manager decided to join the pool of tasks with task similarity level of 80%. At time 1, the task could provide a competition environment with only 1% failure prediction. Starting at time 2, failure prediction hikes to 22%. This policy attracts slow arriving agents to the competition. There are only 6 major agents that arrive in the first 26 days before the task received its first submission. At time 12, the failure ratio increased to 36% from 26%. At time 16 it again was raised to 41% and at time 18 to 47%. At time 26, the first submission occurred and pushed the failure prediction metrics to the submissions state. This task received on average 1 submission out of an average of 23 registrations. This increased the failure prediction up to 63% at time 26, before peaking at 78% failure prediction. Also, 25 times out of 30 runs the task was failed which provides 83% failure.

Interestingly applying this policy will result in a lower level of diversity among participants in terms of ranking belt. During this test, on average, the task attracted 50% of Gray agents, 45% of Green agents and only 5% of Blue agents. Also, task submission was made by Gray and Green agents only.

#### 7.5.1.4 Testing Policy 4: 90% Degree of Task Similarity

When the project manager chose to follow policy 4, the task faced the challenge of very slow agent attraction. However, the failure prediction starts with almost 0% before increasing to 52% in day 24. The task faced two jumps in attracting agents. The first happened at day 4 which raised the failure prediction to 25% from 14%. The second happened on day 16 that pushed failure prediction to 43%. This task received an average of 2 submissions out of 32 registrations. The first submission happened on day 25 and the second one on day 28. Task failure prediction in the submission state increased to an average of 78% in day 30. The result of simulation provides 23 failed tasks of 30 runs which represents 77% task failure.

However, while applying higher level of task similarity did not help with lower failure prediction, it attracted better diversity of agents in terms of ranking belt. Simulation results show that 45% of participants are Gray agents, 27% are Green agents, 18% are Yellow agents and 9% are Blue agents. However, only Gray and Green agents made submissions at an equal rate.

#### 7.5.1.5 Discussion and Finding

It seems that submissions ratio decreases by decreasing the level of task openness in the pool of open tasks (i.e. increasing degree of task similarity). Interestingly, tasks with similarity level of 70% and 60% could attract higher reliable agents in terms of making a submission. Moreover, higher similarity level among tasks resulted in higher chance of task failure. This fact can be attributed due to group of similarly attracted agents in terms of skill set to the task. This result confirms the fact that agents are more interested in lower level of switching concepts among tasks. Also, higher level of similarity among available tasks in the pool of open tasks may provide higher chance of cheap talk [35] happening to a particular task due to the greater number of available options for active agents.

Deeper investigation showed that lowest and highest degree of similarity leads to higher diversity in terms of agents' belt among registrants while middle level degree of task similarity attracts lower diversity and those in lower rated belts. One reason can be lower level of skill set and experience among lower belt agents which makes them to more interested in working on similar tasks [38].

We investigated task 8 of the motivation example with the findings of scenario 1. On the day task 8 was posted in the platform the average similarity among pool of open tasks was 75%, figure 7.5-2. The average similarity of open tasks on April 1st, 5th and 6th is under 70% and under 60% on April 4th.

According to findings of scenario 1, if task 8 was posted with 2 days delay, on April 4th, failure prediction would only drop from 73% to 60%. Interestingly, when the task was successfully completed as task 18, the average task similarity in the pool of open tasks was 54%. This fact could have raised the chance of success to 13% and as a result would have provided 28 days reduction in the project schedule.

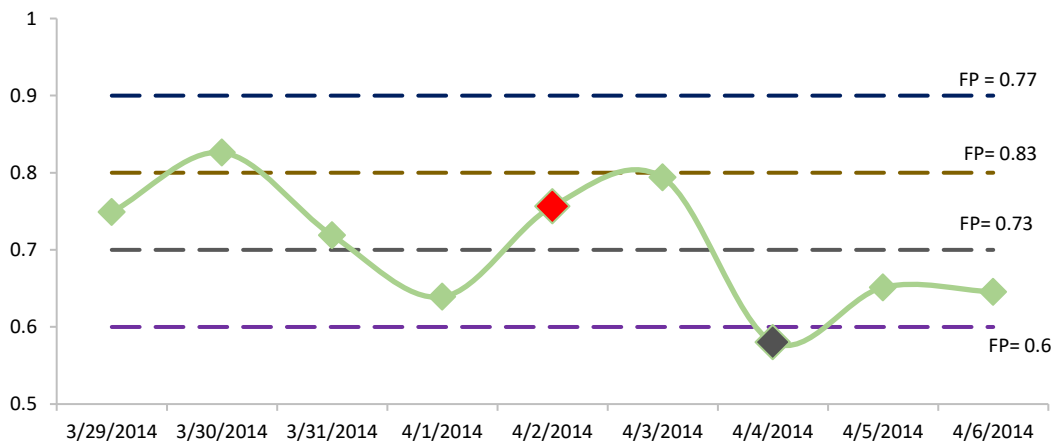

*Figure 7.5-2: Impact of Scenario 1 on task 8*

**_Finding:_** higher level of openness in the pool of tasks provides lower chance of task failure for the crowdsourced Project.

## 7.5.2    *Scenario2, Agents' Diversity*

Reliability in receiving a submission for a task by a registered agent is one of the main factors for measuring the success of the tasks. However, the reliability factor is measured based on number of submissions by an agent and is not related to the quality of the task submission. It is expected that higher rated agents have higher reliability in making a submission in general.

*Table 7-6: Summery of Scenario 2*

| Scenario 2 | | TP1 | TP2 | TP3 | TP4 |
|---|---|---|---|---|---|
| **Task Status** | Fail | 24 | 18 | 22 | 26 |
| | Success | 6 | 12 | 8 | 4 |
| | Failure Prediction | 80% | 60% | 73% | 87% |
| **Agents' Participation** | Gray — Reg | | | | 44% |
| | Gray — Sub | | | | 18% |
| | Green — Reg | | | 53% | 29% |
| | Green — Sub | | | 50% | 45% |
| | Blue — Reg | | 62% | 26% | 17% |
| | Blue — Sub | | 67% | 50% | 36% |
| | Yellow — Reg | 82% | 31% | 16% | 7% |
| | Yellow — Sub | 80% | 33% | 0% | 0% |
| | Red — Reg | 18% | 8% | 5% | 2% |
| | Red — Sub | 20% | 0% | 0% | 0% |

Generally, agents not only attempt to register for a higher number of tasks than their ability to complete [88], but higher rated agents may employ cheap talk strategy to ease the competition level and guarantee their win [35]. In this scenario, we analyze the impact of restricting the availability of tasks to different agents based on their rating belt to address the cheap talk strategy. The simulation ran under 4 different availability policies, and each policy ran 30 times. Figure 7.5-3 presents failure prediction and submissions ratio under different testing policies, and table 7-6 summarized agents' participation in the different testing policies in scenario2.

### 7.5.2.1    Testing Policy 1: Red and Yellow

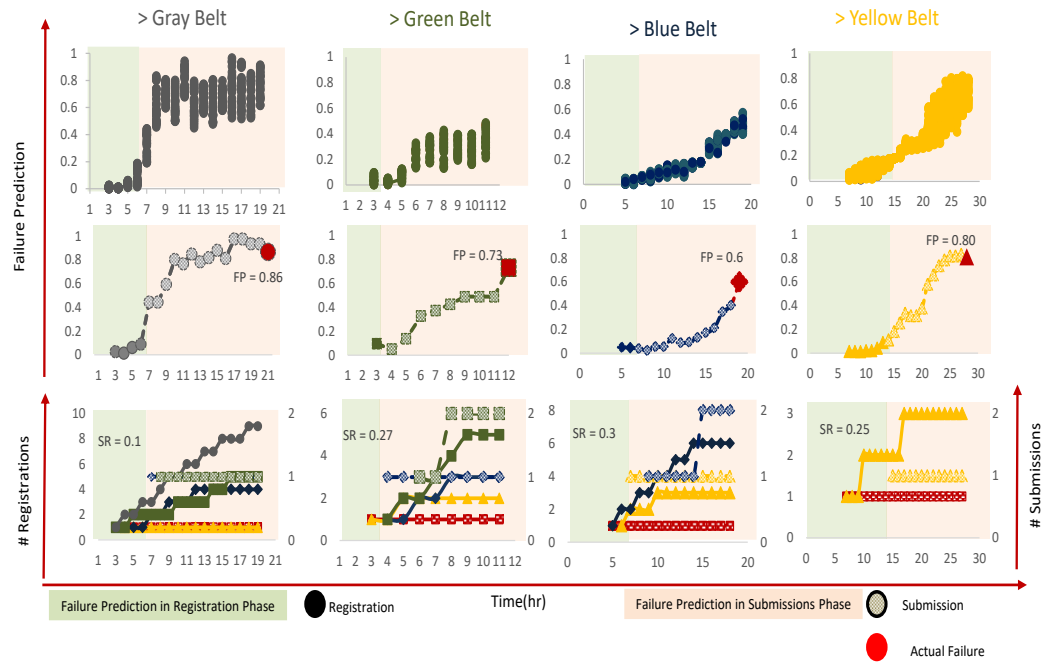

*Figure 7.5-3: Scenario 2, Impact of Agents' Experience on Task Failure*

Under this policy project manager gave access to agents with rating belt more than 1500 to compete on the task. The first registration happened at time 7, with failure prediction ratio around 1% which subsequently rose to 11% at time 14. At time 15, the first submission occurred and increased the failure prediction to 18%. Task submissions deadline continued until time 27 and failure prediction increased up to 83%. On average, 4 agents were interested in this task and only one agent made a submission. Under this policy, the chance of receiving a non-qualifying submission is 80%. 82% of participants in this task were from Yellow agents and 18% from Red agents. Also, submission levels for both Yellow and Red agents are 80% and 20% respectively.

### 7.5.2.1   Testing Policy 2: Red, Yellow and Blue

Under this policy all agents with rating above 1200 points are allowed to compete on the task. The first agent arrived at time 5 after posting the task and task failure prediction was 5% at this time. It seems the second arrived agent is a more reliable one and reduced the failure prediction to 4%. The first submissions occurred at time 3 and reduced the task failure prediction to 3%. However, registration continued, and after this point only one more agent made a submission which increased the failure prediction up to 40%. On average, the task received 10 registrations and 3 submissions under this policy, and the average 60% chance of task failure. 62% of attracted agents were Blue, 31% were Yellow and only 8% were Red agents. Interestingly Red agents did not make any submission, while Yellow agents made 33% of submissions and the rest occurred by Blue agents.

### 7.5.2.2   *Testing Policy 3: All but Gray*

The project manager decided that the task is open to available agents with rating higher than 900points. Higher diversity provides closer competition from the start. At time 3, the failure prediction level is 9%. The first attempt to make a submission occurs at time 4 and decreases the failure prediction to 5%.  While registration was continuing, the last submission for the task was received at time 12 and increased the failure prediction up to 40%. Under this policy, the task could attract 12 registrants and a maximum of 4 submissions. There is a 73% chance that the task did not receive qualified submissions.

Under this policy 53% of registrations were made by Green agents, 26% by Blue agents, 16% by Yellow agents, and only 5% by Red agents. Against the expectation, higher ranked agents did not make any submission. 50% of submissions were made by Green agents while Blue agents made the other half.

### 7.5.2.3   *Testing Policy 4: All Welcome*

All available agents in the platform are welcome to compete on the posted task. Simulation results showed that the first agents arrived at time 3 and task failure prediction was at 2%at this time. The second arrived agent decreased the failure prediction for 1%; however, the third arrival resulted in increasing the failure prediction. At time 6, the first submission was made. While the registration was going on, no more submissions was made. This fact increased the task failure prediction up to 93%. On average, the task could attract 20 registrations and only received 2 submissions. The chance of receiving a non-qualifying submission under this policy is around 86%.

This policy welcomed a wide diversity of available agents in the platform as competitors on this task. The competition level contained 44% of Gray agents, 29% Green Agents, 17% Blue agents, 7% Yellow agents, and 2% red agents. Interestingly while Gray agents were the highest level of participants, they only made 18% of the submissions. Green agents made 45% of submissions and Blue agents made 36% of submissions. Yellow and Red agents did not submit for the tasks under this policy.

### 7.5.2.4  *Discussion and Finding*

It is expected that higher level of diversity in the task competition level in terms of agents' ranking and experience provides lower level of task failure and higher level of receiving submissions; however, the simulation results show that higher level of diversity among agents leads to a higher level of failure prediction per task. Interestingly, the lowest level of diversity will lead to the same result. Attracting middle level rated agents provides the lowest level of task failure prediction. Applying policies numbers 3 and 4 not only provide a higher level of task submission of 30% and 27% but also a lower failure prediction of 60% and 73% respectively. Empirical study confirms this finding [50]. Moreover, this result indicates that middle rated agents are more reliable in making a submission as was reported in empirical studies [88].

We tested the findings of scenario 2 on task 8 from the motivation example, shown in figure 16. As shown in figure 7.5-4, on the day task 8 was posted in the platform, the availability of middle level agents is less than 5%. The level of availability of middle level agents (blue and green) was considerably higher on April 1st, 3rd, 4th and 6th. Interestingly, according to empirical data, the platform hosted the highest level of middle level agents on April 3rd and 4th. It seems that if the task was posted in either of those days, it might have had a better chance of success.

Based on available data, when the task was to successfully complete as task 18, the average of middle level agents' availability is 7%. Therefore, if the task was posted on April 6<sup>th</sup>, the chances of failure would decrease from 86% to 73% and the project schedule would be 26 days shorter.

**_Finding:_** Attracting higher level of middle level experienced agents leads to a lower chance of task failure.

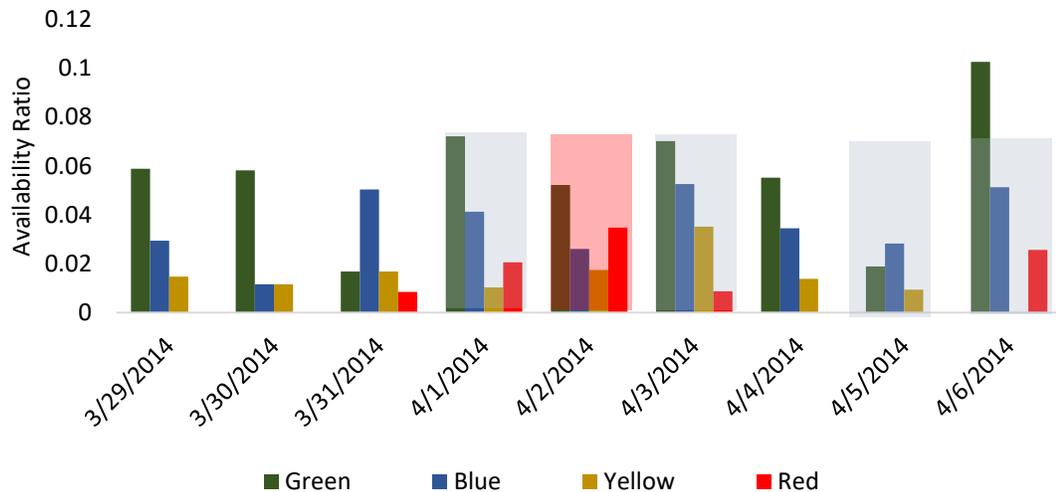

*Figure 7.5-4:  Impact of Scenario 2 on task 8*

## 7.6   Conclusion

It is assumed that task preparation is one of the most challengeable factors in the crowdsourcing development world. Crowdsourced tasks will be performed by an unknown group of workers, who will choose tasks based on personal and mostly unknown utility factors. This issue raises the concern about planning task execution and task failure ratio in the CSD platform. One of the main challenges is the posting time of tasks in the platform to increase the chance of receiving qualified submissions.

A hybrid simulation model was presented in this chapter. The result of the model presents that task openness in the platform in terms of task similarity and agents' experience are two of the most effective factors that should be considered when planning for a more effective task execution. In detail:

1) Pool of open tasks with higher degree of task similarity in the platform leads to higher level of failure prediction,

2) Attracting agents with middle level of experience to compete on the task helps to achieve lower level of failure prediction,

3) Agents with middle level experience provide higher level of task submissions ratio.

The proposed simulation model empowers managers to explore potential outcomes of open software development and impacts of different levels of uncertainties and task configuration strategies.

The user manual of the proposed hybrid simulation model is attached as the appendix to this dissertation.

# Chapter 8    Research Contributions and Future Direction

## 8.1   Generalization of the research

The general aim of this research was to provide a failure awareness framework in open software development via hybrid simulation modeling and simulation. To do so, we have chosen CSD as an example of OMSD, and the model has been created based on TopCoder flow. While the reported result in this dissertation is based on TopCoder data, according to available studies, workers' distribution and task arrival are following the same patterns in most of the crowdsourced platforms [38] [8]. Also, the reliability factor and similarity among tasks may vary in different CSD platforms based on adapted analysis algorithm to each individual platform. Since the degree of task similarity and reliability factor in the proposed hybrid simulation model is following random number, the result of the model can easily be extended to other available competitive CSD platforms.

The workflow of the collaborative CSD platforms are different. The quality of the tasks is the average of the quality of all the workers who are registering on the task. while in competitive platforms the quality of the task is based on the quality of the winner submission. The same concept applies to OSS as well. Therefore, the proposed model may successfully reflect the collaboration level, but requires a new failure prediction model in different task state based on collaboration level.

## 8.2   Research Contribution

To achieve the failure awareness framework, first an overall empirical study to understand the complexity of the system was provided. Then, a three-layered hybrid simulation model systems dynamic simulation, discrete event simulation and agent-based simulation was proposed. The proposed hybrid simulation model enabled managers to explore potential outcomes of open software development and impacts of different levels of uncertainties and task configuration strategies. The main contribution of this study is summarized as followed:

*1. An Empirical Evaluation of Current Task Completion Patterns*

In order to analyze parallel uploaded tasks in a crowd project, we categorized the available data to different cluster per project based on task attributes, workers' performance and task success rate. The analysis results conclude that:

- Crowdsourcing task scheduling follows traditional patterns including prototyping, component development, bug hunt, and assembly and coding.
- Budget phase distribution patterns do not follow traditional cost patterns.
- Higher number of uploading parallel tasks allows greater stability and lower failure rate for the project, however stability and failure rate do not follow the same pattern.
- Higher degree of parallelism would lead to higher demand for competing on tasks and shorter planning schedule to complete the project which consequently leads to better resource allocation and shorter project schedule planning.

*2. An Empirical Evaluation of Impact of Workers' Behavior Pattern*

Monetary prize plays a dual role in this trade-off. Higher monetary prizes will lead to greater motivation, and consequently, to a greater willingness to compete based on motivation. At the same time, the higher price represents a task of high complexity and/or workload, requires high skills and effort to win the competition and leads to a lesser willingness to compete. The overall willingness to compete represents a trade-off between the two variables. On the other hand, there are many uncontrollable distracting factors for the workers to fail to complete the task and make final submissions.

*3. An Empirical Evaluation of Team Performance pattern*

Leveraging crowd workforce in CSD has great potential to increase team elasticity and rapid delivery. The findings of available projects in our data illustrate that average Specific Absorption Rate (SAR) is 1.7. Hence, using CSD will help the project manager achieve a better flexible resource allocation rate in terms of diversity of skills, budget and schedule planning.

Based on the available empirical data and related research, we developed a set of research questions about the impact of worker performance with different skill and experience level. The main results of this study showed that on average:

- More than half of available workers respond to a task call on the first day of task arrival in the platform,
- Almost one-quarter of the workers who registered early will make submissions to tasks.

### 4. *An Empirical Evaluation of Reducing Schedule Acceleration Ratio*

An overall average of 1.82 schedule acceleration rate is observed through organizing mass parallel development in 4 software crowdsourcing projects. Such empirical evidences are beneficial to help exploring resourcing options and improve team elasticity in adaptive software development.

### 5. *A Hybrid Simulation Framework*

Task openness in the platform in terms of task similarity and agents' experience are two of the most effective factors that should be considered when planning for a more effective task execution. In details:

- Pool of open tasks with higher degree of task similarity in the platform leads to higher level of failure prediction. Pool of open tasks with lower level task similarity would provide higher chance of success compared with higher level of task similarity among available tasks.
- Attracting agents with middle level of experience to compete on the task helps to achieve lower level of failure prediction. In general, attracting higher number of Green and Blue belt agents will lead to higher chance of task success in the platform.
- Agents with middle level experience provide higher level of task submissions ratio. Green and Blue belt agents increase the chance of receiving task submissions.

## 8.3  Threats to Validity

Validating a software system is confined to the set of assumptions relating the measure of an internal property to an externally visible attribute [89]. In this dissertation, to create the hybrid simulation model, learning knowledge of three different empirical studies has been used. Each chosen attribute in the empirical studies are interdependent factors, however they have been studied independent of each other. This fact provides some threats to the validity of the presented researches.  In this part treats to validity of each empirical study will be discussed.

### 8.3.1  Workers behavior

In the first empirical study of this dissertation, the impact of monitory prize on workers behavior was studied. It is assumed that the factor of task complexity is reflected in monitory prize, and the crowd may directly perceive task complexity by reading task description. If a task is underpriced or overpriced, experienced workers will easily notice this, and it will have substantial impact o workers' behavior in task selection.

There are other factors that also influence worker's decision in task selection and completion like motivation patterns and trust network. These are not considered in this research. Also, in this research we only used correlation analysis to analyze the general relationship between monitory prize and workers' behavior. Moreover, it is observed that in TopCoder tasks the same worker may make more than 1 submission for a task corresponding to different revision. This implies that the number of registered workers who are really interested in making submissions could be even smaller. If this phenomenon also happens in used data set in this research, the result of submissions maybe slightly affected.

### 8.3.2  Team Performance Pattern

The second empirical study in this dissertation reported team performance patterns. This research only focused on workers' performance to new task calls based on workers experienced level. The impact of trust network among workers, and workers' individual decision on other available workers was relaxed. Also, influence of task owners' strategy on workers' performance was not considered in this part because of data shortage. Applying both trust network among workers as well as task owners' strategy may result on slightly different performance patterns and preferences. In this research we did not consider workers' multi-tasking factor and all the workers analyzed only based on their experience level not individual knowledge. What is reported in this study are overall characteristics on various worker rating belt category.

### 8.3.3  Task Completion

The last reported empirical study in this dissertation focused on task completion patterns. In order to perform this research, we chose 4 largest projects that we had in the available dataset. Also, to define the sequential and parallel tasks we followed start to finish scheduling pattern, due to lack of access to the high-level project information. The estimated schedule was based on a representative total effort by aggregating the crowd worker individual effort for submitting tasks. This assumption may not be valid because the team member relationship is competitive, but in-house team is collaborative. Also, in analyzing SAR, we did not consider management overhead in the Effort. This factor will increase the task scheduling and consequently larger SAR for both traditional and CSD methods.

Moreover, as it is reported in chapter 4, the numbers of members in each rating belt vary significantly. For example, the number of red workers is very few, and the number of gray workers which consists of almost 90% of the total members. While this confirms the typical developer resource pyramid, the data scarcity in the red group may bring limitations in the findings of all three empirical studies.

## 8.4  Future Research Direction

An interesting characteristic of simulation models is that it entices curiosity and enables both researchers and managers to create new ideas. Similarly, this model raises a lot of open questions to answer in future studies.

The first implication of this study was evaluating the reflection of real life behavior of open software development. Investigating such evaluation requires more studies considering an industrial experiment based on different possible task execution scenarios.

Another implication of this research was that task similarity influences the agents' trust network. However, investigating these effects requires more studies considering different decision-making scenarios by both agents and task owners to enhance our ability in understanding this phenomenon.

Furthermore, the researcher of this study suggests simplifying the model to present basic interactions in different task execution models as well as making more complex models, which include all required open software development features. Both of these directions could provide important skills for future research.

# *Chapter 9*     *Publications*

## *9.1   Published*

The reported results were submitted as following publications to different conferences:

1) Saremi, Razieh, and Ye Yang. "Empirical Analysis on Parallel Tasks in Crowdsourcing Software Development", 2015 30th IEEE/ACM International Conference on Automated Software Engineering Workshop (ASEW)

2) Saremi, Razieh, and Ye Yang. "Dynamic Simulation of Software Workers and Task Completion." CrowdSourcing in Software Engineering (CSI-SE), 2015 IEEE/ACM 2nd International Workshop on. IEEE.

3) Yang, Ye, and Razieh Saremi. "Award vs. Worker Behaviors in Competitive Crowdsourcing Tasks." Empirical Software Engineering and Measurement (ESEM), 2015 ACM/IEEE International Symposium on. IEEE.

4) Saremi, Razieh, Ye Yang, Gunther Ruhe, David Messinger, "Leveraging crowdsourcing for team elasticity: an empirical evaluation at TopCoder ", ICSE 2017 SEIP.

5) Ye Yang, Muhammad Rezaul Karim, Razieh Saremi, Guenther Ruhe, "Who Should Take This Task? Dynamic Decision Support for Crowd Workers", 2016 10th ACM/IEEE International Symposium on Empirical Software Engineering and Measurement

6) Razieh Saremi, "A Hybrid Simulation Model for Crowdsourced Software Development", 2018, Proceedings of the 5th International Workshop on Crowd Sourcing in Software Engineering

## *9.2   In Process*

1) Razieh Saremi,Ye Yang, Gregg Vesonder, He Zhang, Guenther Ruhe "Task Failure Awareness: A Hybrid Simulation Model for Crowdsourced Software Development"

# *Appendix*

## Hybrid Simulation Model User Manual

❖ System Overview

The hybrid simulation model is an intelligent aid for project managers:

- A software system based on Anylogic simulation tool
- Graphical scheduler system for failure notification
- System category:
  - provides failure awareness prediction to investigate the task execution model in open software development platforms
- Operational status:
  - Operational

❖ Getting Started

The hybrid model is designed based on three sub models, Platform, Task Completion, and Agent Decision. This part explains the features each sub-model provides.

- Platform:

  - Agent Arrival:
    Agent arrival in the platform follows an event. The event rate follows Poisson distribution. Each agent is randomly associated to a specific experience level based on Topcoder's worker experience metric. There is also a random reliability level assigned to the workers as soon as they arrive in the platform.
    The mean, max and min of the distribution can be manually updated based on different platform situations.

− Task Similarity:

According to empirical study, task similarity follows uniform distribution with mean of 0.33. In this model we provide the option of changing the mean of task similarity based on the chosen execution strategy with minimum of 0 and maximum of 1.

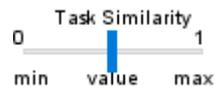

- Task Completion:

− Task schedule:

Task execution in the platform follows the Project scheduling plan provided by the task owner/ project manager. The specific schedule uploads in the Arrival Schedule which will be used for task execution.

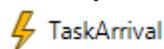

**ArrivalSchedule - Schedule**

| Name: | ArrivalSchedule | ☑ Show name | ☐ Ignore |
|---|---|---|---|
| Visible: | ◉ yes | | |

▼ **Data**

| Type: | real |
|---|---|

The schedule defines: ◉ Intervals (Start, End)  ◯ Moments

Duration type: ◯ Week  ◉ Days/Weeks  ◯ Custom (no calendar mapping)

| Repeat every: | 1 | days |
|---|---|---|

☐ Snap to:   9/27/2010

Default value: 0

☐ Loaded from database

| Start | End | Value |
|---|---|---|
| 3:00 AM | 6:00 AM | 0.5 |
| 6:00 AM | 7:00 AM | 1.8 |
| 7:00 AM | 8:00 AM | 4.0 |
| 8:00 AM | 11:00 AM | 7.5 |
| 11:00 AM | 3:00 PM | 5.2 |
| 3:00 PM | 5:00 PM | 3.1 |
| 5:00 PM | 6:00 PM | 1.3 |

▼ **Action**

− Task Arrival:

Task arrival in the model follows an event. The event is based on the task schedule calendar, starting from the day the entire project starts. Tasks will be uploaded automatically based on their time sequence.

⚡ TaskArrival

- Task Status:

  Task status is following the state chart based on Topcoder task flow at any given time.

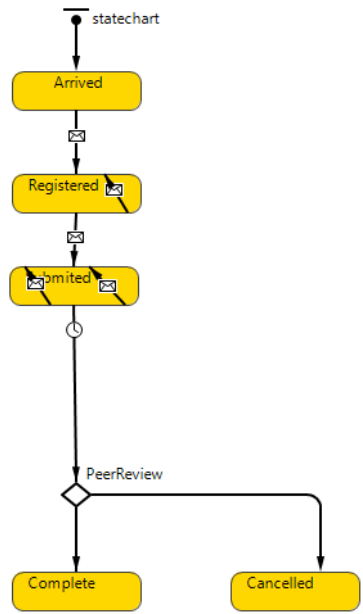

- Competition level:

  A collection list for both task registrants and task submitters is introduced.

  At any given time, the model provides the list of registrants and submitters for the executed task.

- Agent Decision:

    – Registration:

      Agent's decision to register for a task follows an event with specific rate per day. The rate of registration can be manually updated based on the platform strategy. According to empirical results, the average list of tasks per agent is 8, therefore, the maximum registration rate per agent can be 8.

      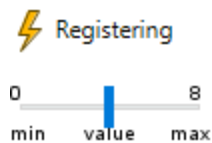

    – Submissions:

      Agent's decision for submission in this model follows a rate of 0.51 per day based on available empirical studies.

      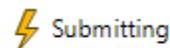

    – Participation Level:

      A collection list for all task registrations and task submissions and task winnings per agent is introduced.

      At any given time, the model provides the list of task registration and task submission and task winning for an available agent.

- Hybrid Simulation Outcome:

  The outcome of the model is based on two levels, the overall platform level and the task level.

    – Platform Level:

      Platform level in this model provides two sets of results:

      1- Distribution of task registration, task submission and task completion patterns in the platform.

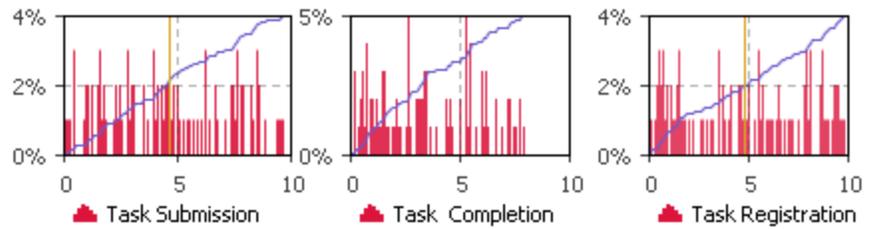

2- Arrival agents in the platform and their performance per experience belt at any given time.

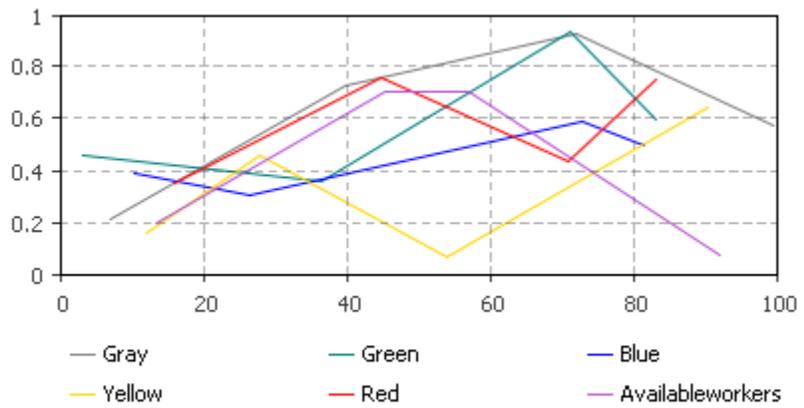

− Task Completion Level:

Task completion level in this model provides three sets of results:

1- Task failure prediction models in both registration and submission state at any given time per task.

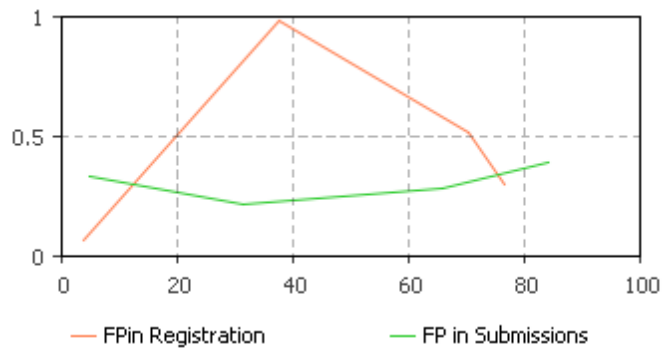

2- Number of registrants and submitters at any given time per task.

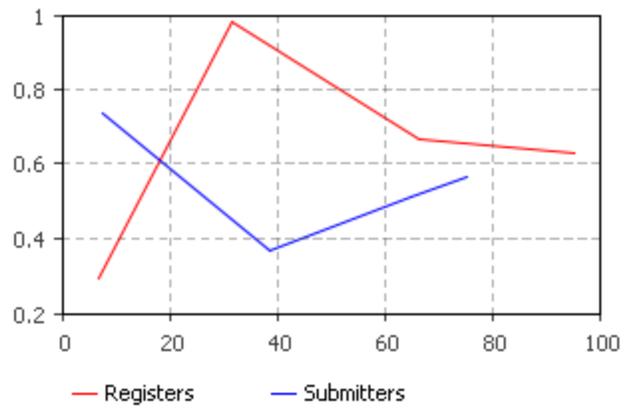

3- Agents' performance per agent experience per task.

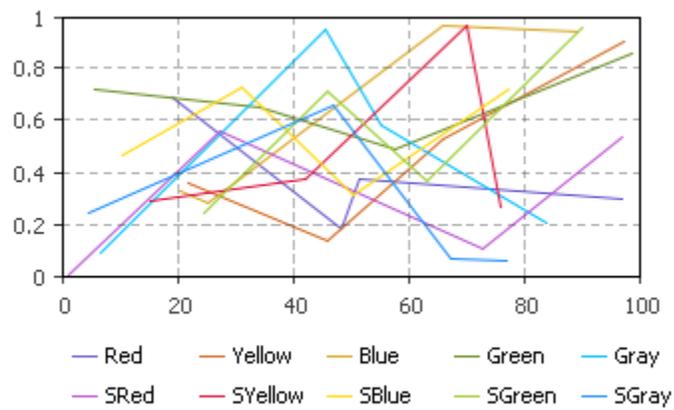

# *VITA*

## *NAME:*

Razieh Lotfalian Saremi

## *DATE AND PLACE OF BIRTH:*

September 16, 1984, Tehran, Iran

## *EDUCATION:*

- *Ph.D. in Systems Engineering*
  Stevens Institute of Technology, Hoboken, NJ      Aug 2014 - Aug 2018
- *Master of Engineering in Systems Engineering*
  Stevens Institute of Technology, Hoboken, NJ      Jan 2012 - May 2013
- *Bachelor of Science in Industrial and Systems Engineering*
  Azad University (Southern Tehran Campus), Tehran, Iran      Sep 2004 - Feb 2009

## *EMPLOYMENT HISTORY:*

- *STEVENS INSTITUTE OF TECHNOLOGY*
  Research Assistant – Software Analytics Lab      Sep 2014- May 2018
  Teaching Assistant – School of Systems and Enterprises      June 2015- May 2018

- *OPENASSEMBLY (START-UP)*
  SaaS Systems Engineer (Part Time)      May 2015 – June 2016

- *CELGENE*
  Supply Planner (Contractor through US-Tech Solution)      Feb 2014 – Sep 2014

- *MERCK & CO, INC*
  SAP Logistics Specialist (Contractor through RCI Technologies)      Feb 2013 – Dec 2013

- *HITT GROUP (START-UP)*
  Senior Planner and Supply Chain Engineer      Feb 2009 – Dec 2011

- *MCTOUGH PHARMACEUTICAL DISTRIBUTION CO.*
  Logistics and Distribution Planner      June 2004 – Feb 2009

## *HONOR:*

ASEM Outstanding Teaching Assistant Award, Stevens Institute of Technology      May 2018